\documentclass[11pt]{article}
\usepackage{a4}
\usepackage{amsmath}
\usepackage{amssymb}
\usepackage[dvips]{graphicx}
\usepackage{color}
\def\beq{\begin{equation}}
\def\eeq{\end{equation}}
\def\beqa{\begin{eqnarray}}
\def\eeqa{\end{eqnarray}}
\def\n{\nonumber \\}

\newcommand {\al}{\alpha}
\newcommand {\be}{\beta}

\begin{document}

\begin{flushright}
{SAGA-HE-281}\\
{KEK-TH-1705}
\end{flushright}
\vskip 0.5 truecm

\begin{center}
{\Large{\bf 
Evolution of vacuum fluctuations generated\\
during and before inflation
}}\\
\vskip 1cm

{\large Hajime Aoki$^{a}$, 
 Satoshi Iso$^{bc}$ 
 and Yasuhiro Sekino$^{b}$ 
}
\vskip 0.5cm

$^a${\it Department of Physics, Saga University, Saga 840-8502,
Japan  }\\
$^b${\it KEK Theory Center, 
High Energy Accelerator Research Organization (KEK), }\\
$^c$
{\it  Graduate University for Advanced Studies (SOKENDAI), \\
Ibaraki 305-0801, Japan}
\end{center}

\vskip 1cm
\begin{center}
\begin{bf}
Abstract
\end{bf}
\end{center}

We calculate the time evolution of the expectation value of the energy-momentum tensor 
for a minimally-coupled massless scalar field
in cosmological spacetimes, with an application to dark energy  in mind. 
We first study the evolution 
from inflation until the present,
fixing the Bunch-Davies initial condition. The energy density of
a quantum field evolves as
$\rho \sim  3(H_I H)^2 /32 \pi^2 $ in the matter-dominated (MD)  period,
where $H_I$ and $H$ are the Hubble parameters during inflation
and at each moment. Its equation of state, $w=\rho/p$, 
changes from a negative value to $w=1/3$ in the 
radiation-dominated period, and from $1/3$ to $w=0$ in the MD period.
We then consider possible effects of a Planckian universe, which may
have existed before inflation, by assuming there was
another inflation with the 
Hubble parameter $H_P (> H_I)$.  
In this case, modes with wavelengths longer than the
current horizon radius are mainly amplified, and  
the energy density of a quantum field grows with time 
as $\rho \sim (a/a_0)(H_P H)^2/32$ in the MD period,
where $a$ and $a_0$ are the scale factors at each time and at present.
Hence, if $H_P$ is of the order of the Planck scale $M_P$, 
$\rho$  becomes comparable to 
the critical density $3(M_P H)^2$ at the present time.
The contribution to $\rho$ from the long wavelength fluctuations generated
before the ordinary inflation has $w=-1/3$ in the free field approximation.
We mention a possibility that interactions further amplify the 
energy density and change the equation of state.

\newpage
\section{Introduction}
\label{sec:introduction}   
\setcounter{footnote}{0}
\setcounter{equation}{0}

Our universe is well described by the so-called 
spatially-flat $\Lambda$CDM model. Energy density of our
universe is close to the critical density. 
According to the PLANCK 2013 results~\cite{PlanckCosm}, 
only 5.1\% of the energy density is attributed to
a known form of baryonic matter, while 26.8\% 
is attributed to cold dark matter (weakly interacting 
non-relativistic matter), and 68.3\% to
 dark energy (or the cosmological constant).
Explaining the origin of these unknown ingredients, 
dark matter and dark energy, is one of the 
biggest challenges in modern physics.

There are various proposals for dark matter, such as
supersymmetric particles, axions, and so on. Dark matter 
may well be one of these. However, the origin of
dark energy is totally unclear. Although its equation 
of state, $w(=p/\rho)=-1$, seems like that of vacuum energy 
of quantum fields, 
there is no reasonable explanation for its magnitude,
$\rho_{\rm{DE}}= 3 (M_P H_0)^{2} \Omega_\Lambda \sim (2.2{\rm meV})^4$ 
with meV=$10^{-3}$eV. 
Here, $M_P = (8 \pi G_N)^{-1/2} \sim 2.4\times 10^{30} {\rm meV}$ is 
the (reduced) Planck scale
and $H_0 \sim 1.4\times 10^{-30} {\rm meV}$ is the current 
Hubble parameter. $\rho_{\rm{DE}}$ is far smaller than 
the expected magnitude of the vacuum energy $m^4$ in a theory 
with an ultra-violet (UV) cutoff, $m$, with any reasonable 
choice for $m$. If we take $m$ to be $M_P$, $\rho_{\rm{DE}}$ 
is smaller than $m^4$ by more than 120 orders of magnitude. 
Even if we take $m$ to be the supersymmetry breaking scale, 
electroweak scale, or any other natural scale in high-energy 
physics, $\rho_{\rm{DE}}$ is still much smaller than $m^4$.  
This is the cosmological constant problem~\cite{WeinbergCC}. 
It may turn out that the solution to this problem 
is  given by the anthropic principle~\cite{WeinbergAP}, but 
attempts at a dynamical explanation of dark energy are
undoubtedly important. 

In this paper, we study some aspects of vacuum energy
of quantum fields in cosmological spacetimes, 
with an application to the cosmological constant problem 
in mind. 
Let us first note that the order of magnitude of 
the energy density of quantum fields is not necessarily given by the value of 
the UV cutoff.
Expectation values of the energy-momentum tensor
should be renormalized by subtracting the cutoff-dependent 
(divergent) term. There is a well-defined
method for regularization and renormalization, which
yields finite expectation values for energy-momentum tensors,
which are covariant and conserved~\cite{Birrell:1982ix}. 
The terms to be subtracted are a combination of
spacetime curvature tensors. Typical terms in the
renormalized expectation value (such as the terms
responsible for the Weyl anomaly) are of the order of
the background curvature. 

In particular, vacuum energy
for fluctuations in the de Sitter background with the Hubble 
parameter $H$ is of the order $H^4$ (for massless fields), 
and has $w=-1$~\cite{BunchDavies, Dowker}. 
Since our present universe is close to de Sitter space,
one may wonder if dark energy can be explained as 
vacuum energy in de Sitter with the current Hubble 
parameter $H_{0}$, but this does not seem to be plausible. 
Dark energy that we observe, $M_P^2H_{0}^{2}$, is much larger 
than the expected contribution from a single field, $H_{0}^4$. 

However, local curvature is not the only dimensionful quantity 
that affects the renormalized energy-momentum tensor. To compute
expectation values of fluctuations, we need to specify 
the vacuum state. This may depend on global properties 
of the geometry and the whole history of the universe;
thus, different scales might be
introduced in the problem. 

There is by now strong evidence~\cite{PlanckInf} that there has 
been a period of inflation in our past. The spacetime during 
inflation is nearly de Sitter space with the Hubble parameter 
$H_I$ being much larger than $H_0$. 
Inflation should have lasted long enough 
(e-foldings $N_{e} \gtrsim  60$) to solve the flatness and 
horizon problems. 
It would be reasonable to take the vacuum to be 
the Bunch-Davies vacuum~\cite{BunchDavies} 
for de Sitter space with $H_I$.  
The Bunch-Davies vacuum is the one obtained by the Euclidean prescription,
and reduces to the Minkowski vacuum in the short wavelength limit. 
Even if different initial states are taken, correlation functions
will be attracted to those taken with respect to the Bunch-Davies vacuum,
as shown quite generally \cite{HabibMottola}.  

Fluctuations of the massless scalar in the de Sitter background 
(in four spacetime dimensions) is of the order $H_I$, as 
is clear from dimensional analysis. Massless fluctuations 
are frozen (remain constant) outside the Hubble radius
(see, {\it e.g.}, \cite{Mukhanov:2005sc}); thus, infra-red (IR) modes
could have a large value in the universe after inflation. 
In fact, these fluctuations are considered to be the 
origin of the fluctuations of the cosmic microwave background 
(CMB) that is observed today~\cite{PlanckCosm, PlanckInf}. 

The purpose of this paper is to understand how the IR modes
of quantum fields could affect the energy density 
in the present universe. 
We will mostly consider a massless minimally-coupled scalar 
field, and study the time evolution of the energy-momentum 
tensor in detail. Our result will shed light on the effect
of almost massless and non-interacting fields, such as axions.
Our work is related to the studies of the fluctuations of 
the graviton or the inflaton~\cite{Abramo1, Abramo2, Abramo3}, but 
further modifications are necessary since
energy-momentum tensors for gravitons have different
tensor structures from those for scalars and those for 
the inflaton have 
contributions from the classical value of the field. 
We believe our work serves as a starting point for 
the study of the time evolution of those fields.\footnote{
There is a paper \cite{Brown:2009cy}, in which the authors claim that 
back reactions of gravitational waves produced during inflation 
can serve as a source for the acceleration of the present universe, 
by taking an approach somewhat different from ours.
}

We will find that the magnitude of the present energy-momentum
tensor of a massless field is of the order $H_I^{2} H_0^{2}$,
with a possible factor logarithmic in the scale factor. 
The contribution from each momentum mode $k$ can be 
renormalized separately before integration 
over $k$ is performed. The equation of state for the 
(un-renormalized)
contribution from each $k$ has $w>-1/3$: 
The IR mode $k\to 0$ has $w\to -1/3$; for larger
$k$, $w$ is larger, and the UV limit has 
$w\to 1/3$, which is the same as radiation. The equation of 
state may become $w<-1/3$ due to renormalization, but
the terms arising from renormalization will be of the order 
of the background curvature. In the present universe,
these will be of the order $H_0^4$, and are negligible.\footnote{
After submitting our paper, we became aware of a recent paper~\cite{Glavan:2013mra},
which has substantial overlap with the first part of our paper. 
In \cite{Glavan:2013mra}, the authors study vacuum fluctuations of a massless 
minimally coupled test field in a background that evolves from inflation to 
radiation domination to matter domination. They use the approximation of 
sharp transition between these eras, just like we do. The results in the 
first part of our paper (which is summarized in this paragraph) is completely 
consistent with the ones in \cite{Glavan:2013mra}.
}

The value $H_I^{2} H_0^{2}$ is smaller than dark energy 
(or total energy density) in our universe, since the scale 
of inflation is smaller than the Planck scale,  
$H_{I}/M_P<3.6 \times 10^{-5}$, as suggested by the observations
of the CMB~\cite{PlanckInf}. 
But, in this paper, we point out a possibility that the energy 
density of a quantum field
takes a larger value.\footnote{This is different from proposals 
based on secular growth of fluctuations in 
inflation~\cite{Kolb, Barausse}. We 
are not assuming the expectation value of energy density
in pure de Sitter grows with time.} Even though the e-folding of 
inflation has to be large enough to make the region inside 
our horizon smooth, it does not have to be infinite. 
Finiteness of the e-foldings means the long wavelength modes
have not been in causal contact during the inflationary
era. There is no reason to expect that those modes are in
the Bunch-Davies vacuum. 
It would not be too surprising if there are large 
fluctuations in the far IR. After all, the point of inflation 
is to push away inhomogeneities beyond our horizon. In particular,
in the context of eternal inflation~\cite{CDL, HM, Linde}, 
our universe is generally
surrounded by the region with a larger Hubble parameter where
there are large fluctuations.

It is difficult to know what happened before inflation.
In this paper, as an explicit example, we consider a double 
inflation model. We assume there was an inflation 
with the Hubble parameter $H_P$ of the order of the Planck scale $M_P$ 
(such as Starobinsky's inflation~\cite{Starobinsky}, 
for example), followed by either a radiation-dominated or
curvature-dominated transition period, before 
the usual inflation with $H_I$ starts. 
We fix the initial condition of the fields in the
Planckian inflation period by taking the Bunch-Davies vacuum
for de Sitter with the Hubble $H_P$, 
and study the time evolution afterwards. 
In this case, IR mode is enhanced to 
$H_P$. 

By computing the evolution of the energy-momentum tensor 
for the double inflation model, we find the present 
value of vacuum energy to be of the order of $H_{P}^{2} H_{0}^{2}$. 
One may worry that the fluctuations become so large that 
they contradict the observed value of CMB fluctuations.
This will depend on what field we are considering, and 
needs careful study. In this paper, we argue that it is 
possible to enhance the modes that have a longer wavelength 
than the scales that are observed in CMB, leaving shorter 
wavelength modes essentially unaffected. 

The message of this paper is that vacuum energy of the order of
the energy density of our present universe may arise, 
due to the enhancement of the IR fluctuations generated
in the very early universe.
In our analysis of free fields, we were able to
obtain only $w>-1/3$, which cannot drive acceleration. 
However, we should mention that the free field approximation 
is not likely to be valid in the far IR where large fluctuations
generated before the ordinary inflation exist. 
At the end of the paper, we will mention possible directions for 
future study to take interactions into account.

The paper is organized as follows. 
We review the basics of quantization 
of scalar fields in curved spacetimes in Section~\ref{sec:review}.
After specifying
the cosmic history of background  geometry in Section~\ref{sec:history},
we obtain the wave function of a massless minimally-coupled scalar field,
with the initial condition fixed in the inflationary era 
in Section~\ref{sec:wavefunc}. 
Then we calculate the energy-momentum tensor, paying special
attention to the contributions from the IR modes 
in Section~\ref{sec:emt}.
We explain the prescription for treating the UV divergence,
and mention subtleties associated with a  physical interpretation 
of cutoff depenent terms, and present the renormalized 
energy-momentum tensor in Section~\ref{sec:renormalization}.
We consider time evolution of the energy-momentum tensor 
from the inflationary era until the present 
in Section~\ref{sec:timeevo}.
In Section~\ref{sec:double}, 
we consider a double inflation model, and discuss the effects 
of a period that may have preceded the ordinary inflation.
In section \ref{sec8.2}, we summarize  time evolution of 
energy densities generated in the inflation period
and in the pre-inflation period. They are shown in Figure~\ref{fig:timeevorhos}.
Section~\ref{sec:discussion} is devoted to conclusions and discussion.
In Appendix~\ref{sec:chasmmodel}, we consider a double inflation model with a different 
intermediate stage.
In Appendix~\ref{sec:IRbhvwf}, we investigate IR behaviors of the wave functions.
 
\section{Scalar field in curved spacetimes}
\label{sec:review}   
\setcounter{equation}{0}

In this section we briefly review some basics
of the scalar field on curved spacetimes (see, for instance,  ref.~\cite{Birrell:1982ix}).

A scalar field $\phi$ with a mass $m$ and a coupling $\xi$ to the scalar curvature $R$
in $n$-dimensional spacetime is described by the action
\beq
\int d^n x~(-g)^{1/2}\frac{1}{2}
\left[g^{\mu\nu}\partial_\mu \phi \partial_\nu \phi-(m^2+\xi R) \phi^2\right] \ ,
\eeq
which gives the equation of motion
\beq
\left[\Box+m^2+\xi R \right] \phi =0 \ .
\label{eomphi}
\eeq
The energy momentum tensor is given by
\beqa
T_{\mu\nu}&=&(1-2\xi)\phi_{,\mu}\phi_{,\nu} 
+(2\xi-\frac{1}{2})g_{\mu\nu}g^{\rho\sigma}\phi_{,\rho}\phi_{,\sigma} 
-2\xi \phi_{;\mu\nu} \phi\n 
&&+\frac{2}{n} \xi g_{\mu\nu}\phi\Box\phi 
-\xi\left[R_{\mu\nu}-\frac{1}{2}R g_{\mu\nu}+\frac{2(n-1)}{n}\xi R g_{\mu\nu} \right]\phi^2 \n
&&+2\left[\frac{1}{4}-(1-\frac{1}{n})\xi\right] m^2 g_{\mu\nu} \phi^2 \ ,
\label{emtgen}
\eeqa
where $\phi_{,\mu}=\partial_\mu \phi$ and $\phi_{;\mu\nu}=\nabla_\nu \partial_\mu \phi$.
The conformally coupled scalar  is described by $\xi=(n-2)/(4(n-1))$.
In this paper we will study the minimally coupled scalar with $\xi=0$. 

For background geometries, we consider Robertson-Walker spacetimes, 
which enjoy homogeneous and isotropic spaces,
with the metric
\beq
ds^2 = a(\eta)^2\left[d\eta^2 -(dx^i)^2\right] \ .
\label{RWmetric}
\eeq
$a(\eta)$ is the scale factor, and $\eta$ and $x^i$ are the conformal time and spacial coordinates.
An explicit form of $a(\eta)$ will be specified in
Section~\ref{sec:history}.

For quantizing the field, one  expands the field as
\beq
\phi(\eta,x^i) = \int \frac{d^{n-1}k}{(2 \pi)^{n-1}}
\left[a_{\bf k} u_{\bf k}(\eta) + a_{-{\bf k}}^\dagger u_{-{\bf k}}(\eta)^*\right]e^{i{\bf k}\cdot {\bf x}} \ ,
\label{phiexpau}
\eeq
where the mode functions
$u_{\bf k}(\eta)$ with the comoving momentum ${\bf k}$ 
are the solutions of the equation of motion (\ref{eomphi}), and are
chosen to asymptote to positive-frequency modes in the remote past.
A vacuum is then defined by $a_{\bf k} |0\rangle =0$.
The vacuum $|0 \rangle$, which is an in-state, evolves as $\eta$ increases, and
if an adiabatic condition is  broken the state gets excited above an adiabatic ground
state at each moment, $\eta$.

In the Robertson-Walker spacetime (\ref{RWmetric}),
the wave equation (\ref{eomphi}) is written as
\beq
\left[\partial_\eta^2 +k^2 
+a^2\left(m^2+\left(\xi-\frac{n-2}{4(n-1)}\right)R\right)\right]\chi_{\bf k} =0 \ ,
\label{eomchi}
\eeq
where $k=\sqrt{{\bf k^2}}$ and 
\beq
u_{\bf k} = a^{(2-n)/2} \chi_{\bf k} \ .
\label{uchigen}
\eeq

The expectation value of the energy momentum tensor in the state $| 0\rangle$
is given by
\beq
\langle 0 | T_{\mu\nu}(\eta,x^i) |0 \rangle_{\rm un-ren}
=\int \frac{d^{n-1}k}{(2 \pi)^{n-1}}~{\cal D}_{\mu\nu}~ u_{\bf k}(\eta) u_{\bf k}(\eta)^* \ .
\label{evemt}
\eeq
The right-hand side is obtained by inserting (\ref{phiexpau}) 
into (\ref{emtgen}),
and  ${\cal D}_{\mu\nu}$ represents the differential operator
that acts on  $\phi^2$ in (\ref{emtgen}). The subscript `un-ren'
indicates that the UV divergence has not been subtracted yet. Regularization
and renormalization will be discussed in Section 6.
Note  that (\ref{evemt}) is independent of the space coordinates $x^i$,
due to the spacial homogeneity of the Robertson-Walker spacetime.

\section{The background geometry}
\label{sec:history}   
\setcounter{equation}{0}

Our universe  is well approximated by
the Robertson-Walker spacetime (\ref{RWmetric}) in four spacetime
dimensions, $n=4$. It experienced 
 the inflation, radiation-dominated (RD), and matter-dominated (MD) periods.
 We describe the three stages of the cosmic history
  by the following scale factor $a(\eta)$:
\beq
a(\eta)=\left\{ \begin{array}{lll}
a_{\rm Inf}(\eta)=-\frac{1}{H_I \eta} &(-\infty<\eta<\eta_1<0) &(\mbox{Inflation})  \\
a_{\rm RD}(\eta)=\al \eta &(0<\eta_2<\eta<\eta_3) &(\mbox{RD})  \\
a_{\rm MD}(\eta)=\be \eta^2 &(\eta_4<\eta<\eta_0) &(\mbox{MD}) 
\end{array} \right. \ ,
\label{a_caldera}
\eeq  
which is specified by the eight parameters 
$(H_I, \alpha,\beta, \eta_1, \eta_2,\eta_3,\eta_4, \eta_0)$.
$H_I$ is the Hubble parameter in the inflation period.
We show below that both $a$ and $a'=\partial_\eta a$ must be continuous
at the boundaries of the inflation-RD and the RD-MD periods. 
Then only four of the eight parameters
are independent. As the four independent parameters, we will use
$(H_I, a_0, H_0, z_{\rm eq})$ where  $a_0$ and  $H_0$ are
the present scale factor\footnote{We could set $a_0=1$, 
but we do not fix the value here.} 
and the Hubble parameter, respectively, and 
$z_{\rm eq}$ is the red-shift factor at the matter-radiation equality. 
All the eight parameters are written in terms of them.

The continuity condition of $a'=\partial_\eta a$ 
can be easily understood as follows.
The scale factor satisfies the Friedmann equation
\beqa
&&H^2=\frac{1}{3 M_p^2}\rho \ , \label{friedmann} \\
&&\rho'+3aH(\rho+p)=0 \ , \label{conservationlaw}
\eeqa
where $\rho$ and $p$ are energy and pressure densities.
The second equation demands that $\rho$ should be continuous unless 
the second term has a singularity.
Then, according to the first equation, $H=a'/a^2$ is continuous
and so is $a'$ as well as $a$. 

The continuity conditions between the inflation and RD periods are given by
\beqa
a_{\rm Inf}(\eta_1)=a_{\rm RD}(\eta_2)&:& -\frac{1}{H_I \eta_1}=\al \eta_2 \ , \label{macoa1a2}\\
a'_{\rm Inf}(\eta_1)=a'_{\rm RD}(\eta_2)&:& \frac{1}{H_I \eta_1^2}=\al \ . \label{macoa1pa2p}
\eeqa
They give the relation
\beqa
\eta_1&=&-\eta_2 \ . \label{eta1eta2} 
\eeqa
Similarly, the condition between the RD and MD periods is given by
\beqa
a_{\rm RD}(\eta_3)=a_{\rm MD}(\eta_4)&:& \al \eta_3 = \be\eta_4^2 \ , \label{macoa3a4}\\
a'_{\rm RD}(\eta_3)=a'_{\rm MD}(\eta_4)&:& \al = 2\be \eta_4 \ , \label{macoa3pa4p}
\eeqa
which lead to 
\beqa
\eta_3&=&\frac{1}{2}\eta_4  \ . \label{eta3eta4} 
\eeqa

Now we determine $(a_0,H_0,z_{\rm eq})$. 
The third equation of (\ref{a_caldera}) gives 
\beqa
a_0&=&\be \eta_0^2 \ , \label{a0Beta} \\
H_0&=&\frac{2}{\be\eta_0^3} \ , \label{H0Beta} 
\eeqa
and we can write $\eta_0$ and $\be$ in terms of $a_0$ and $H_0$ as
\beqa
\eta_0 &=& 2H_0^{-1} a_0^{-1} \ , \label{eta0H0a0}\\
\be &=&\frac{1}{4} H_0^2 a_0^3 \ . \label{BH0a0}
\eeqa
Also, the same equation gives 
\beq
\frac{\eta_0}{\eta_4} = \sqrt{\frac{a_0}{a_4}}=\sqrt{1+z_{\rm eq}} \ 
\label{eta0eta4}
\eeq
and $\eta_4$ is written in terms of $z_{\rm eq}$ and $\eta_0$ (hence, $a_0, H_0$).
The other parameters can be similarly solved in terms of the four parameters.
Here, we note a relation:
\beq
\left(\frac{\eta_3}{\eta_2}\right)^2 \left(\frac{\eta_0}{\eta_4}\right)^3
= \frac{H_I}{H_0} \ . \label{eta3204HIH0} 
\eeq
It can be proved straightforwardly from the above equations,
and understood from the evolution of the Hubble parameters,
$H \propto \eta^{-2}$ and $H \propto \eta^{-3}$ in the RD and MD periods, respectively,
which are obtained by $H=a'/a^2$ and (\ref{a_caldera}).

Finally, we estimate the numerical values of various parameters.
$a_0 \eta_0$ is determined by (\ref{eta0H0a0}) with the use of the 
present Hubble:
\beq
H_0 \sim 67 {\rm km}~{\rm s}^{-1}{\rm Mpc}^{-1} \sim 1.4\times10^{-30}{\rm meV} \ .
\label{H0obsdata}
\eeq
Note that, as seen from (\ref{RWmetric}), definition of $\eta$ has a rescaling ambiguity that can be 
absorbed into $a$, and only the combination of $\eta$ and $a$ has a physical meaning.
$\eta_4$, and hence $\eta_3=\eta_4/2$,  is determined by  (\ref{eta0eta4}) as
\beq
\frac{\eta_4}{\eta_0} \sim \frac{1}{58} \sim 1.7\times 10^{-2} \ ,
\label{e4/e0obs}
\eeq
where we used
\beq
z_{\rm eq} \sim 3.4 \times 10^3 .
\eeq
Similarly, $\eta_2$, and hence $\eta_1=-\eta_2$,  is determined by  (\ref{eta3204HIH0}) as
\beq
\frac{\eta_2}{\eta_0} 
=\frac{\eta_3}{\eta_4}\sqrt{\frac{\eta_0}{\eta_4}\frac{H_0}{H_I}}
>
 \frac{1}{2}\sqrt{58\frac{1.4\times10^{-30} {\rm meV}}{8.8\times10^{25}{\rm meV}}}
\sim 4.8\times10^{-28} \ ,
\label{e2/e0obs}
\eeq
where we have  used the constraint from the CMB fluctuations:
\beq
H_I < 3.6\times 10^{-5} M_P \sim 8.8\times 10^{25}{\rm meV} \ .
\eeq

As a final comment, we note that the higher derivatives of $a$ with respect to $\eta$
are not continuous. Consequently, we will see that
the Bogoliubov coefficients  have a long UV tail as a function of the momentum $k$.
Such a long tail is an artifact of the rapid change of the scale factor and 
can be removed by smoothing the connections between the stages.
It will be discussed in Section~\ref{sec:renormalization}.

\section{Time evolution of wave functions}
\label{sec:wavefunc}   
\setcounter{equation}{0}

\begin{figure}
\begin{center}
\includegraphics[height=5.5cm]{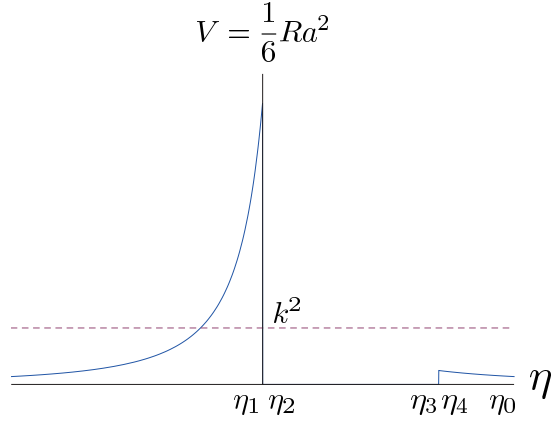}
\end{center}
\caption{Potential for the wave equation (\ref{waveeq}) with $m=0$.
The solid line depicts the potential (\ref{potIRM}), while the
dashed line represents the right-hand side of  (\ref{waveeq}).
$\eta_1$ to $\eta_4$ are the boundaries of the inflation, RD, and MD periods.
$\eta_0$ is the present time.
Their numerical values are specified at the end of Section~\ref{sec:history}.
In particular,
heights of the potential peaks are determined by
(\ref{e4/e0obs}) and (\ref{e2/e0obs}). 
If we normalize the height of 
the potential at the present $\eta_0$ as 1, 
the heights at $\eta=\eta_1$ and $\eta_4$ are given by $\lesssim 4.3 \times10^{54}$
and $3.4\times 10^3$, respectively. 
}
\label{fig:caldera_pot}
\end{figure}

We now solve the wave equation (\ref{eomchi})
to obtain the wave function $\chi_k(\eta)$  in the cosmic history (\ref{a_caldera}).
In this paper we consider the minimally coupled case, $\xi=0$,
in four dimensions, $n=4$. Then the wave equation (\ref{eomchi}) becomes
\beq
\left[ -\partial_\eta^2 + \frac{1}{6} R a^2 - m^2 a^2 \right] \chi_{\bf k}
= k^2 \chi_{\bf k}
\label{waveeq}
\eeq
with
\beq
\frac{1}{6}Ra^2 = \frac{a''}{a}
=\left\{\begin{array}{lll}
2/\eta^2 & ( \eta < -|\eta_1|)  & \mbox{(Inflation)} \\
0 &  (|\eta_1| <\eta <\eta_4/2 ) & \mbox{(RD)} \\
2/\eta^2 &  (\eta_4 <\eta<\eta_0 ) & \mbox{(MD)}
\end{array}
\right. \ .
\label{potIRM}
\eeq
The relation with $u_k$  (\ref{uchigen}) becomes
\beq
u_{\bf k}= \chi_{\bf k} /a \ .
\label{uchi4}
\eeq
Eq. (\ref{waveeq}) is interpreted as the time-independent
Schr\"{o}dinger equation for a one-dimensional quantum system 
with a potential $V=Ra^2/6-m^2 a^2$, by regarding $\eta$ as
the spatial position.
Figure~\ref{fig:caldera_pot} shows the potential  (\ref{potIRM})
for the $m=0$ case. Note that it is discontinuous at the boundaries of
the inflation, RD, and MD periods. 

In the massless case,
the solutions of (\ref{waveeq}) are given as
\beqa
\chi_{{\rm Inf}, {\bf k}} &=& \chi_{{\rm BD}, {\bf k}}
\  \label{chiInf}\\
\chi_{{\rm RD},{\bf k}} &=& A(k)~\chi_{{\rm PW}, {\bf k}}+B(k)~\chi_{{\rm PW},-{\bf k}}^*
 \ \label{chiRD}\\
\chi_{{\rm MD} ,{\bf k}} &=& C(k)~\chi_{{\rm BD} ,{\bf k}}+ D(k)~\chi_{{\rm BD},-{\bf k}}^* 
\  \label{chiMD}
\eeqa
in the inflation, RD, and MD periods, respectively, where the wave functions are
\beqa
\chi_{{\rm BD},{\bf k}} &=& \frac{1}{\sqrt{2k}}\left(1-\frac{i}{k \eta} \right) e^{-i k\eta}
 \ , \label{chiBD}\\
\chi_{{\rm PW} ,{\bf k}}&=& \frac{1}{\sqrt{2k}} e^{-i k\eta}  \ . \label{chiPW}
\eeqa
The wave function (\ref{chiInf}) contains only the positive frequency mode,
which corresponds to  taking the Bunch-Davies vacuum in the de Sitter spacetime. 
The constants $A$, $B$, $C$, and $D$ are easily determined by using
the junction conditions of the wave functions,
{\it i.e.}, the continuity of $\chi$ and $\chi'$.
They become
\beq
\begin{pmatrix} A(k) \cr B(k) \end{pmatrix}
=\begin{pmatrix} \left(1-\frac{i}{k \eta_1}-\frac{1}{2k^2 \eta_1^2}\right)e^{i k \eta_2} \cr
 \frac{1}{2k^2 \eta_1^2} e^{-i  k \eta_2} \end{pmatrix}
 e^{-i k \eta_1} \ ,
 \label{AB}
 \eeq
 \beq
\begin{pmatrix} C (k)\cr D (k)\end{pmatrix}
=\begin{pmatrix}\left(1+\frac{i}{k \eta_4}-\frac{1}{2k^2 \eta_4^2}\right)e^{ik (\eta_4-\eta_3)}
&-\frac{1}{2k^2 \eta_4^2}e^{i k (\eta_4+\eta_3)}\cr
-\frac{1}{2k^2 \eta_4^2}e^{-i k (\eta_4+\eta_3)}
&\left(1-\frac{i}{k \eta_4}-\frac{1}{2k^2 \eta_4^2}\right)e^{-ik(\eta_4-\eta_3)} \end{pmatrix}
\begin{pmatrix} A(k) \cr B(k) \end{pmatrix} \ .
\label{ABCD} 
\eeq

These constants give the coefficients of the  Bogoliubov  transformations,
and the Bunch-Davies vacuum is interpreted as an excited state on the adiabatic vacuum defined
in  the RD and MD periods, respectively.\footnote{In a recent
paper~\cite{Padmanabhan}, particle content and the degree of 
classicality were studied in a background similar but slightly
different from ours (de Sitter in inflation, followed by 
radiation dominance, and a late time de Sitter).}
Let us look at the particle creation in the RD period. 
We expand $\phi(\eta,x^i)$ in terms of the adiabatic wave functions $\chi_{\rm PW}$ 
at each moment in the RD period as
\beq
\phi(\eta,x^i) = \frac{1}{a} 
\int \frac{d^{3}k}{(2 \pi)^{3}}
\left[a_{\bf k}^{\rm ad} \chi_{{\rm PW}}(\eta,k) + a_{-{\bf k}}^{{\rm ad} \dagger} 
\chi_{{\rm PW}}(\eta,  k)^*\right]e^{i{\bf k}\cdot {\bf x}} \ .
\label{phiexpad}
\eeq
The adiabatic vacuum is defined by $a_{\bf k}^{\rm ad}|0_{\rm ad} \rangle =0$.
If we start from the Bunch-Davies vacuum defined by $a_{\bf k}|0 \rangle=0$,
the state $|0\rangle$ evolves into a highly excited state on the adiabatic vacuum $|0_{\rm ad}\rangle$.
By using the solutions of the wave equation and comparing (\ref{phiexpau}) and (\ref{phiexpad}),
we have
\beqa
\begin{pmatrix} 
a_{\bf k}^{\rm ad} \cr a_{-{\bf k}}^{{\rm ad} \dagger}
\end{pmatrix}
&=&
\begin{pmatrix} A(k) &B^*(k)\cr
B(k) &A^*(k)
\end{pmatrix}
\begin{pmatrix} a_{\bf k} \cr a_{-{\bf k}}^\dagger \end{pmatrix} \ .
\label{Bog} 
\eeqa 
The coefficients $A(k),B(k)$ satisfy the relation $|A(k)|^2-|B(k)|^2=1$.
Furthermore, we can always make $A(k)$ real by redefining the phase of $a_{\bf k}$.
So the coefficients can be parametrized as 
\beq
A(k)=\cosh \theta_k ,  \ \ \  B(k)= - \sinh \theta_k e^{ i \phi_k} \ .
\eeq
The inverse of the transformation (\ref{Bog}) is generated by using the squeezing operator
\beq
G_k =a_{\bf k}^{\rm ad} a_{-{\bf k}}^{\rm ad} e^{i \phi_k} - a_{\bf k}^{{\rm ad} \dagger}
a_{-{\bf k}}^{{\rm ad} \dagger}  e^{-i \phi_k} \ ,
\eeq
as
\beqa
\begin{pmatrix} a_{\bf k} \cr a_{-{\bf k}}^\dagger \end{pmatrix}
&=&
e^{\theta_k G_k} 
\begin{pmatrix} 
a_{\bf k}^{\rm ad} \cr a_{-{\bf k}}^{{\rm ad} \dagger}
\end{pmatrix}
 e^{-\theta_k G_k} \ .
\label{Bog2} 
\eeqa 
The Bunch-Davies vacuum is then written as 
\beqa
|0 \rangle = \prod_k e^{\theta_k G_k} |0_{\rm ad} \rangle \ .
\eeqa
 Since the squeezing operator $G_k$ is bilinear in the creation operators, 
 the Bunch-Davies vacuum is a collection of excited states with multiple pairs of particles.
 Concretely, $|0 \rangle$ can be expanded as 
 \beqa
 |0 \rangle &=& \prod_k \frac{1}{\cosh \theta_k}  e^{- \tanh \theta_k e^{-i \phi_k} a_{\bf k}^{{\rm ad} \dagger}
a_{-{\bf k}}^{{\rm ad} \dagger}} |0_{\rm ad} \rangle \\
 &=& 
 \prod_k  \frac{1}{\cosh \theta_k} \sum_{n=0}^\infty (-\tanh \theta_k e^{i \phi_k})^n 
 | n \rangle_{k} \otimes |n \rangle_{-k} \ .
 \eeqa
The number of created particles is calculated as
\beqa
\langle 0| a_{\bf k}^{{\rm ad} \dagger} a_{\bf k}^{{\rm ad}} | 0\rangle = ( \sinh \theta_k)^2
= |B(k)|^2 = \frac{1}{4 k^4 \eta_1^4} \ .
\eeqa
Hence, it becomes very large for $ k \ll 1/ \eta_1$. 
The same calculation is  performed in the MD period, and the number of created 
particles is given by $|D(k)|^2$. 

Now, let us investigate the IR, {\it i.e.}, small-$k$, behavior of the wave functions.
When
$k |\eta| \ll 1$, (\ref{chiBD})  
behaves as $\chi_{\rm Inf} =\chi_{\rm BD} \sim k^{-3/2}$.
Terms of ${\cal O}(k^{-1/2})$ cancel 
and the next term starts with $k^{1/2}$.
By using (\ref{uchi4}) with (\ref{a_caldera}), 
the IR behavior of $u$ becomes
\beq
u_{\rm Inf}^{\rm IR} = \frac{i}{\sqrt{2}} H_I k^{-3/2} + {\cal O} (k^{1/2}) 
\label{u_IR_scaleinv}
\eeq
in the inflation period.
This IR behavior of the wave function is kept until the RD and MD periods.
It  generally holds that the leading part of the superhorizon modes of 
a massless field is time-independent, {\it i.e.}, frozen.
It gives the seeds of the CMB fluctuations.
(See, for instance, sections 7.3.2, 8.4, 
and 9.9 of ref. \cite{Mukhanov:2005sc}.)

Let us confirm the above IR behavior of the wave function. 
Since $A(k)$ and $B(k)$ in (\ref{AB})  are proportional to $k^{-2}$ in the IR region for $k \ll |\eta_1|^{-1}$,
 the wave function in the RD period (\ref{chiRD})  seems to behave 
$\chi_{\rm RD}(k) \sim k^{-5/2}$ for $k \ll \eta^{-1},  |\eta_1|^{-1}$.
Similarly, the behavior $C, D \sim k^{-4}$ indicates much more violent 
IR behavior, $\chi_{\rm MD} \sim k^{-11/2}$,
for $k \ll \eta^{-1}, |\eta_1|^{-1}, \eta_4^{-1}$. 
However, lots of cancellations occur and the IR behavior becomes milder.
Indeed,  in the RD period,
 (\ref{AB}) is rewritten as
\beq
\begin{pmatrix} A \cr B \end{pmatrix}
=\frac{-1}{2k^2 \eta_1^2}
\begin{pmatrix} 
e^{ik(\eta_1+\eta_2)} \left[1-e^{-2ik\eta_1}\sum_{n=3}^\infty \frac{1}{n!}(2ik\eta_1)^n \right] \cr
-e^{-ik(\eta_1+\eta_2)} \end{pmatrix}
 \ ,
\label{ABrew}
\eeq
where the terms with $k$ and the terms with $k^2$ cancel in the square bracket.
Then the wave function (\ref{chiRD}) behaves as
\beqa
\chi_{\rm RD}^{\rm IR} &=& 
\frac{1}{\sqrt{2k}}\left(\frac{2i}{2k^2\eta_1^2}\sin(k(\eta-\eta_1-\eta_2))
+{\cal O}(k\eta_1)\cdot e^{-ik(\eta-\eta_1-\eta_2)} \right) \ \nonumber \\
&=& \frac{i}{\sqrt{2}}\frac{\eta-\eta_1-\eta_2}{\eta_1^2}k^{-3/2}+ {\cal O}(k^{1/2}) 
\label{chiRDIR}
\eeqa
in the IR regions.
Note  that the terms with $k^{-1/2}$ cancel each other 
and the next terms start with $k^{1/2}$.
Moreover, by using $\eta_1=-\eta_2$ in (\ref{eta1eta2}), 
the leading term in (\ref{chiRDIR})
gives
\beq
u^{\rm IR}_{\rm RD} =\frac{\chi^{\rm IR}_{\rm RD}}{a_{\rm RD}} 
=\frac{i}{\sqrt{2}}\frac{\eta}{\eta_1^2} k^{-3/2}\frac{H_I \eta_1^2}{\eta}
=\frac{i}{\sqrt{2}}H_I k^{-3/2} \ ,
\label{uIRRD}
\eeq
where  we used (\ref{a_caldera}) and (\ref{macoa1pa2p}).
Hence, the IR behavior (\ref{u_IR_scaleinv}) is shown to be 
maintained in the RD period as well. 

As we show in Appendix \ref{sec:wfIRMD}, the same
IR behavior  holds in the MD period.

\section{Energy-momentum tensors}
\label{sec:emt}   
\setcounter{equation}{0}

For a minimally-coupled massless scalar in four dimensions, {\it i.e.}, $\xi=0$, $m=0$, and $n=4$, 
the energy density $\rho = \langle T^\eta_{\ \eta} \rangle$
and the pressure density $p=-\langle T^i_{\ i} \rangle$ ($i$ is not summed over)
are given from  (\ref{emtgen}) by
\beqa
\rho(\eta)^{\rm un-ren} 
&=& \frac{1}{a^2} \int \frac{d^{3}k}{(2 \pi)^{3}} \frac{1}{2}
\left[ |u'_k(\eta)|^2 + k^2 |u_k(\eta)|^2 \right] 
\label{rhogenu}\\
&=& \frac{1}{4 \pi^2 a^4} \int dk~k^2
\left[|\chi'|^2-\frac{a'}{a}\left(\chi^*\chi'+\chi'^*\chi\right)
+\left(\left(\frac{a'}{a}\right)^2+k^2\right)|\chi|^2 \right] \ ,\n
\label{rhogenchi}
\eeqa
\beqa
p(\eta)^{\rm un-ren} 
&=& \frac{1}{a^2} \int \frac{d^{3}k}{(2 \pi)^{3}} \frac{1}{2}
\left[ |u'_k(\eta)|^2 -\frac{1}{3} k^2 |u_k(\eta)|^2 \right] 
\label{pgenu}\\
&=& \frac{1}{4 \pi^2 a^4} \int dk~k^2
\left[|\chi'|^2-\frac{a'}{a}\left(\chi^*\chi'+\chi'^*\chi\right)
+\left(\left(\frac{a'}{a}\right)^2-\frac{1}{3}k^2\right)|\chi|^2 \right] \ .\n
\label{pgenchi}
\eeqa
Here, the expectation values are taken in the Bunch-Davies vacuum.
 The superscript `un-ren' means that the UV divergences
are not yet subtracted.

We first examine the contributions from the IR modes.
By substituting the IR behavior of the wave function (\ref{u_IR_scaleinv}),
which is kept until the RD and MD periods, into
(\ref{rhogenu}) and (\ref{pgenu}), one obtains
\beqa
\rho^{\rm IR} &=& \frac{H_I^2}{8\pi^2 a^2} \int_0 dk \left[k+{\cal O}(k^3)\right] \ , 
\label{rhoIR}\\
p^{\rm IR} &=& 
\frac{H_I^2}{8\pi^2 a^2} \int_0 dk \left[-\frac{1}{3} k+{\cal O}(k^3)\right] \ .
\label{pIR}
\eeqa
Since the IR wave function (\ref{u_IR_scaleinv}) is frozen and time-independent,
only the spacial derivative terms  $k^2 |u_k(\eta)|^2$  contribute to 
$\rho^{\rm IR}$ and $p^{\rm IR}$.\footnote{ 
The absence of  ${\cal O}(k^{-1/2})$ term in the IR behavior of wave function
(\ref{u_IR_scaleinv})
is important to assure the robustness of   (\ref{rhoIR}) and (\ref{pIR}). 
 If ${\cal O}(k^{-1/2})$ term existed, it would change the coefficient 
 of ${\cal O}(k)$ term in (\ref{rhoIR})  
through the $|u_k^{\prime}(\eta)|^2$ term in  (\ref{rhogenu}). 
 }
Due to the amplification of the wave function (\ref{u_IR_scaleinv}),
$\rho^{\rm IR}$ and $p^{\rm IR}$ are enhanced to the order of $H_I^2$.
The leading term in (\ref{rhoIR}) and (\ref{pIR}) gives 
$w^{\rm IR}=p^{\rm IR}/\rho^{\rm IR}=-1/3$.

The two-point correlation function of massless fields
receives a logarithmic IR growth:
\beqa
\lim_{x\to y} \langle 0| \phi(\eta, x) \phi(\eta,y) |0\rangle
&=&\lim_{x\to y} \int \frac{d^{3}k}{(2 \pi)^{3}} u_k(\eta) u_k(\eta)^* e^{ik \cdot (x-y)} \n
&\sim& \frac{H_I^2}{4\pi^2}  \int dk \frac{1}{k} \ .
\label{phiphi}
\eeqa
In contrast,  the energy and pressure densities, 
(\ref{rhoIR}) and  (\ref{pIR}), are IR finite since
they involve derivatives, and the IR divergences are canceled. 
This is due to the fact that for an exactly massless field, 
the constant part of the field (which gives rise to the logarithmic 
divergence) does not have a physical meaning.

However, if we take the massless limit of a massive field,
the mass term in the energy-momentum tensor
gives a non-vanishing contribution, since 
$\langle \phi^2 \rangle\sim H_I^4/m^2$. 
This can be computed by using the exact wave function for massive
fields, which is written in terms of Hankel 
functions in de Sitter space \cite{Birrell:1982ix, BunchDavies}, or can be seen as
follows. We compute 
\beq
\rho(\eta)_{\rm mass} 
= \int \frac{d^{3}k}{(2 \pi)^{3}} \frac{m^2}{2}
  |u_k(\eta)|^2 \ .
  \label{rhomass}
\eeq
By introducing a small mass,  
the potential in the inflation period changes from $2/\eta^2$ to $(2-(m/H_I)^2)/\eta^2$, and
the IR behavior of the massless wave function
(\ref{u_IR_scaleinv}) is modified to
\beq
u_{\rm Inf}^{\rm IR} (\eta) \sim \frac{i}{\sqrt{2}} H_I k^{-3/2} (k |\eta|)^{\frac{(m/H_I)^2}{3}} \ .
\label{IRbehavior}
\eeq
Inserting it into (\ref{rhomass}), we obtain
\beq
\rho(\eta)_{\rm mass} 
= \frac{m^2}{2} \int_0^{1/|\eta|} \frac{dk}{2 \pi^2}  \frac{H_I^2}{2 k} (k |\eta|)^{\frac{2(m/H_I)^2}{3}}
=\frac{3H_I^4 }{16 \pi^2} .
\label{rhomass2}
\eeq
Here, the integration is performed up to 
$k=1/|\eta|$, which corresponds to the horizon scale
$k_{\rm phys}=k/a_{\rm Inf}=H_I$, since 
the IR behavior \eqref{IRbehavior} is valid below this
scale.
The result \eqref{rhomass2} is independent of the mass. 
Hence, the massless limit of a massive theory gives an additional 
contribution to the expectation value of the energy-momentum tensor.
But we should note that,
in order to obtain the contribution, 
we implicitly assumed that there is no other
IR cutoff.  
If there exists a physical IR cutoff, $\Lambda_{\rm IR}$, 
the logarithmic IR divergence is automatically cured by $\Lambda_{\rm IR}$
and no such singular behavior with $1/m^2$ appears.
Such an IR cutoff may be given, {\it e.g.}, by the initial time 
of the inflation period. In that case, 
the expectation value of a 
massless scalar $\langle \phi(\eta, x)^2\rangle$
will be proportional to the physical time interval 
since the initial time~\cite{Linde1982, Starobinsky2, VilenkinFord,
Linde2}, and will 
not be infinite.\footnote{See \cite{Finelli1, Finelli2, Finelli3,
Marozzi} for recent studies on this linear growth of perturbation
in inflation and the effect of the initial time cutoff.}
Then $\rho_{\rm mass}$ vanishes in the  $m \rightarrow 0$ limit and
the massless limit of a massive theory gives the same answer 
as the purely massless theory. 
In the following analysis of the present paper,
 we assume an existence of such an IR physical cutoff.
A further possible contribution of $\rho_{\rm mass}$ to the 
vacuum energy will be investigated 
in a separate paper.

\section{Renormalized energy-momentum tensors}
\label{sec:renormalization}   
\setcounter{equation}{0}

The energy and pressure densities (\ref{rhoIR}) and  (\ref{pIR}) are UV divergent,
and must be regularized and renormalized.
This has been studied extensively in the past
(see, for instance, section 6 of ref. \cite{Birrell:1982ix}),
so we will keep the description brief and just present the result.

We first make the integral finite by using one of the regularization 
prescriptions, such as the dimensional regularization or the covariant 
point-splitting. We then perform renormalization by subtracting
the terms that can be absorbed by redefinitions of  
coupling constants in the gravity action that have 
the dimensions four (the cosmological constant), two (the Newton constant inverse),
and zero (the coefficients for curvature
tensor squared terms). In general, the renormalized 
energy-momentum tensor takes the form
\beq
\langle T_{\mu\nu} \rangle_{\rm ren}
= {\cal D}_{\mu\nu} \left[G - G^{(A)}\right] \ ,
\label{emtren}
\eeq
where ${\cal D}_{\mu\nu}$ stands for the derivative
operators in (\ref{evemt}).
$G$ is the two-point correlation function
obtained in the previous sections.   
$G^{(A)}$ is the subtraction term, which is the  
two-point function obtained using the first few orders of 
the DeWitt-Schwinger expansion~\cite{DeWitt}.
This is an expansion around the flat spacetime, and the expansion
depends only on the information of local geometry, namely, the subtraction terms
are written in terms of curvature tensors. 

Another way of obtaining subtraction terms, which has been shown
to be equivalent to the DeWitt-Schwinger expansion in various cases,
including the Friedmann-Robertson-Walker (FRW) universe, is the adiabatic
regularization~\cite{ParkerFulling,Bunch}.
In this scheme, $G^{(A)}$ is obtained by using 
the WKB expansion of the mode functions up to the adiabatic 
order four. With this prescription, the subtraction can be 
performed at each $k$ separately, and we can obtain 
finite integrals without the need of an explicit regularization. 
In this sense,
the adiabatic regularization is  a prescription for
subtraction, not a regularization method.

The subtraction term contains an IR divergent term for massless theories;
thus, the renormalized energy-momentum tensor should be
defined by taking the massless limit of a massive theory. 
The pieces that remain finite as a result of this procedure
are given, {\it e.g.}, in Eqs. (3.14) and (3.15) of \cite{Bunch}
for the general FRW universe. The Weyl anomaly arises from this procedure. 
However, since these contributions are smaller
than the terms that we are mainly interested in,
we will ignore these contributions
in the analyses of the paper.

Before starting the analysis of the renormalized energy-momentum
tensor in our background, we would like to  make a side remark about 
the UV divergent (cutoff-dependent) terms. 
We emphasize here that a physical interpretation of these
terms is a very subtle issue. 
For example, consider the quartically 
divergent term in the effective action, which is the leading 
divergence in four spacetime dimensions, 
\begin{equation}
 {\cal L}_{\rm eff, 4}=-\sqrt{-g}\; k_{\rm UV}^4
\label{Leff4}
\end{equation}
where $k_{\rm UV}$ is the UV cutoff for the momentum. This term  
is often interpreted as a contribution from quantum fields to
the cosmological constant. 

On the other hand, if one computes
$\langle T_{\mu\nu}\rangle$, the quartically divergent term 
has $w=1/3$, which is  the equation of state for radiations,
not the cosmological constant.
This is indeed expected from the fact that in the limit of
high momentum, 
there is no particle creation, and the field behaves as 
a collection of radiations. 
One can also see $w=1/3$ directly
from the coefficients of the $k^3$ terms in the integrand 
of \eqref{rhoRDAB} and \eqref{pRDAB} below. 

It is intuitively unclear how these two different equations
of state for the quartically divergent terms are consistent
with each other. 
The argument in the literature~\cite{Bunch} is based
on the dimensional regularization:
The quartic divergence in $\langle T_{\mu\nu}\rangle$ is 
regularized, then one looks at the contribution at the 
pole $1/(n-4)$ and finds that it is proportional 
to $g_{\mu\nu}$. Thus, it is removed by renormalizing the 
cosmological constant.

The above difference of equation of state in the two approaches
may be attributed to the fact that the UV cutoff $k_{\rm UV}$
itself depends on the metric. 
One can introduce a UV cutoff by putting the fields at
two points separated by a coordinate time interval $\Delta t$ 
(though we expect that the details of the regularization will not 
affect the conclusion).
Then $k_{\rm UV}$ will depend on $g_{00}$; we assume it is 
of the noncovariant form $k_{\rm UV}^2=(g_{00}(\Delta t)^2)^{-1}$.
By taking this into account,\footnote{A similar term 
appears in a different context~\cite{KK1}, where a time-dependent
cosmological constant due to
infrared effects in de Sitter space is discussed~\cite{KK2}.
}
the energy-momentum tensor for
\eqref{Leff4} becomes
\begin{align}
\langle T_{\mu\nu}\rangle_{4}
&={2\over \sqrt{-g}}{\delta {\cal L}_{\rm eff, 4}\over 
\delta g^{\mu\nu}}\nonumber\\
&=k_{\rm UV}^4 g_{\mu\nu} -4 k_{\rm UV}^4\; 
\frac{g_{0\mu}g_{0\nu}}{g_{00}} \ ,
\label{Tmn4}
\end{align}
where the second term comes from the variation of $k_{\rm UV}^4$
with respect to $g^{00}$. The expression \eqref{Tmn4} indeed 
has $w=1/3$ when the background metric is diagonal.

In our opinion, 
it could be misleading
to discuss the cosmological constant problem by looking at 
cutoff-dependent quantities with UV power divergences\footnote{
In \cite{Aoki:2012xs}, two of the present authors discussed
quadratic divergences of the Higgs boson mass term from the Wilsonian renormalization group point of view.
Since they are always absorbed in the definition of  
critical surfaces, we argued that cutoff-dependent quadratic divergences are unphysical
and should be simply subtracted. 
The noncovariance of quartic divergences of the cosmological constant 
suggests the same thing. In our opinion, power divergences
 should always be subtracted from the beginning, as is automatically realized in
 the dimensional regularization. 
}
since the subtracted term is not generally covariant.
In this paper, we will always consider 
the renormalized quantities when we discuss physical effects.

\subsection{RD period}
\label{sec:renemtRD}

Let us now evaluate $\rho$ and $p$ for the state (\ref{chiRD}) in the RD period.
By substituting the wave function (\ref{chiRD}) into 
 (\ref{rhogenchi}) and (\ref{pgenchi}), one obtains
\beqa
\rho_{\rm RD}^{\rm un-ren}&=&\frac{1}{8\pi^2 a^4}\int_0^\infty dk 
\Biggl[(|A|^2+|B|^2)\left(2k^3+\frac{k}{\eta^2}\right) \n
&&+A^* B\left(-2i\frac{k^2}{\eta}+\frac{k}{\eta^2} \right)e^{2ik\eta} 
+AB^* \left(2i\frac{k^2}{\eta}+\frac{k}{\eta^2} \right)e^{-2ik\eta} \Biggr] \ , \ \ \
\label{rhoRDAB}
\eeqa
\beqa
p_{\rm RD}^{\rm un-ren}&=&\frac{1}{8\pi^2 a^4}\int_0^\infty dk 
\Biggl[(|A|^2+|B|^2)\left(\frac{2}{3}k^3+\frac{k}{\eta^2}\right) \n
&&+A^* B\left(-\frac{4}{3}k^3-2i\frac{k^2}{\eta}+\frac{k}{\eta^2} \right)e^{2ik\eta} 
+AB^* \left(-\frac{4}{3}k^3+2i\frac{k^2}{\eta}+\frac{k}{\eta^2} \right)e^{-2ik\eta} \Biggr] \ .\n
\label{pRDAB}
\eeqa
The terms  with $A^*B$ and $AB^*$ represent  interference between the 
positive and negative frequency modes.

We now perform the subtraction of the UV divergences 
following the adiabatic regularization procedure. 
Since the potential for the wave equation vanishes in the RD period,
the adiabatic wave function, or the WKB wave function,
agrees with the  plane-wave solution (\ref{chiPW}). 
The terms to be subtracted are given by 
the first line in (\ref{rhoRDAB}) and (\ref{pRDAB})
with $A=1$, $B=0$. Thus, the renormalized expression is obtained
by simply replacing $(|A|^2+|B|^2)$ by $(|A|^2+|B|^2-1)$ in 
(\ref{rhoRDAB}) and (\ref{pRDAB}).

The subtraction term has been obtained for the FRW 
universe with the general scale factor $a$.
See, for instance, eqs.~(2.30) and (2.35) in ref.~\cite{Bunch}
(see also (2.10) in \cite{AndersonParker}).
Substituting the present form of $a$ in (\ref{a_caldera}),
one indeed obtains the above mentioned subtraction term. 
Apart from that term, there are finite terms of the order $H^4$
(where $H$ is the Hubble parameter at each moment),
which arise as a result of performing the $k$ integration in
massive theory, and taking the massless limit in the end.
This procedure is necessary since some of the terms
in the adiabatic expansion are IR divergent, and 
the massless theory cannot be studied directly.
These terms are important 
since they give the Weyl anomaly, and also contribute to the vacuum 
energy of pure de Sitter space. However, at late times, these terms
of the order $H^4$ are much smaller than the
total contribution that we are studying. The latter is 
enhanced due to the IR behavior of wave functions, as we have 
seen in (\ref{rhoIR}) and (\ref{pIR}). Thus, we will ignore this 
finite contribution of the order $H^4$ in this paper.

We now consider the consequences of our approximation,
in which the scale factor (\ref{a_caldera}) changes its functional
forms instantaneously and its higher derivatives with respect to $\eta$
are not continuously connected at the boundaries of the stages.
Due to this approximation, we have non-differentiable sharp peaks in the 
potential at the boundaries (see Figure~\ref{fig:caldera_pot}). 
In this potential, the 
reflection coefficient decreases only with a power, $B(k) \propto k^{-2}$,
as given in (\ref{AB}).
Even at high $k$, over-the-barrier scattering by the sharp potential
is not strongly suppressed. This is different from the general behavior 
for the scattering by a smooth potential, where the reflection coefficients 
fall off exponentially at large $k$. Namely, if wavelengths 
are smaller than the typical curvature radius at the peak of the
potential, over-the-barrier scattering cannot occur. (See, for instance, 
section 52 of ref.~\cite{Landau}.) 

The power law tail of $B(k)$ for the sharp potential produces problems
in the UV behavior of the $k$-integral in (\ref{rhoRDAB}) and (\ref{pRDAB}).
The coefficient in the first line of (\ref{rhoRDAB}) and (\ref{pRDAB}) after the UV subtraction 
behaves as $(|A|^2+|B|^2-1)=2|B|^2 \propto k^{-4}$, 
where the unitarity relation $|A|^2-|B|^2=1$ was used.
It then follows that  the leading term $(|A|^2+|B|^2-1)k^3 \propto k^{-1}$  
gives rise to a UV logarithmic divergence.
A problem also arises from the interference terms.
In the IR region of $k<\eta_2^{-1}$,
the coefficients behave as
$A,B \propto k^{-2}$, as given by (\ref{AB}),
and the term $A^*Bk^3$ in (\ref{pRDAB}) decreases as $1/k$
while oscillating.
The integrand $p(k)$ will be seen below in Figure~\ref{fig:rhopintegrandRD}.
However, in the UV region of $k>\eta_2^{-1}$, 
the coefficients behave as $A \propto 1$ and $B \propto k^{-2}$,
and the term $A^*Bk^3$ grows as a linear function of $k$ though it oscillates.
 
These problems are caused by our setting where
the scale factors are not sufficiently smoothly connected at the 
boundaries of the stages.
The subtraction term obtained by the DeWitt-Schwinger expansion or adiabatic regularization, 
which is based on the adiabatic expansion of the background geometry, 
could not cancel all the UV divergences in such cases.
Since modes with infinitesimally small wavelengths are affected by the sharp potential at 
the boundary, the UV behaviors at later times become dependent on the past history
and cannot be controlled only by the local quantities. 

We assume that the scale factors in the realistic settings are
smoothly connected, so that the reflection coefficients
decay quickly when $k\gtrsim \eta_2^{-1}$.
To take this behavior into account, we will introduce a
mask function $f_2(k)$, which takes a value close to 1 
for $k \lesssim \eta_2^{-1}$,
and falls off rapidly for $k \gtrsim \eta_2^{-1}$.
The scattering coefficients have to satisfy the 
unitarity relation $|A|^2-|B|^2=1$. The simplest way to apply 
a mask function, $f_2(k)$, that is consistent with this
relation 
would be to make the following replacements in (\ref{AB}):
\beq
\frac{1}{\eta_1} \to \frac{f_2(k)}{\eta_1}, \ \ \frac{1}{\eta_1^2} \to \frac{f_2(k)^2}{\eta_1^2}  \ .
\label{repe1tofe1}
\eeq
In the actual analysis, we will take 
\beq
f_2(k) = \left\{\begin{array}{ll}
1 & (k \le \eta_2^{-1}) \\
0 & (k > \eta_2^{-1})
\end{array} \right. \ .
\label{f2step}
\eeq
Note that, after the subtraction of UV divergences, 
all the terms in $(|A|^2+|B|^2-1)$, $A^*B$, and $AB^*$ contain
a factor of $1/\eta_1^n$  ($n \geq 1$). Hence, owing to $f_2^n=f_2$, the effect of 
introducing the mask function is to set the upper bound of the $k$ integration at $1/\eta_2$.
As we will see at the end of Section~\ref{sec:timeevoRD},
the final results do not depend much on the explicit form of the mask function.\footnote{
The UV divergence caused by the sharp changes of the background has been 
noticed also in \cite{Glavan:2013mra}. The authors of \cite{Glavan:2013mra} introduced a 
regulator function, which decays exponentially at large $k$, to model the 
behavior of a smooth transition. Our masking procedure is very similar
to this in spirit, and will give essentially the same result.
}

To summarize, after the subtraction of the UV divergences and the smoothing of the potential,
$\rho_{\rm RD}$ and  $p_{\rm RD}$  are given by
 (\ref{rhoRDAB}) and (\ref{pRDAB}) with the following replacements:
\beqa 
 (|A|^2+|B|^2)    &\longrightarrow & (|A|^2+|B|^2-1) \ , \n 
  \int_0^\infty dk &\longrightarrow&  \int_0^{\eta_2^{-1}} dk \ .
\eeqa

\subsection{MD period}
\label{sec:renemtMD}

In the MD period,
by substituting the wave function (\ref{chiMD}) into 
(\ref{rhogenchi}) and (\ref{pgenchi}), one obtains the energy and pressure densities:
\beqa
\rho_{\rm MD}^{\rm un-ren}&=&\frac{1}{8\pi^2 a^4}\int_0^\infty dk 
\Biggl[(|C|^2+|D|^2)\left(2k^3+4\frac{k}{\eta^2}+9\frac{1}{k\eta^4}\right) \n
&&+C^* D~2k^3 \left(7(k\eta)^{-2}-\frac{9}{2}(k\eta)^{-4}
+\left(-2(k\eta)^{-1}+9(k\eta)^{-3}\right)i \right)e^{2ik\eta} \n
&&+C D^*~2k^3 \left(7(k\eta)^{-2}-\frac{9}{2}(k\eta)^{-4}
-\left(-2(k\eta)^{-1}+9(k\eta)^{-3}\right)i \right)e^{-2ik\eta} \Biggr] \ , \n
\label{rhoMDCD}
\eeqa
\beqa
p_{\rm MD}^{\rm un-ren}&=&\frac{1}{8\pi^2 a^4}\int_0^\infty dk 
\Biggl[(|C|^2+|D|^2)\left(\frac{2}{3}k^3+\frac{8}{3}\frac{k}{\eta^2}+9\frac{1}{k\eta^4}\right) \n
&&+C^* D~2k^3 \left(-\frac{2}{3}+\frac{23}{3}(k\eta)^{-2}-\frac{9}{2}(k\eta)^{-4}
+\left(-\frac{10}{3}(k\eta)^{-1}+9(k\eta)^{-3}\right)i \right)e^{2ik\eta} \n
&&+C D^*~2k^3 \left(-\frac{2}{3}+\frac{23}{3}(k\eta)^{-2}-\frac{9}{2}(k\eta)^{-4}
-\left(-\frac{10}{3}(k\eta)^{-1}+9(k\eta)^{-3}\right)i \right)e^{-2ik\eta} \Biggr] \ .\n
\label{pMDCD}
\eeqa

The subtraction of UV divergences is similarly performed by replacing 
 $(|C|^2+|D|^2)$ by $(|C|^2+|D|^2-1)$ in (\ref{rhoMDCD}) and (\ref{pMDCD}). 
In the MD period, the adiabatic expression is given by
the first line in (\ref{rhoMDCD}) and (\ref{pMDCD}) with 
$C=1, D=0$. One can see this by expanding the wave function 
\eqref{chiBD} in $1/(k\eta)$, or by substituting our scale factor $a(\eta)$
into the general expression in \cite{Bunch, AndersonParker}. 
As we mentioned in the previous subsection, we ignore the finite terms of 
the order $H^4$ arising from renormalization, 
since they are smaller than the terms of interest at late times.

For the masking procedure, 
in addition to the replacement of (\ref{repe1tofe1})  with (\ref{f2step}) in (\ref{AB}),
we can also make the replacement
\beq
\frac{1}{\eta_4} \to \frac{f_4(k)}{\eta_4}, \ \ \frac{1}{\eta_4^2} \to \frac{f_4(k)^2}{\eta_4^2}  
\label{repe4tofe4}
\eeq
with
\beq
f_4(k) = \left\{\begin{array}{ll}
1 & (k \le \eta_4^{-1}) \\
0 & (k > \eta_4^{-1})
\end{array} \right. 
\label{f4step}
\eeq
in (\ref{ABCD}).
The latter is introduced for smoothing the potential peak at 
the matter-radiation equality $\eta=\eta_4$.
Then the scattering amplitudes $C(k)$ and $D(k)$  
reduce to $A(k)$ and $B(k)$ 
for $k >\eta_4^{-1}$.
They further become $C(k)=1$ and $D(k)=0$ for $k >\eta_2^{-1}$,
which truncates the $k$-integral at $k=\eta_{2}^{-1}$, as in the RD period case.
However, the results do not depend much on
whether or not we replace  $C$ and $D$ by $A$ and $B$ 
in the region $k \in [\eta_4^{-1} , \eta_2^{-1}]$.
Since the leading
contributions to the scattering coefficients $C$ and $D$
for the modes  $k \in [\eta_4^{-1} , \eta_2^{-1}]$ 
come from the waves that are scattered by the potential in the inflation 
period but pass through the potential in the MD period,
over-the-barrier scatterings by the potential in the MD period 
give only sub-leading contributions and do not affect the results much.
We therefore will not perform the replacement of
$C$ and $D$ by $A$ and $B$ in the region $k \in [\eta_4^{-1} , \eta_2^{-1}]$.

To summarize, our formula in the MD period is as follows.
$\rho_{\rm MD}$ and $p_{\rm MD}$ are given by
 (\ref{rhoMDCD}) and (\ref{pMDCD}), with 
 the replacement of 
 $(|C|^2+|D|^2)$ by $(|C|^2+|D|^2-1)$, 
modifying the subtraction term in the IR region properly,
and introducing an upper bound
 of the $k$ integration at $\eta_2^{-1}$.

\section{Time evolution of energy and pressure densities}
\label{sec:timeevo}   
\setcounter{equation}{0}

We are now ready to study the time evolution of the
energy and pressure densities in the cosmic history after the inflation period.

\subsection{RD period}
\label{sec:timeevoRD}
In the RD period, the energy and pressure densities are given by
(\ref{rhoRDAB}) and (\ref{pRDAB}) with modifications
by the UV subtraction and masking procedures
addressed at the end of Section~\ref{sec:renemtRD}.
With ({\ref{AB}) plugged in, they become
\beqa
\rho_{\rm RD}&=&\frac{1}{8\pi^2 a^4 \eta_1^4} \int_0^{\eta_2^{-1}}\frac{dk}{k}
\Big[1-(k\eta)^{-1}\left(s-2k\eta_1 c-2(k\eta_1)^2 s\right) \n
&&+(k\eta)^{-2}\left(\frac{1}{2}-\frac{1}{2}c-k\eta_1 s+(k\eta_1)^2 c \right)\Big] \ ,
\label{rhorenRD}
\eeqa
\beqa
p_{\rm RD}&=&\frac{1}{8\pi^2 a^4 \eta_1^4} \int_0^{\eta_2^{-1}}\frac{dk}{k}
\Big[\frac{1}{3}+\frac{2}{3}c+\frac{4}{3}k\eta_1 s-\frac{4}{3}(k\eta_1)^2 c \n
&&-(k\eta)^{-1}\left(s-2 k\eta_1 c-2(k\eta_1)^2 s \right) \n
&&+(k\eta)^{-2}\left(\frac{1}{2}-\frac{1}{2}c-k\eta_1 s+(k\eta_1)^2 c \right)\Big] \ ,
\label{prenRD}
\eeqa
where $s=\sin{[2k(\eta-\eta_2)]}$ and $c=\cos{[2k(\eta-\eta_2)]}$.
The prefactor in (\ref{rhorenRD}) and (\ref{prenRD}) is rewritten as
\beq
\frac{1}{8\pi^2 a^4\eta_1^4}=\frac{1}{8\pi^2a^4} \left(a_{\rm Inf}(\eta_1) H_I\right)^4 
=\frac{1}{8\pi^2}H_I^4 \left(\frac{H}{H_I}\right)^2 = \frac{1}{8\pi^2} (H_I H)^2 \ ,
\label{prefct}
\eeq
where we have used (\ref{a_caldera}) for the first equality,
and the relation $H^2 \propto a^{-4}$ in the RD period for the second equality.
Here, $H_I$ and $H$ are the Hubble parameters at the inflation period and 
at the time of interest.
As we have noticed before, $\rho$ and $p$ are enhanced to 
the order of (\ref{prefct}).
The modes with $k <\eta_2^{-1}$ are enhanced due to the non-adiabatic
evolution in the inflation period.

The integrands (without the prefactor $1/8\pi^2a^4 \eta_1^4$)
of (\ref{rhorenRD}) and (\ref{prenRD})
 are shown in Figure~\ref{fig:rhopintegrandRD}.
They oscillate and decrease as $k$ increases. 
The oscillation period is  $\Delta k= \pi/\eta$.
Under the conditions  of $\eta \gg \eta_2=|\eta_1|$ and $k \ll 1/\eta_2$,
the height of each peak increases in proportion to $\eta$.  
This behavior can easily be seen from the fact that under the above
conditions,  $\eta_2=|\eta_1|$ can be dropped from the integrands without the
prefactor, and they are invariant under 
$k \rightarrow \lambda k$ and $\eta \rightarrow \eta/\lambda$.
Thus, the integration over each peak remains almost constant irrespective of $\eta$. 

Time evolution of $\rho$ and $p$, besides the time dependence of  the prefactor (\ref{prefct}),
comes from the finite integration range of the $k$-integral, $ k \in [0,\eta_2^{-1}]$.
Since the oscillation period is  $\Delta k= \pi/\eta$,
the number of peaks within the interval is given by $\eta/(\pi \eta_2)$,
which grows with time.
For small values of $\eta/\eta_2$, {\it i.e.}, just after inflation,
the integration is mainly contributed to by 
the first few peaks of the integrands in the IR regions.
On the contrary,  for the later stage with large values of $\eta/\eta_2$, 
the integration is dominated by the UV tail of the integrands.

\begin{figure}
\begin{center}
\begin{minipage}{.45\linewidth}
\includegraphics[width=1.1 \linewidth]{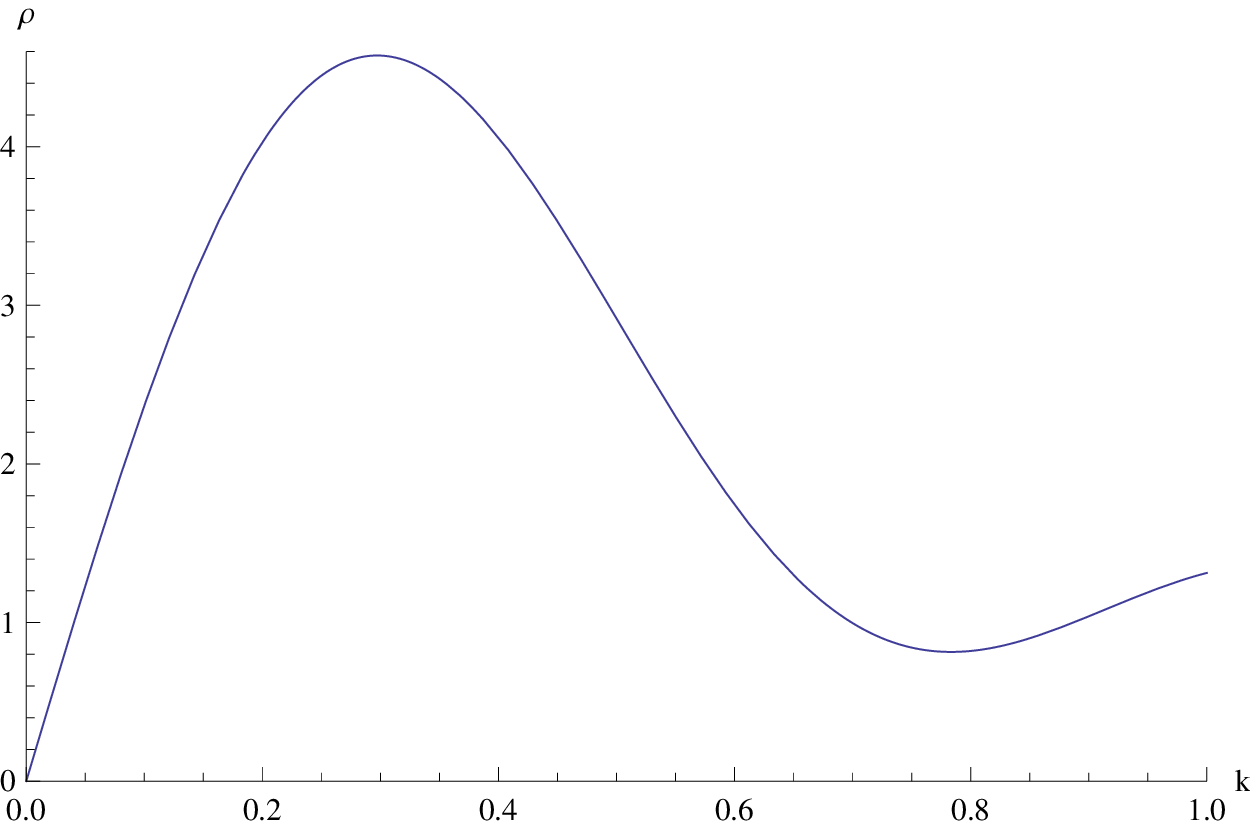}
  \end{minipage}
  \hspace{1.0pc}
\begin{minipage}{.45\linewidth}
\includegraphics[width=1.1 \linewidth]{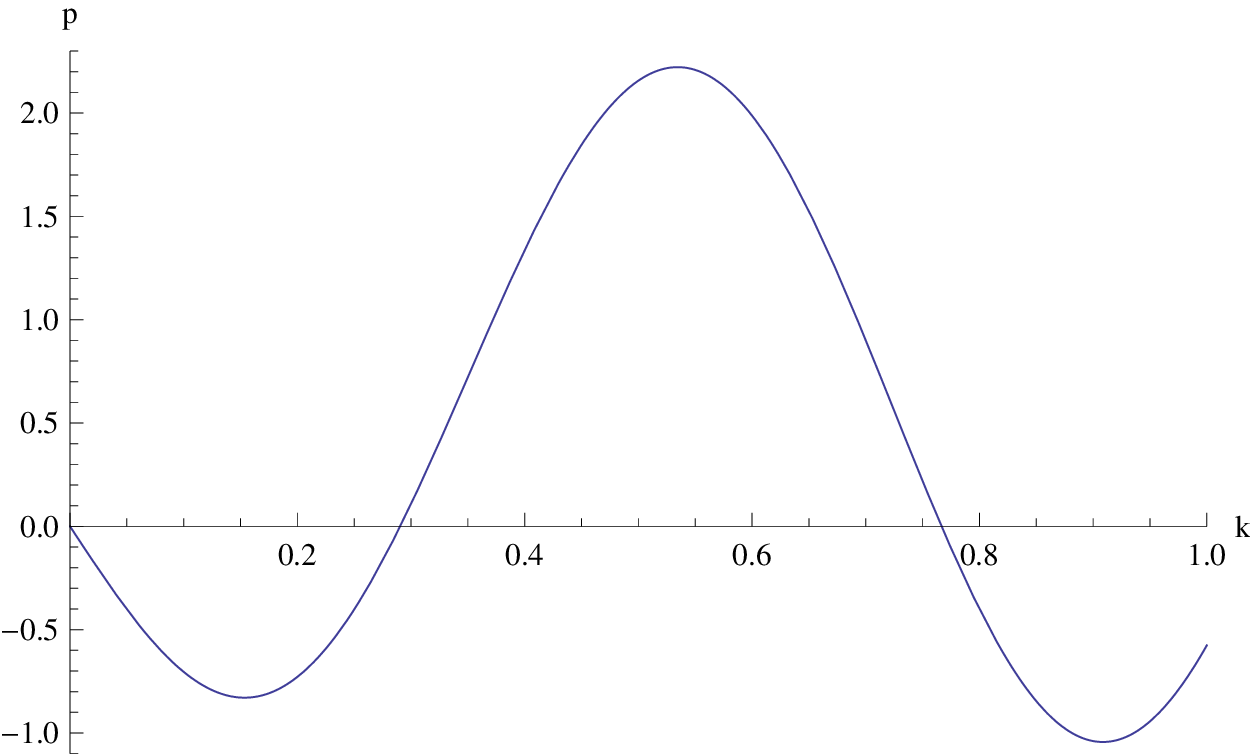}
  \end{minipage}
\begin{minipage}{.45\linewidth}
\includegraphics[width=1.1 \linewidth]{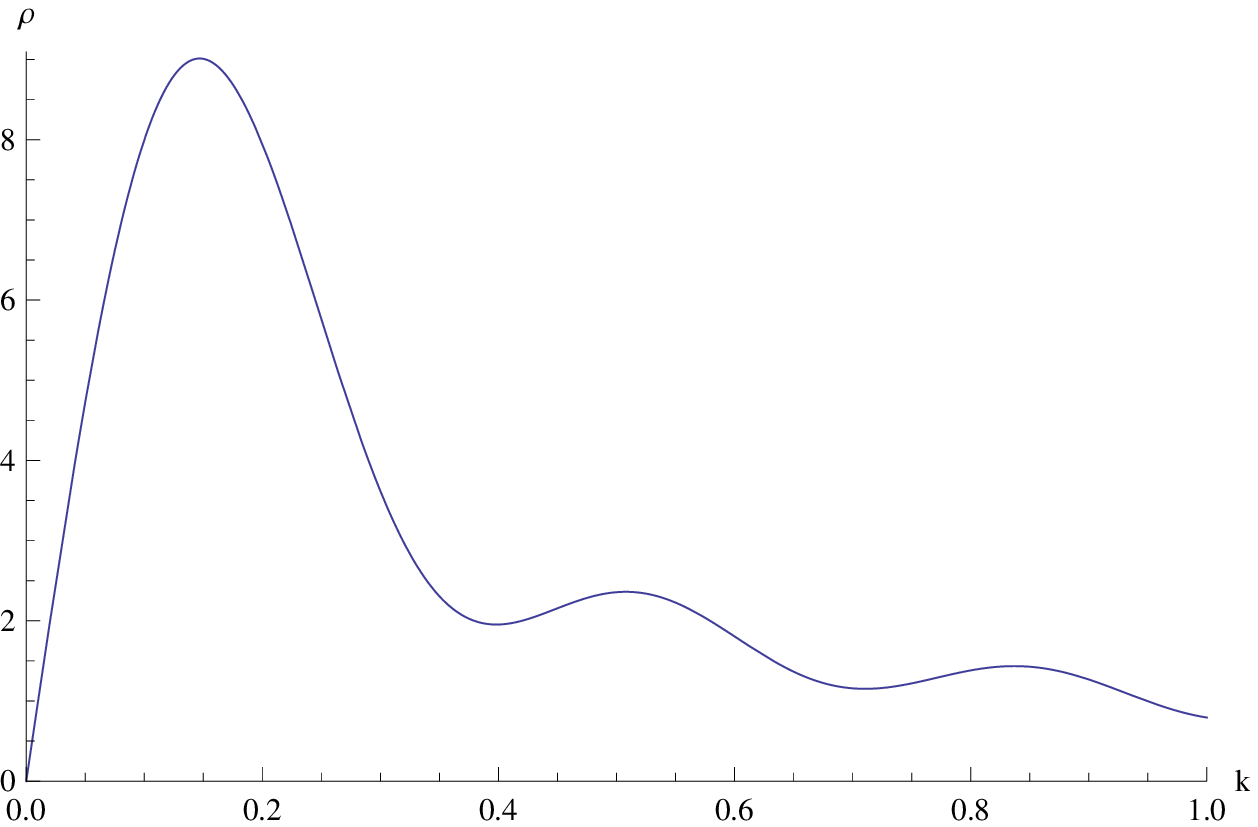}
  \end{minipage}
  \hspace{1.0pc}
\begin{minipage}{.45\linewidth}
\includegraphics[width=1.1 \linewidth]{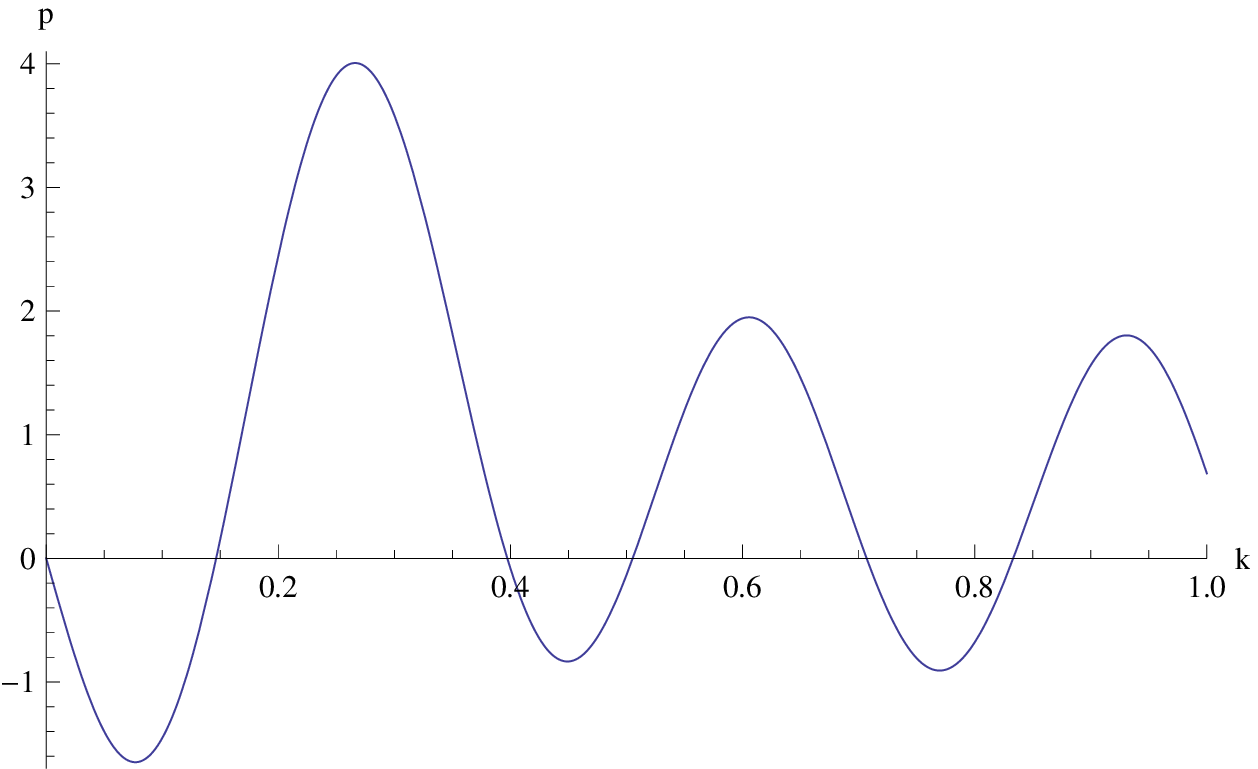}
  \end{minipage}
\begin{minipage}{.45\linewidth}
\includegraphics[width=1.1 \linewidth]{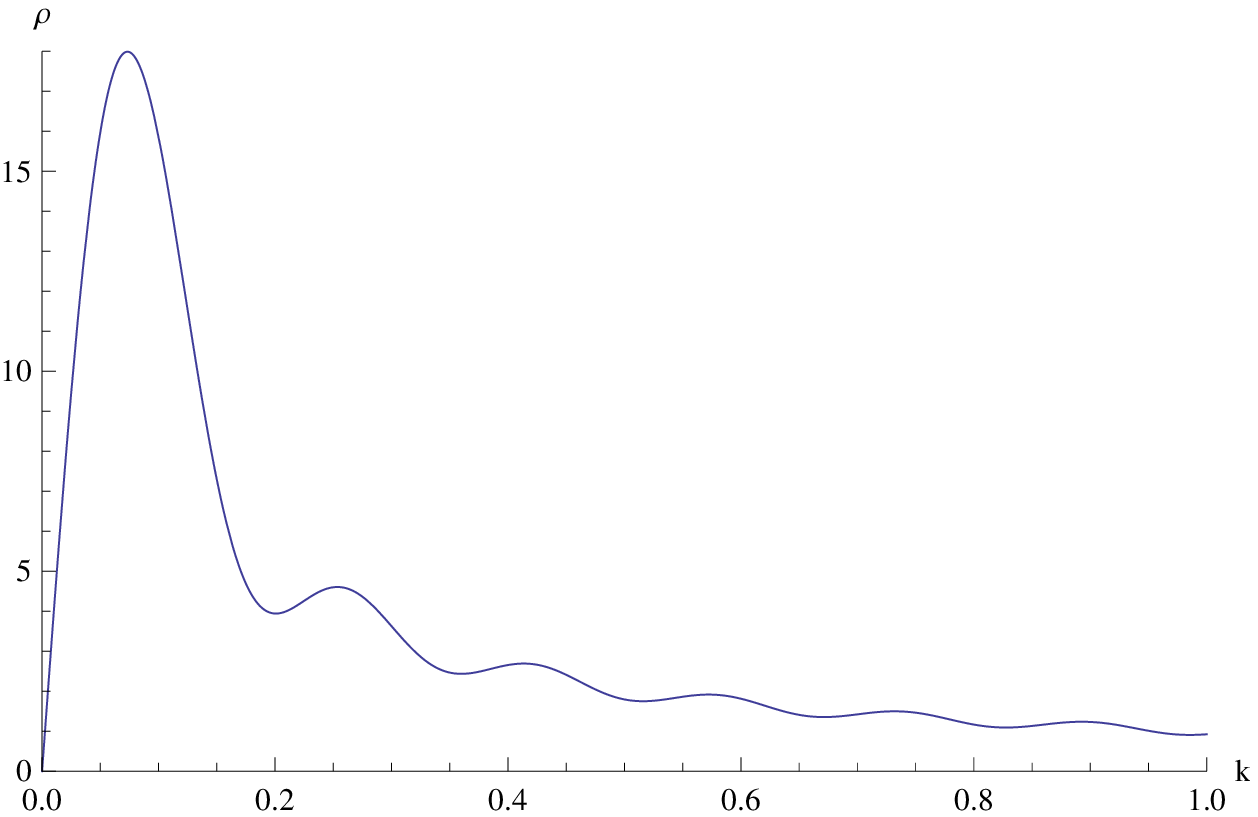}
  \end{minipage}
  \hspace{1.0pc}
\begin{minipage}{.45\linewidth}
\includegraphics[width=1.1 \linewidth]{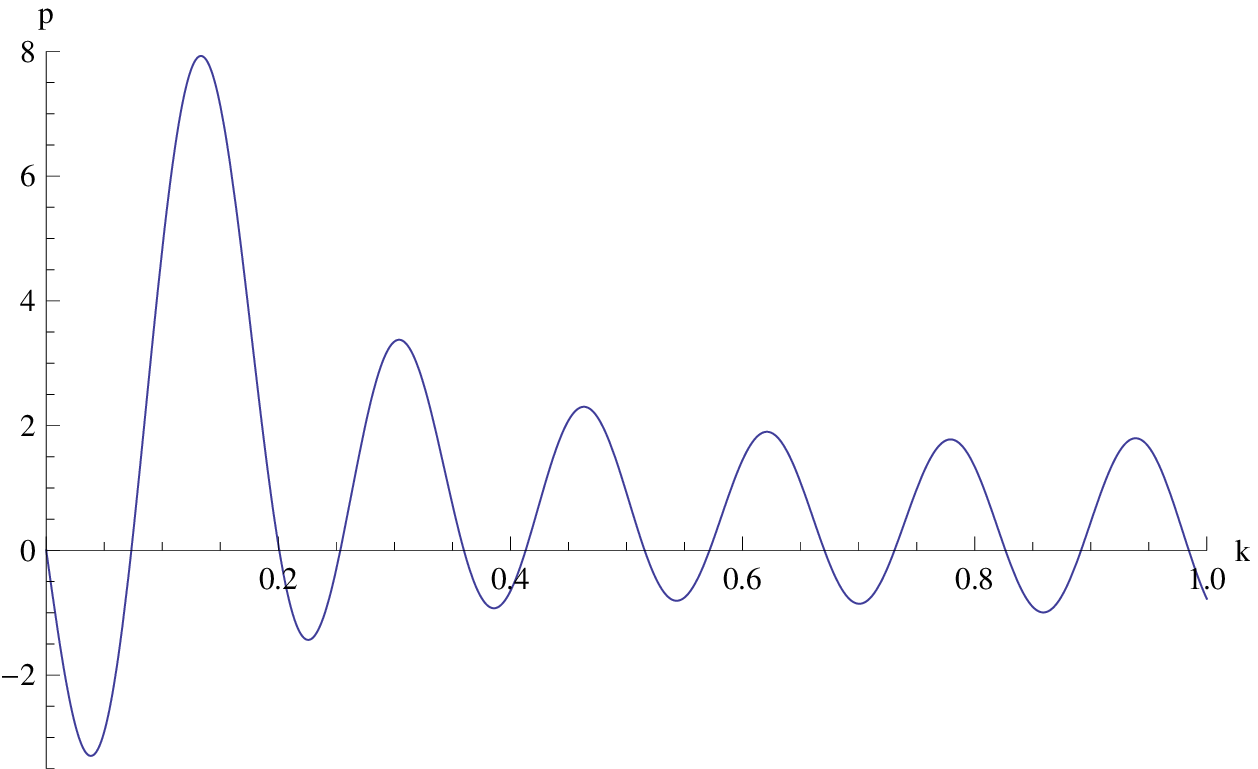}
  \end{minipage}  
\begin{minipage}{.45\linewidth}
\includegraphics[width=1.1 \linewidth]{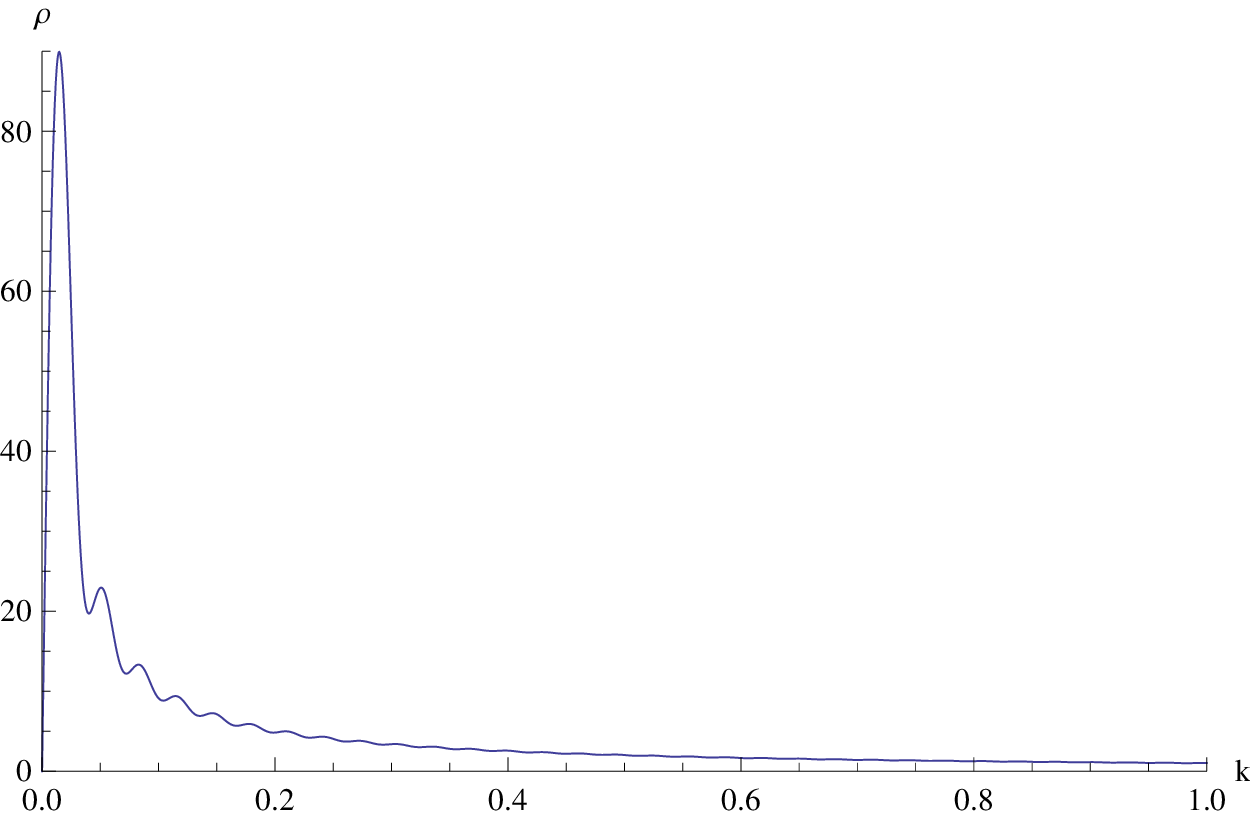}
  \end{minipage}
  \hspace{1.0pc}
\begin{minipage}{.45\linewidth}
\includegraphics[width=1.1 \linewidth]{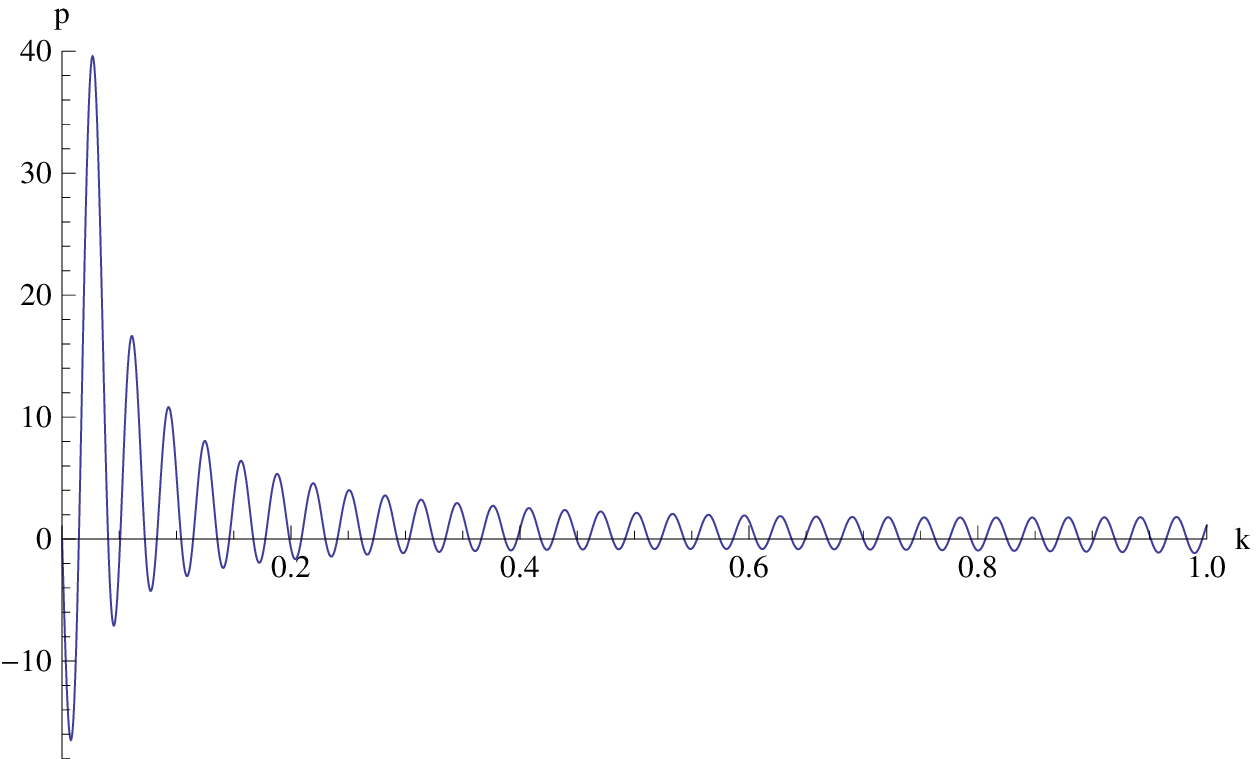}
  \end{minipage}  
\end{center}
\caption{
Integrands $\rho(k)$ of (\ref{rhorenRD}) and $p(k)$ of (\ref{prenRD}),
without the prefactor $1/8\pi^2a^4 \eta_1^4$,
are shown over the integration interval  $k \in [0,\eta_2^{-1}]$.
The parameters are taken to be
 $\eta_2=-\eta_1=1$ and $\eta=5, 10, 20, 100$ from the top to the bottom figures.
}
\label{fig:rhopintegrandRD}
\end{figure}

Before discussing the time evolution in more detail, 
we first investigate the analytical behaviors of the integrals 
in the IR and UV regions separately.
In the IR region, the integrals of (\ref{rhorenRD}) and (\ref{prenRD}) seem to diverge,
but, as we saw before, cancellations among various terms occur and they become IR finite. 
Indeed, (\ref{rhorenRD}) and (\ref{prenRD}) are estimated
in the IR region as
\beqa
\rho_{\rm RD}^{\rm IR}&=&\frac{1}{8\pi^2 a^4 \eta_1^4} \int_0 
dk\left[\left(1-\left(\frac{\eta_2}{\eta}\right)^4\right)\eta^2 k +{\cal O}(k^3) \right] \ ,
\label{rhoRDrenIR} \\
p_{\rm RD}^{\rm IR}&=&\frac{1}{8\pi^2 a^4 \eta_1^4} \int_0 
dk\left[-\frac{1}{3}\left(1+3\left(\frac{\eta_2}{\eta}\right)^4\right)\eta^2 k +{\cal O}(k^3) \right] \ .
\label{pRDrenIR}
\eeqa
The second term in the parenthesis, which is proportional to $(\eta_2/\eta)^4$, 
comes from the subtraction term.
The leading terms reproduce eqs.~(\ref{rhoIR}) and (\ref{pIR})
by using (\ref{prefct})
and the relation $a\eta=H^{-1}$ in the RD period.
Therefore, the contribution to the integral (\ref{rhorenRD}) from the IR region below the first peak
in Figure~\ref{fig:rhopintegrandRD} is evaluated as
\beq
\rho^{\rm IR}_{\rm RD} \sim 
\frac{1}{8\pi^2 a^4 \eta_1^4} ~ \int_0^{\pi/2\eta} dk ~ \eta^2 k
=\frac{1}{8\pi^2}(H_I H)^2 \eta^2 \frac{1}{2}  
 \left(\frac{\pi}{2\eta}\right)^2
=\frac{1}{64}(H_I H)^2 \ .
\eeq
The first peak is located approximately at $\pi/2\eta$.
In the second equality, (\ref{prefct}) was used.
If we further perform the integration up to the first minimum of $\rho$, 
it becomes almost doubled:
\beq
\rho^{\rm IRpeak}_{\rm RD} \sim (H_I H)^2/32 \ . \label{rhobump}
\eeq
For $k<\pi/(4 \eta)$, the pressure density $p$ is approximated by (\ref{pRDrenIR}) and
behaves as $p \sim -\rho/3$. 
However, as shown in Figure~\ref{fig:rhopintegrandRD},
the integrand of $p$ 
changes its sign from negative to positive  around $k=\pi/(2 \eta)$, and the
equation of state $w=\rho/p$ gradually changes.

On the other hand, in the UV region with $k\eta \gg 1$, 
the energy and pressure densities are approximately given by
the first term in the square bracket of
 (\ref{rhorenRD}) and  (\ref{prenRD}).\footnote{
The other terms in (\ref{rhorenRD}) may
give similar contributions to (\ref{rhoRDUV}), but, compared to the logarithmic
factor $\ln(\eta/\eta_2)$, they are at most of the order of one
and thus negligible.
For instance,  the second term in (\ref{rhorenRD}) becomes
 $\int^{\eta^{-1}_2}_{\eta^{-1}}dk~k^{-2}\eta^{-1}\sin{(2k\eta)}\sim 1$,
since only the region around $k \sim \eta^{-1}$ contributes.
The third term in the first parenthesis, an oscillating but non-decreasing function of $k$, 
becomes
 $\int^{\eta^{-1}_2}_{\eta^{-1}}dk~\eta^{-1}\eta_1^{2}\sin{(2k\eta)}\sim (\eta_1/\eta)^2 \ll 1$.
The first term in the second parenthesis, which is not oscillating, becomes
$\int^{\eta^{-1}_2}_{\eta^{-1}}dk~k^{-3}\eta^{-2}\sim 1$.
The most worrisome term is the fourth term in (\ref{prenRD}), which is 
oscillating but increasing. 
As we mentioned before, this is an artifact of the potential with a sharp peak and
 cured by introducing the mask function. 
This term gives
$\int^{\eta^{-1}_2}_{\eta^{-1}}dk~k \eta_1^2 \cos{(2k\eta)} \sim |\eta_1|/\eta \ll 1$,
since only the region around $k\sim \eta_2^{-1}=|\eta_1|^{-1}$ with the width $\eta^{-1}$ contributes.
}
We then obtain\footnote{
These expressions agree with the ones obtained in \cite{Glavan:2013mra}.
}
\beqa
\rho_{\rm RD}^{\rm UV}
&\simeq& \frac{1}{8\pi^2 a^4 \eta_1^4} \int_{\sim \eta^{-1}}^{\eta_2^{-1}}\frac{dk}{k} 
=\frac{1}{8\pi^2}(H_I H)^2\ln{\left(\frac{\eta}{\eta_2}\right)} \ ,
\label{rhoRDUV}
\\
p_{\rm RD}^{\rm UV}
&\simeq& \frac{1}{8\pi^2 a^4 \eta_1^4} \int_{\sim \eta^{-1}}^{\eta_2^{-1}}\frac{dk}{k}\frac{1}{3}
=\frac{1}{3} \rho_{\rm RD}^{\rm UV}
 \ ,
\label{pRDUV}
\eeqa  
where (\ref{prefct}) is used in the second equality in (\ref{rhoRDUV}).
By comparing (\ref{rhobump}) and (\ref{rhoRDUV}), we find that
the UV tail contribution becomes larger than the first IR peak contribution
when $\eta/\eta_2 >e^{\pi^2/4}\sim 12$. 

We now show the 
numerical evaluations of the time evolution of $\rho$ and $p$ in the RD period.
Figure~\ref{fig:rhopRDnearBB} shows the early time behavior 
for $1<\eta/\eta_2<10$. 
The position of the first minimum of $\rho(k)$ is at $k=\pi/ \eta$ and 
the integration is taken over the interval $k \in [0, \eta_2^{-1}]$.
Hence, for $\eta \sim \pi \eta_2$, $\rho$ is approximated by the first peak contribution
 (\ref{rhobump}), which is  depicted by the line in the figure.
As $\eta/\eta_2$ increases, the energy density $\rho$ gradually grows 
by taking the second and the following peak contributions.
The behavior of $p$ is more complicated.
The integrand $p(k)$ is negative for $k <\pi/(2 \eta)$ and changes its sign around $k=\pi/(2\eta)$.
Hence, the $k$-integral of $p$ is negative until $\eta \sim 2 \eta_2$, and then becomes positive.
Reflecting the oscillating behavior of the integrand in Figure~\ref{fig:rhopintegrandRD},
$p$  gradually increases and oscillates in Figure~\ref{fig:rhopRDnearBB}.
The data in Figure~\ref{fig:rhopRDnearBB} for $\eta/\eta_2 \lesssim 1.5$ 
cannot be taken seriously\footnote{For $\eta/\eta_2<2$,
the contribution to $p$ is mainly given by the first negative peak,
and $w$ becomes negative.
As seen below (\ref{rhoIR}) and (\ref{pIR}), 
$w \to -1/3$ is expected for $\eta/\eta_2 \to 1$.
However, $w$ takes even smaller values
because  the second term in (\ref{rhoRDrenIR}) and (\ref{pRDrenIR}),
which comes from the subtraction term,
becomes the same order as the first term.},
since the spacetime curvature is not small in this region,
and the corrections due to finite renormalization terms from 
the UV subtraction, 
which were discussed in the third paragraph
of Section~\ref{sec:renemtRD},
cannot be ignored.  
Figure~\ref{fig:rhopRDfarBB} shows the late time behaviors of
 (\ref{rhorenRD}) and (\ref{prenRD}) 
for $10<\eta/\eta_2<10^{30}$.
 As expected, the results agree well with Eqs.~(\ref{rhoRDUV}) and (\ref{pRDUV}).

\begin{figure}
\begin{center}
\begin{minipage}{.45\linewidth}
\includegraphics[width=1.1 \linewidth]{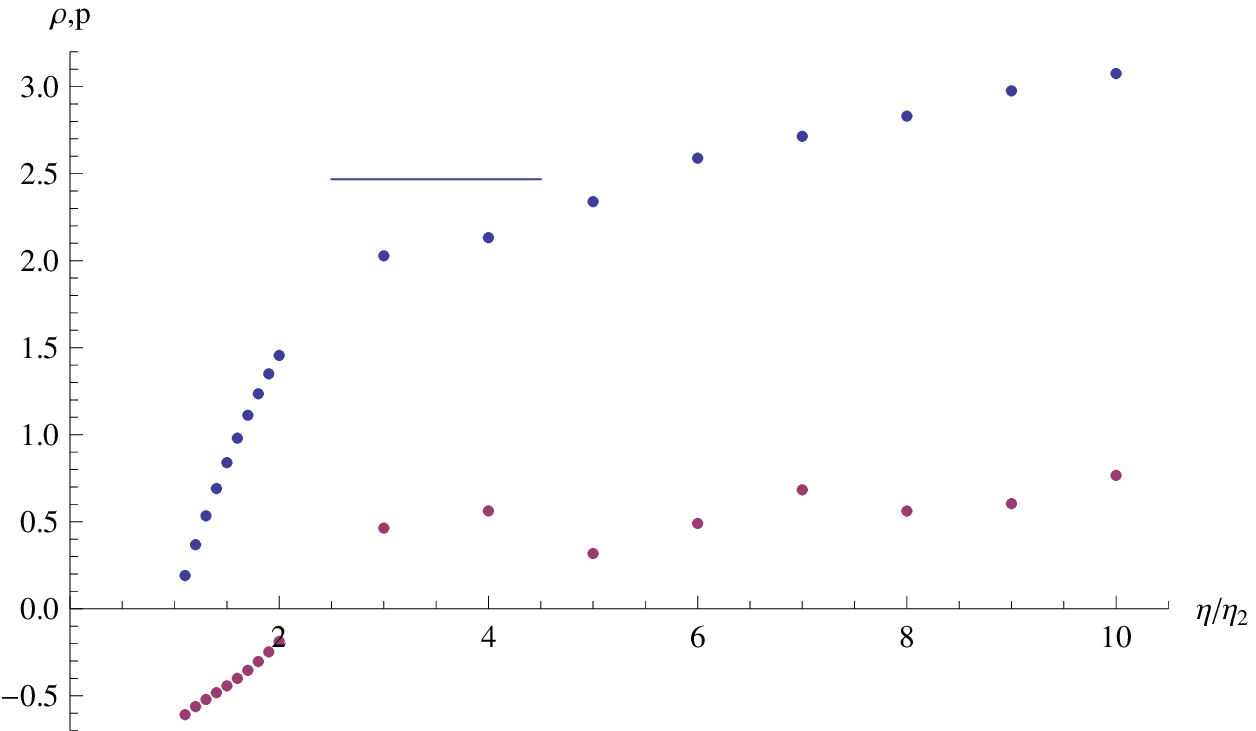}
  \end{minipage}
  \hspace{1.0pc}
\begin{minipage}{.45\linewidth}
\includegraphics[width=1.1 \linewidth]{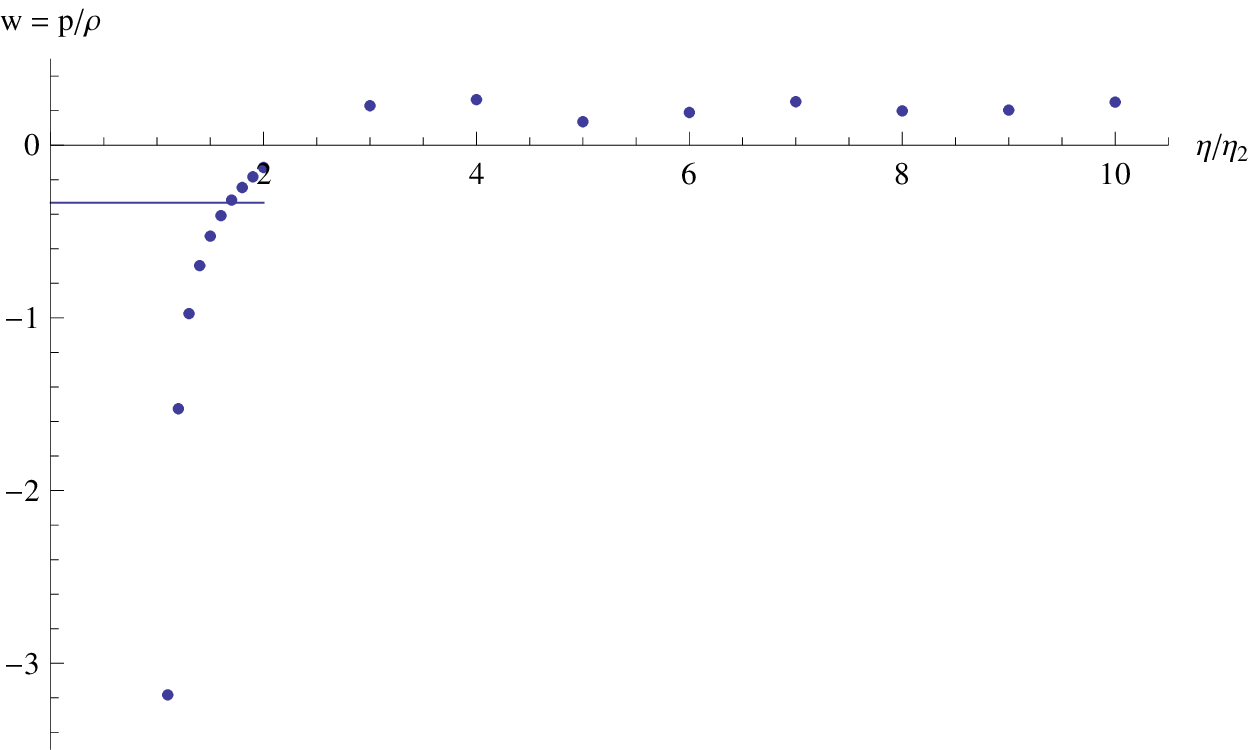}
  \end{minipage}
\end{center}
\caption{Early time behavior of the time evolution of $\rho$ and $p$ in the RD period.
We take  $1<\eta/\eta_2<10$, {\it i.e.},
the time $\eta$ is close to the big bang, $\eta_2$.
$\rho$ gradually increases while $p$ changes its sign from negative to positive as
it oscillates. The very early time behavior below $\eta/\eta_2 < 1.5$
is an artifact of the UV subtraction 
and cannot be taken at face value.
(Left) The upper and lower data represent $\rho$ and $p$, 
{\it i.e.}, (\ref{rhorenRD}) and (\ref{prenRD})
without the prefactor $1/8\pi^2a^4 \eta_1^4$.
The line represents the first IR peak contribution (\ref{rhobump}).
(Right) The equation of state $w=p/\rho$ is shown.
The line represents $w=-1/3$, a value given by  (\ref{rhoIR}) and (\ref{pIR}).
}
\label{fig:rhopRDnearBB}
\end{figure}

\begin{figure}
\begin{center}
\begin{minipage}{.45\linewidth}
\includegraphics[width=1.1 \linewidth]{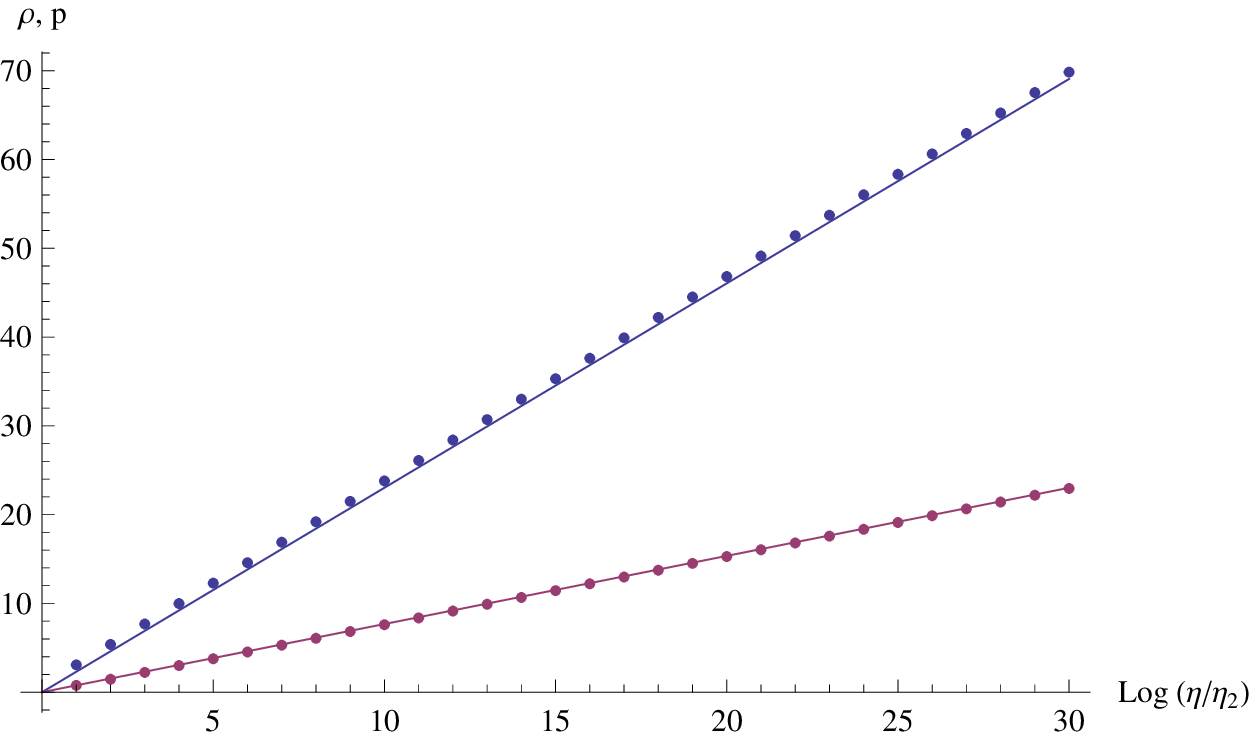}
  \end{minipage}
  \hspace{1.2pc}
\begin{minipage}{.45\linewidth}
\includegraphics[width=1.1 \linewidth]{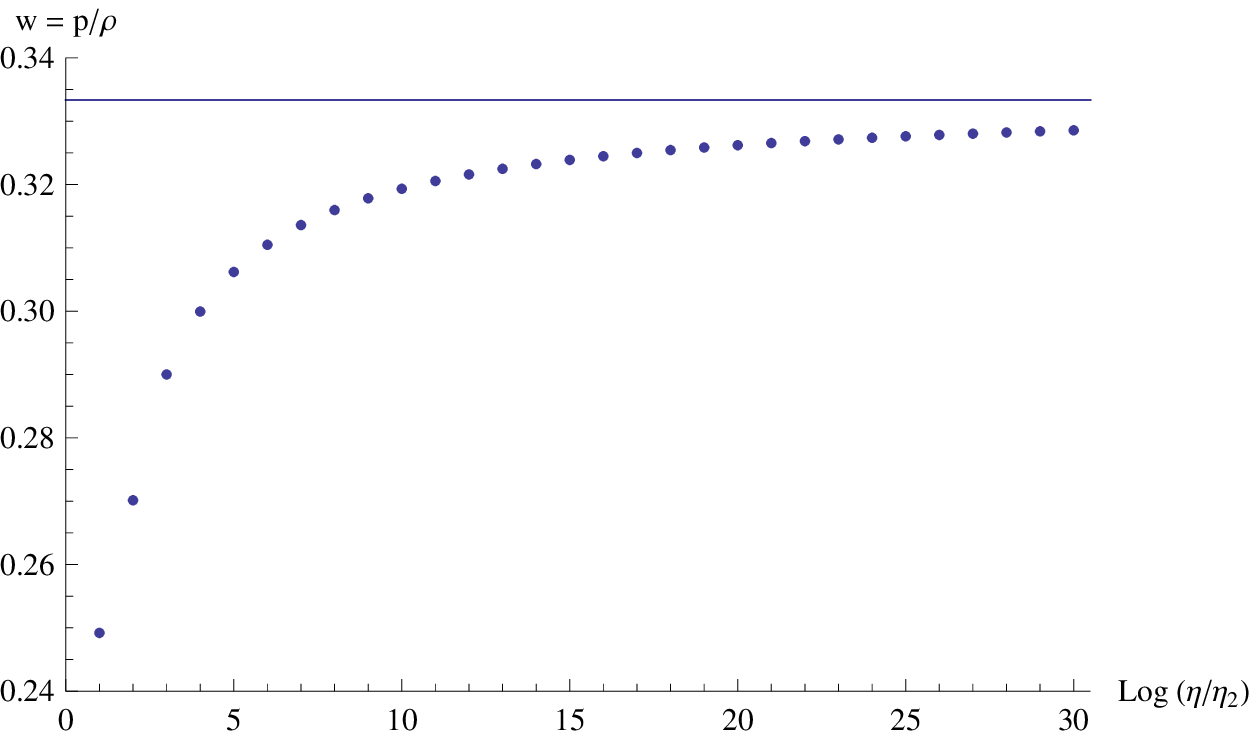}
  \end{minipage}
\end{center}
\caption{Late time behavior of the time evolution of $\rho$ and $p$ in the RD period.
We take $10<\eta/\eta_2<10^{30}$, {\it i.e.},
the time $\eta$ is far from the big bang, $\eta_2$.
(Left) The upper and lower data represent $\rho$ and $p$, 
{\it i.e.}, (\ref{rhorenRD}) and (\ref{prenRD})
without the prefactor $1/8\pi^2a^4 \eta_1^4$.
The lines represent the analytical approximations
(\ref{rhoRDUV}) and (\ref{pRDUV}), which agree well with the numerical evaluations.
(Right) The equation of state $w=p/\rho$ is shown.
The line represents $w=1/3$, a value predicted by (\ref{rhoRDUV}) and (\ref{pRDUV}).}
\label{fig:rhopRDfarBB}
\end{figure}

Finally, we will see that the integrals of $\rho$ and $p$ are not  
sensitive to the details of the masking function introduced to express the smooth potential.
For example, if we change the upper bound of the $k$-integral from 
$1/\eta_2$ to $r /\eta_2$, 
the  logarithmic factor in (\ref{rhoRDUV}) is changed from $\ln(\eta/\eta_2)$
to $(\ln(\eta/\eta_2) +\ln r)$. The difference is very small as long as $r \ll (\eta/\eta_2).$

\subsection{MD period}
\label{sec:timeevoMD}

In the MD period, the energy and pressure densities are given
by the formula (\ref{rhoMDCD}) and (\ref{pMDCD}),
modified by the subtraction and masking procedures
explained at the end of Section~\ref{sec:renemtMD}.

Figure~\ref{fig:rhopintegrandMD} shows the integrands in the IR region $k \in [0, 0.1 \eta_2^{-1}]$,
where $\eta_2^{-1}$ is the upper bound of the $k$-integration.
The top figures correspond to the time when the MD period begins at $\eta=\eta_4$.
They indeed agree with the figures when the RD period ends at $\eta=\eta_3$,
which are shown by the bottom figures in Figure~\ref{fig:rhopintegrandRD}. 
As can be seen from (\ref{ABCD}), 
the oscillation period is given by $\Delta k =\pi/(\eta-\eta_4+\eta_3)$,
and decreases as $\eta$ increases.
A notable feature in Figure~\ref{fig:rhopintegrandMD} is that 
the relative height of the peaks in the IR region to those in the UV region becomes larger as 
$\eta$ evolves.

\begin{figure}
\begin{center}
\begin{minipage}{.45\linewidth}
\includegraphics[width=1.1 \linewidth]{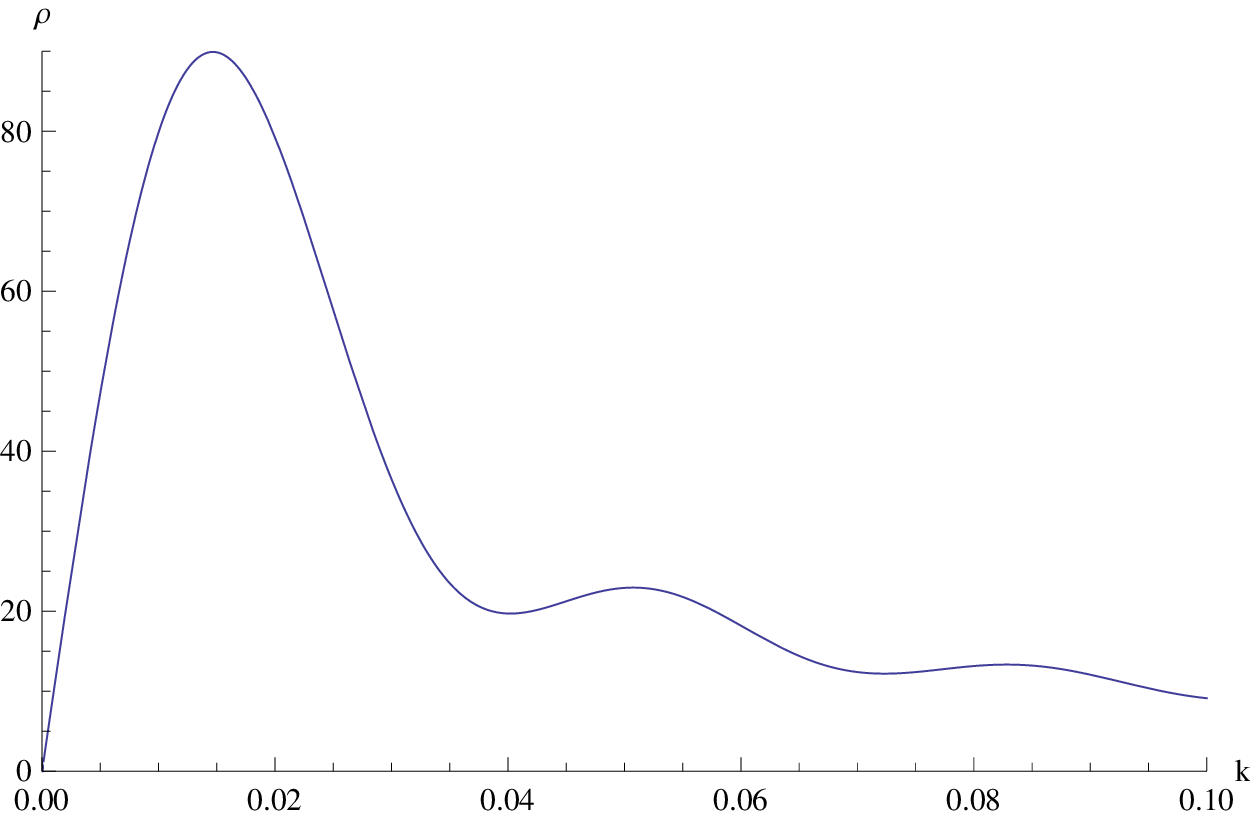}
  \end{minipage}
  \hspace{1.2pc}
\begin{minipage}{.45\linewidth}
\includegraphics[width=1.1 \linewidth]{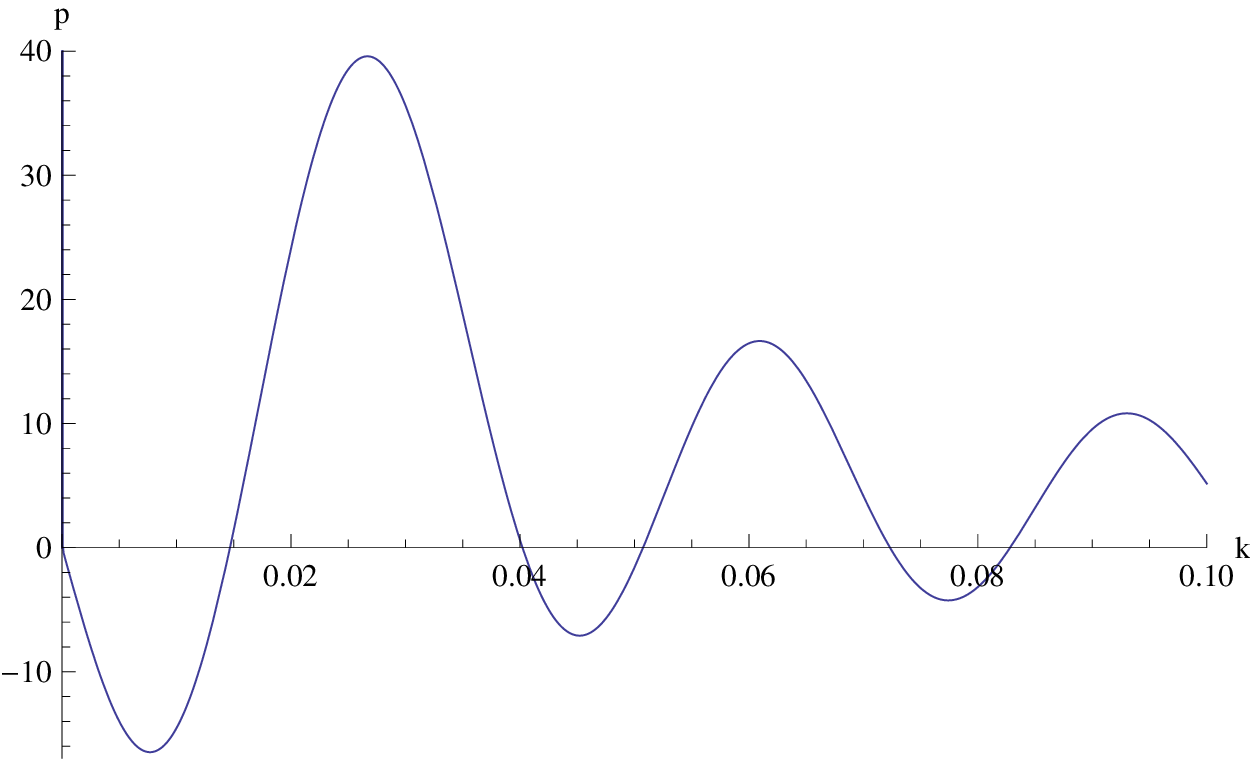}
  \end{minipage}
\begin{minipage}{.45\linewidth}
\includegraphics[width=1.1 \linewidth]{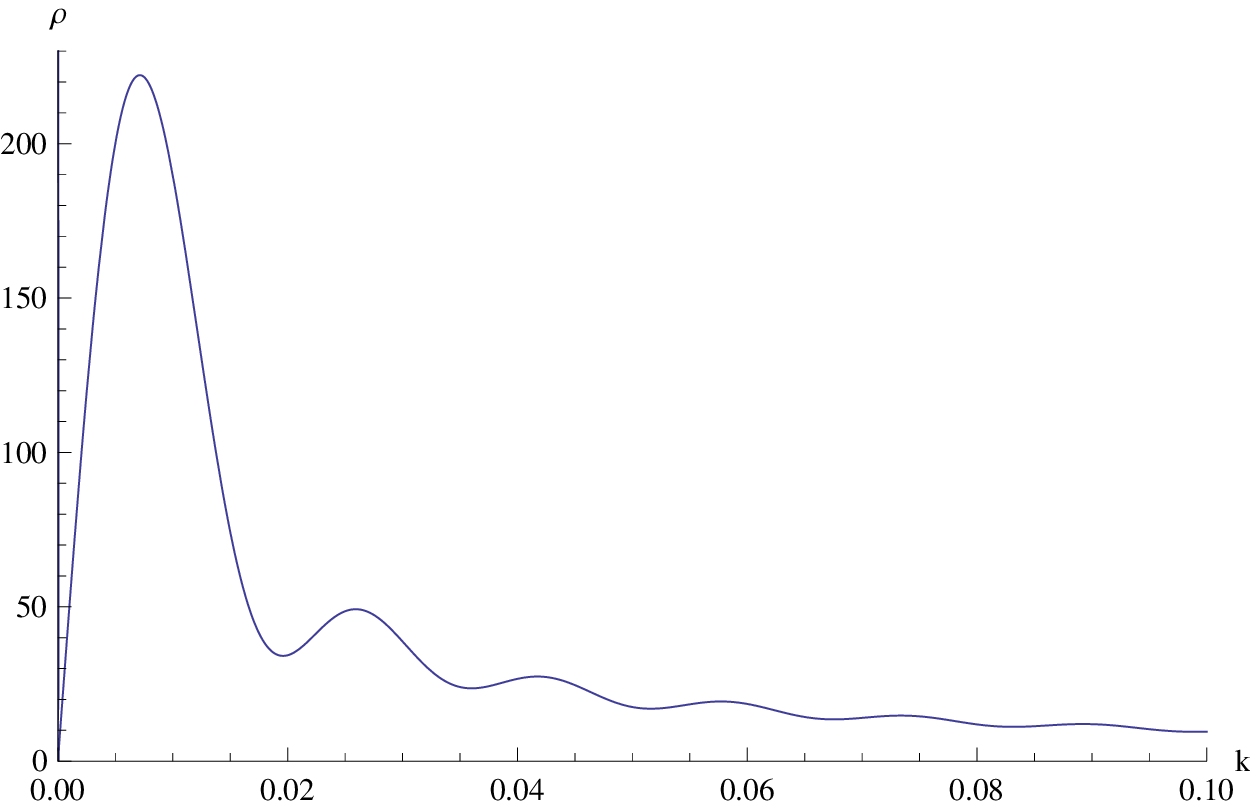}
  \end{minipage}
  \hspace{1.2pc}
\begin{minipage}{.45\linewidth}
\includegraphics[width=1.1 \linewidth]{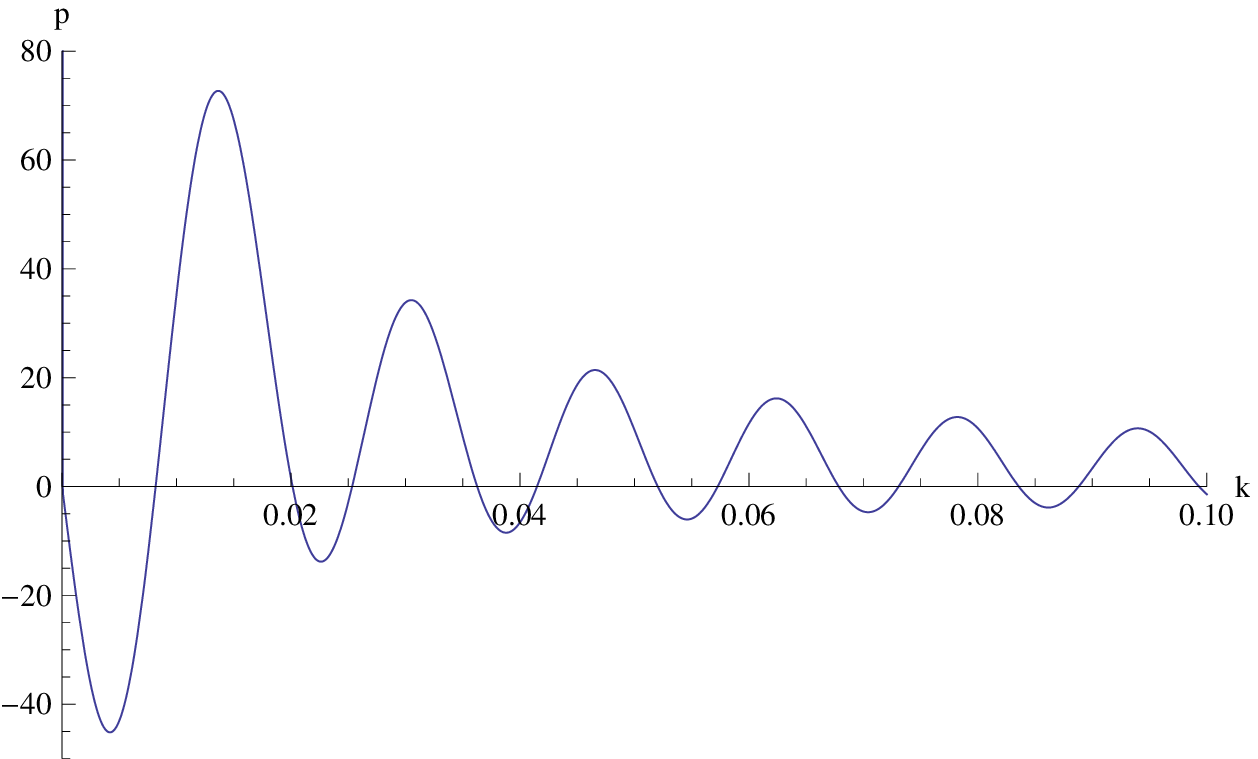}
  \end{minipage}  
  \begin{minipage}{.45\linewidth}
\includegraphics[width=1.1 \linewidth]{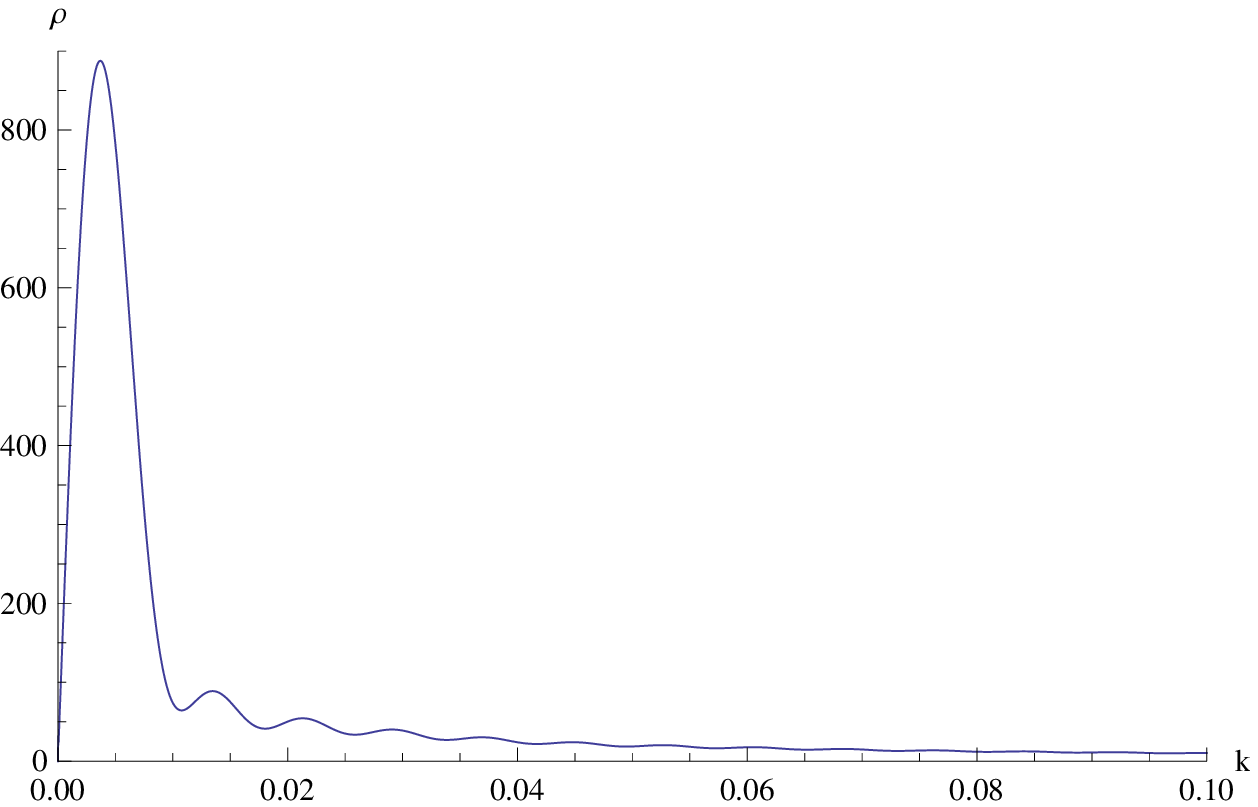}
  \end{minipage}
  \hspace{1.2pc}
\begin{minipage}{.45\linewidth}
\includegraphics[width=1.1 \linewidth]{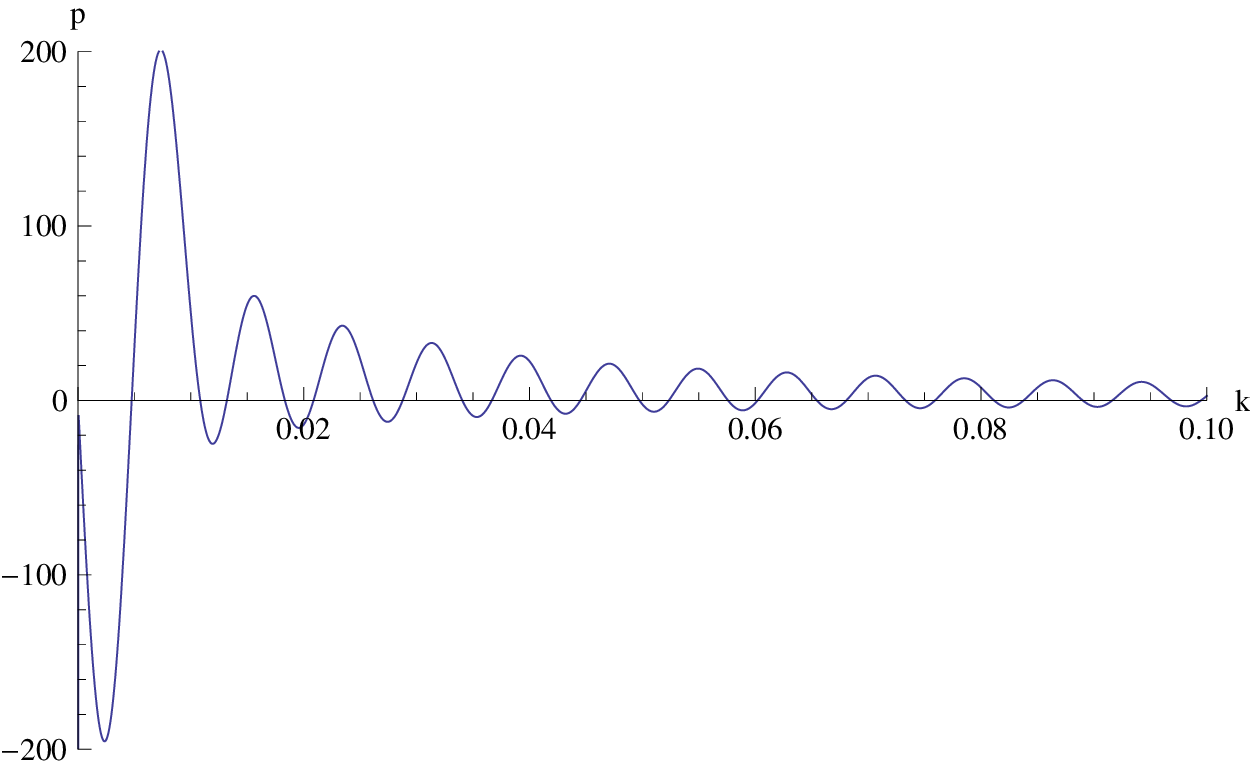}
  \end{minipage}
  \begin{minipage}{.45\linewidth}
\includegraphics[width=1.1 \linewidth]{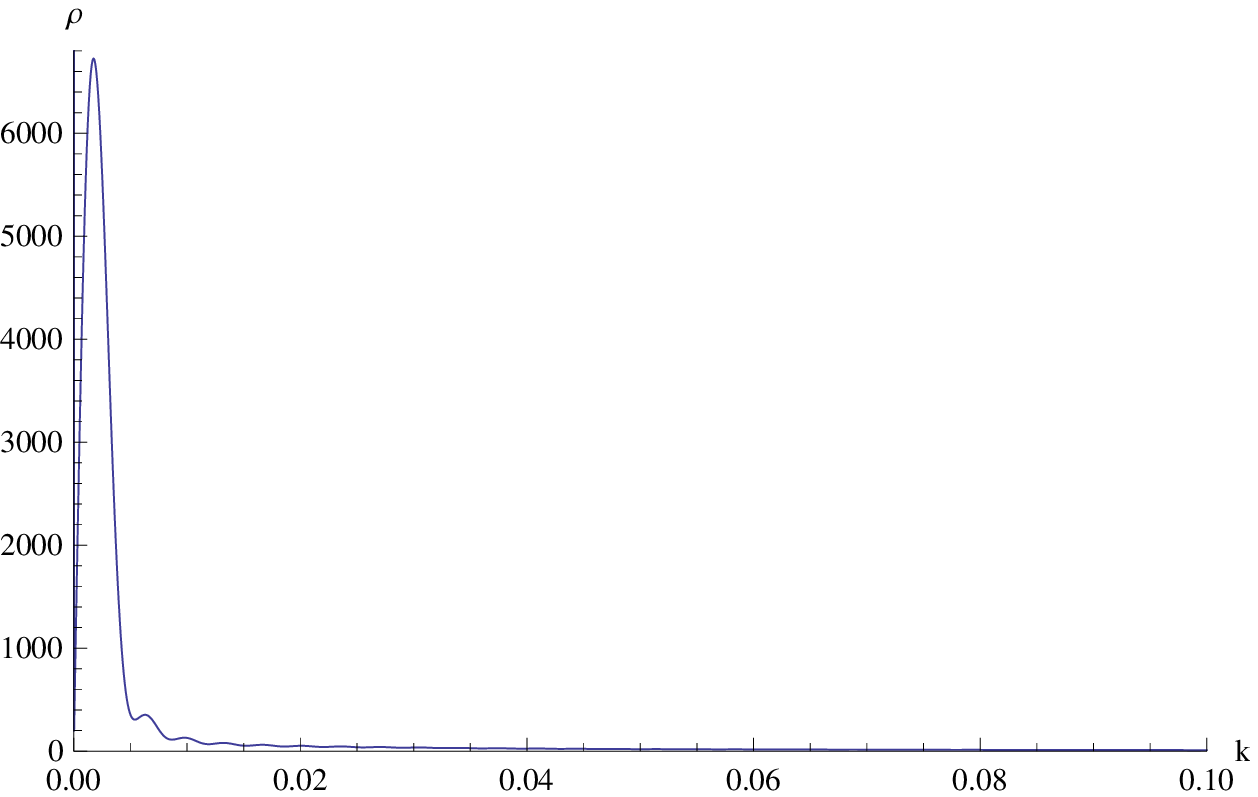}
  \end{minipage}
  \hspace{1.2pc}
\begin{minipage}{.45\linewidth}
\includegraphics[width=1.1 \linewidth]{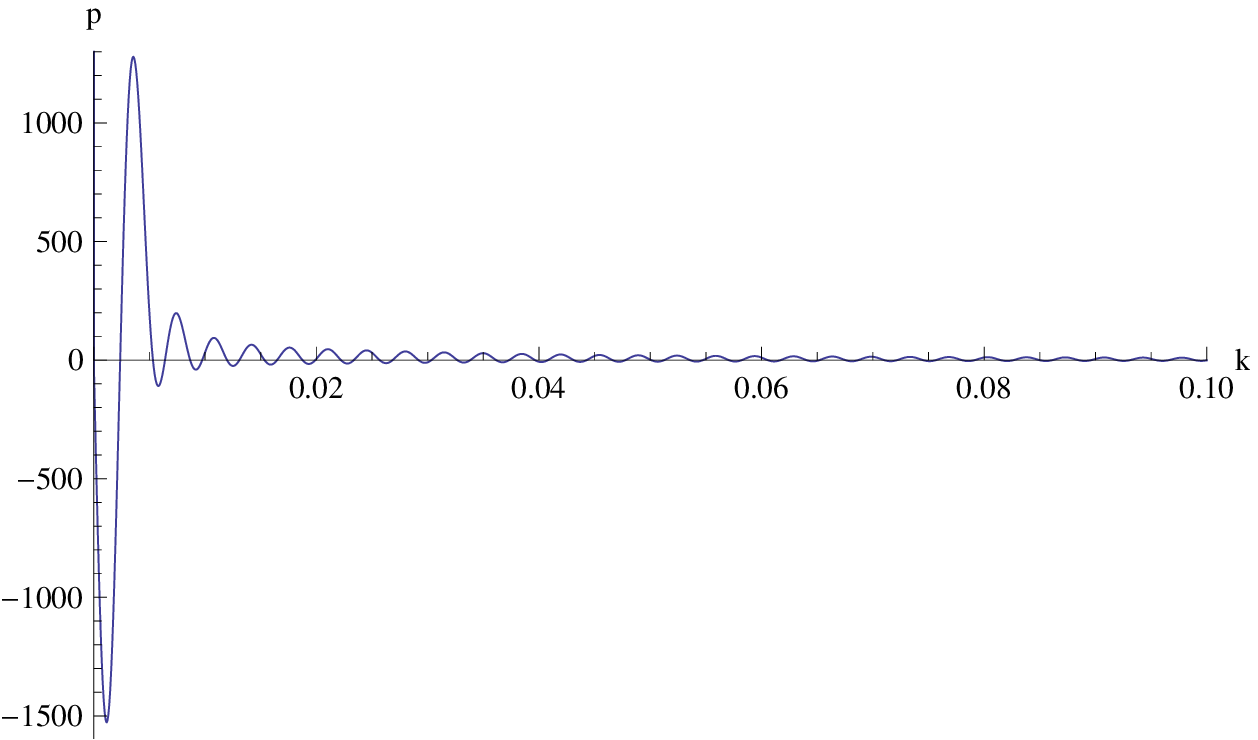}
  \end{minipage}
\end{center}
\caption{Integrands $\rho(k)$ and $p(k)$ in the MD period.
(\ref{rhoMDCD}) and (\ref{pMDCD}), 
multiplied by $8\pi^2 a^4 \eta_1^4$, are depicted 
 for $k \in [0, 0.1 \eta_2^{-1}]$. 
The parameters are taken to be $\eta_2=-\eta_1=1$, $\eta_4=2\eta_3=200$, 
and $\eta=200, 300, 500, 1000$
from the top to the bottom figures. 
The top figures correspond to the time when the MD period starts,
which is the same as the bottom figures of  Figure~\ref{fig:rhopintegrandRD},
corresponding to the time  when the RD period ends.
}
\label{fig:rhopintegrandMD}
\end{figure}

Before discussing the time evolution, we first investigate the integrals in
the IR and UV regions separately.
For the IR region with $k < \eta_4^{-1}$, the coefficients of the wave function 
$C$ and $D$ are approximated by (\ref{Cexp}) and (\ref{Dexp}). 
Substituting them into (\ref{rhoMDCD}) and (\ref{pMDCD}), we obtain
\beqa
\rho_{\rm MD}^{\rm IR}&\simeq&\frac{1}{8\pi^2 a^4 \eta_1^4}\left(\frac{\eta}{\eta_4}\right)^2
\int_0 dk~\frac{9}{8} k^{-3}\eta^{-2} \Big[2+4 (k\eta)^{-2}+9 (k\eta)^{-4} \n
&&+\left(14(k\eta)^{-2}-9(k\eta)^{-4}\right) \cos{(2k\eta)} \n
&&+\left(4(k\eta)^{-1}-18(k\eta)^{-3}\right)\sin{(2k\eta)}\Big] \ ,
\label{rhoMDIRregion} 
\eeqa
\beqa
p_{\rm MD}^{\rm IR}&\simeq&\frac{1}{8\pi^2 a^4 \eta_1^4}\left(\frac{\eta}{\eta_4}\right)^2
\int_0 dk~\frac{3}{8} k^{-3}\eta^{-2} \Big[2+8 (k\eta)^{-2}+27 (k\eta)^{-4} \n
&&+\left(-4+46(k\eta)^{-2}-27(k\eta)^{-4}\right) \cos{(2k\eta)} \n
&&+\left(20(k\eta)^{-1}-54(k\eta)^{-3}\right)\sin{(2k\eta)}\Big] \ .
\label{pMDIRregion}
\eeqa

Note that an extra prefactor  $(\eta/\eta_4)^2$ appears.
This is the reason for the special enhancement of the peaks in the IR region.
By using the relation  $a \propto \eta^2$ in the MD period, it is written as
\beq
\left(\frac{\eta}{\eta_4}\right)^2
=\frac{a}{a_{\rm eq}} \ ,
\label{extraprefacMD}
\eeq
where eq stands for the matter-radiation equality.
However, unlike (\ref{prefct}) in the RD period, the following relation holds in the MD period:
\beq
\frac{1}{8\pi^2 a^4 \eta_1^4} 
=\frac{1}{8\pi^2}(H_I H)^2 \left(\frac{a_{\rm eq}}{a}\right) \ .
\label{prefctMD}
\eeq
Here, we used 
\beq
\left(\frac{a_{\rm Inf}(\eta_1)}{a}\right)^4
=\left(\frac{a_{\rm Inf}(\eta_1)}{a_{\rm eq}}\right)^4 
\left(\frac{a_{\rm eq}}{a}\right)^3 \left(\frac{a_{\rm eq}}{a}\right)
=\left(\frac{H}{H_I}\right)^2 \left(\frac{a_{\rm eq}}{a}\right) \ ,
\eeq
which follows the relations $H^2 \propto a^{-4}$, $a^{-3}$ in the RD and MD periods, respectively.
Then the prefactor including the extra factor (\ref{extraprefacMD})
becomes
\beq
\frac{1}{8\pi^2 a^4 \eta_1^4} \left(\frac{\eta}{\eta_4}\right)^2
=\frac{1}{8\pi^2}(H_I H)^2 \ .
\label{totprefctMD}
\eeq
Hence, in the unit of $(H_I H)^2$, the heights of the IR peaks are the same as in the RD period.
The heights of UV peaks, as we see below in (\ref{rhoMDUV}),
 do not have such an extra enhancement factor, and thus
 are reduced by the factor $a_{\rm eq}/a$.

If we further expand $\cos(2k\eta)$ and $\sin(2k\eta)$ 
in Eqs.~(\ref{rhoMDIRregion}) and (\ref{pMDIRregion}), 
we obtain IR finite integrals:
\beqa
\rho_{\rm MD}^{\rm IR}&=&\frac{1}{8\pi^2 a^4 \eta_1^4} \left(\frac{\eta}{\eta_4}\right)^2 
\int_0 dk\left[\frac{\eta^2}{4} k +{\cal O}(k^2) \right] \ ,
\label{rhoMDIR}\\
p_{\rm MD}^{\rm IR}&=&\frac{1}{8\pi^2 a^4 \eta_1^4} \left(\frac{\eta}{\eta_4}\right)^2 
\int_0 dk\left[-\frac{\eta^2}{12} k +{\cal O}(k^2) \right] \ .
\label{pMDIR}
\eeqa
They indeed reproduce (\ref{rhoIR}) and (\ref{pIR})
by using (\ref{totprefctMD}) and the relation $a\eta=2 H^{-1}$ in the MD period.

We now perform the $k$-integration of (\ref{rhoMDIRregion}) and  (\ref{pMDIRregion})
over the interval $k \in [0,\eta_4^{-1}]$, in which the approximation is valid.
The integrands are shown in Figure~\ref{fig:rhopintegrandMDlowappr}.
They damp quickly as $k^{-3}$, and take negligibly small values for $k  \gtrsim 5\eta^{-1}$.
Hence, for $\eta \gtrsim 5\eta_4$, the integral over $k \in [0,\eta_4^{-1}]$ becomes almost constant
and equal to that over $k \in [0,\infty]$:
\beqa
\rho_{\rm MD}^{\rm IR}&\simeq&\frac{3}{4}\frac{1}{8\pi^2}(H_I H)^2  \ , 
\label{rhoMDIRreginted} \\
p_{\rm MD}^{\rm IR}&\simeq& 0 \ . \label{pMDIRreginted}
\eeqa
The equation of state becomes $w^{\rm IR}=p^{\rm IR}/\rho^{\rm IR}=0$ in the IR region.

\begin{figure}
\begin{center}
\begin{minipage}{.45\linewidth}
\includegraphics[width=1.1 \linewidth]{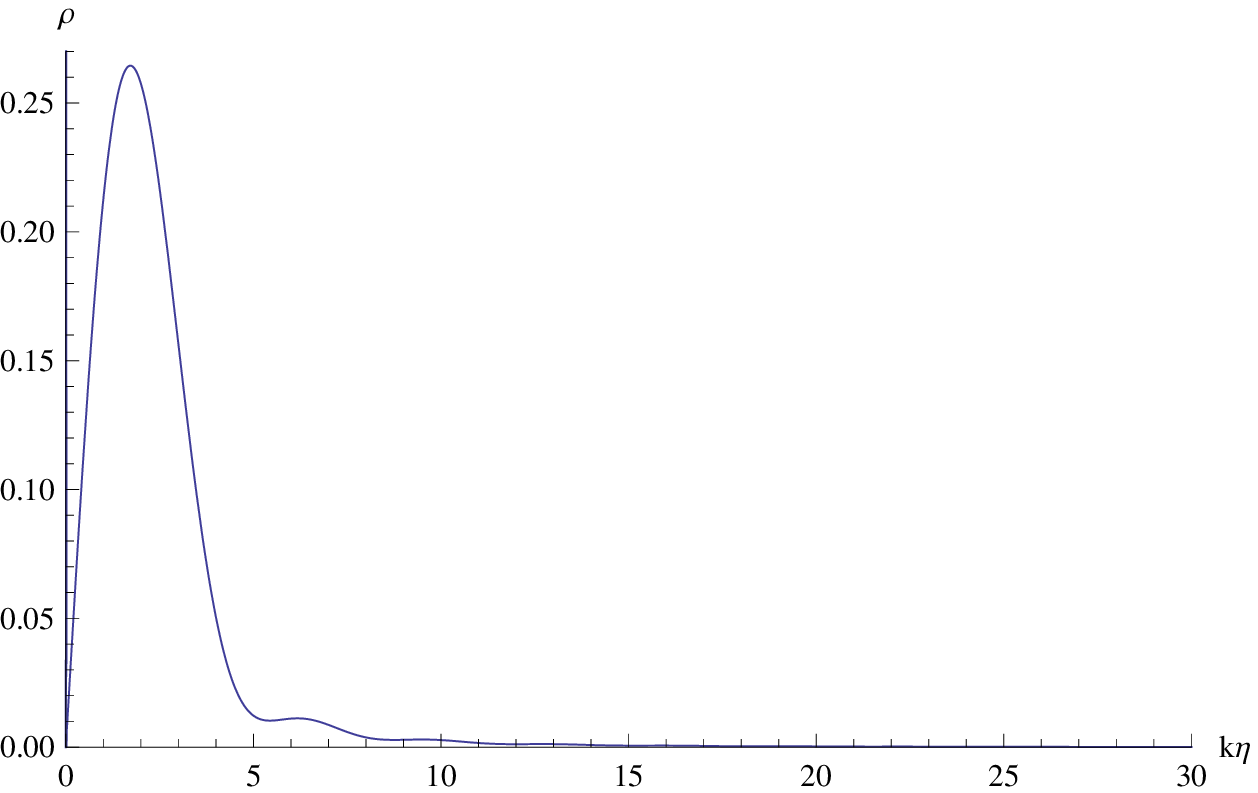}
  \end{minipage}
  \hspace{1.0pc}
\begin{minipage}{.45\linewidth}
\includegraphics[width=1.1 \linewidth]{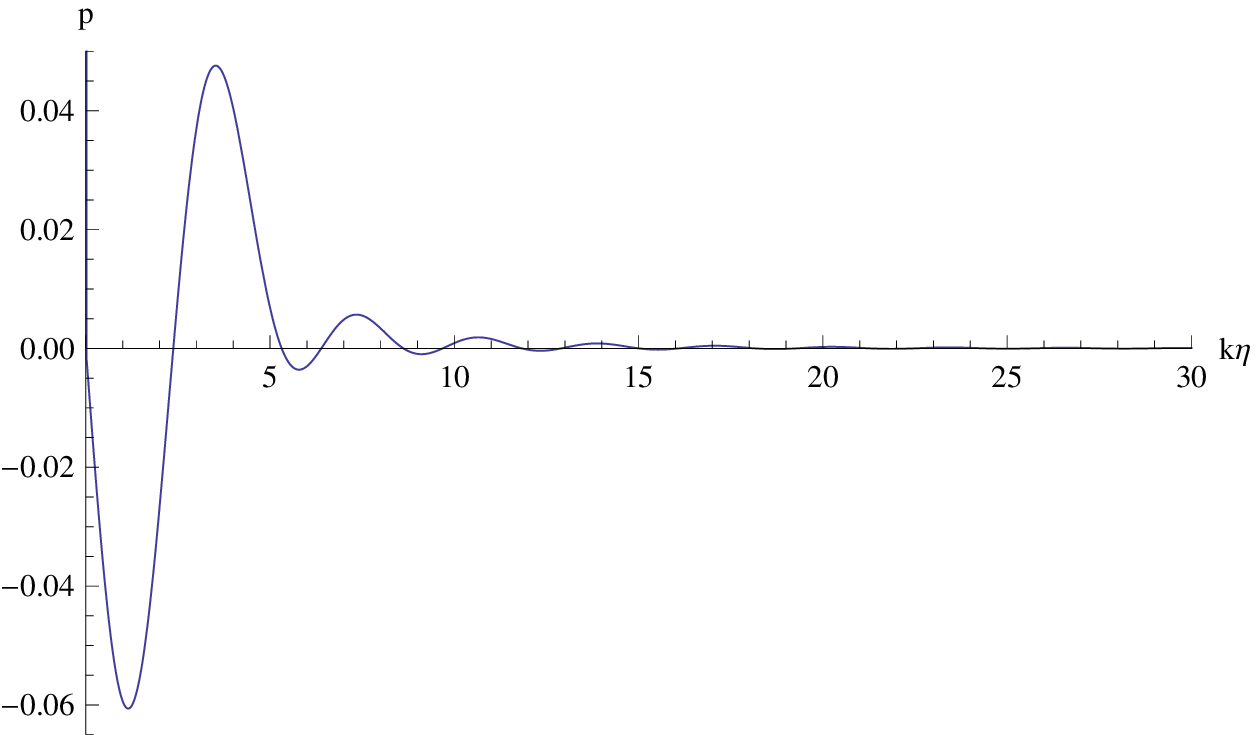}
  \end{minipage}
\end{center}
\caption{
Integrands of (\ref{rhoMDIRregion}) and  (\ref{pMDIRregion}),
without the prefactor $(8\pi^2 a^4 \eta_1^4)^{-1} (\eta/\eta_4)^2$, are plotted 
as a function of $k\eta$.
}
\label{fig:rhopintegrandMDlowappr}
\end{figure}

We next examine the behaviors of the UV region.
For $k > \eta_4^{-1}$, the coefficients $C$ and $D$ in the wave function
become close to $A$ and $B$ since those waves do not scatter in the MD period.
Furthermore, the difference of the integrands
(\ref{rhoMDCD}) and (\ref{pMDCD}) 
from those in the RD period in (\ref{rhoRDAB}) and (\ref{pRDAB})
are higher orders of $1/(k \eta)$ and negligible.
Then one can perform the same analysis mentioned below (\ref{rhoRDUV}) and (\ref{pRDUV}),
and obtain
\beqa
\rho_{\rm MD}^{\rm UV}
&\simeq& \frac{1}{8\pi^2 a^4 \eta_1^4} \int_{\eta_4^{-1}}^{\eta_2^{-1}}\frac{dk}{k} 
=\frac{1}{8\pi^2}(H_I H)^2 \left(\frac{a_{\rm eq}}{a}\right)
\ln{\left(\frac{\eta_4}{\eta_2}\right)} \ , 
\label{rhoMDUV}\\
p_{\rm MD}^{\rm UV}
&\simeq& \frac{1}{8\pi^2 a^4 \eta_1^4} \int_{\eta_4^{-1}}^{\eta_2^{-1}}\frac{dk}{k}\frac{1}{3}
=\frac{1}{3} \rho_{\rm MD}^{\rm UV}
\label{pMDUV}
\eeqa 
where (\ref{prefctMD}) has been used in the second equality.
The UV contribution gives  the equation of state $w^{\rm UV}=p^{\rm UV}/\rho^{\rm UV}=1/3$.
As we will see shortly, this does not mean $w$ approaches $w^{\rm UV}=1/3$ as $\eta \rightarrow \infty$.

We now show the time evolution of $\rho$ and $p$ in the MD period.
The sum of the IR contribution (\ref{rhoMDIRreginted})
and the UV contribution (\ref{rhoMDUV}) gives the total amount of $\rho$.
Similarly, the sum  of (\ref{pMDIRreginted}) and (\ref{pMDUV}) gives $p$.
They are depicted in Figure~\ref{fig:rhopMDtimeevo}.
When the MD period begins,
the UV contribution (\ref{rhoMDUV}) dominates  the IR contribution (\ref{rhoMDIRreginted})
because of  the extra factor
$\ln{(\eta_4/\eta_2)}  \lesssim \ln{(3.5\cdot 10^{25})} \sim 59$ in the UV contribution.
Here, the explicit values (\ref{e4/e0obs}) and (\ref{e2/e0obs}) are used.
The equation of state is then $w \simeq1/3$, which agrees with the late time behavior in the RD 
period. 
This should be the case since $\rho$ and $p$ are connected continuously.
As time passes, however, another factor,  $a_{\rm eq}/a=(\eta_4/\eta)^2$, in (\ref{rhoMDUV}) 
reduces the UV contribution,
and eventually the IR contribution (\ref{rhoMDIRreginted}) dominates over (\ref{rhoMDUV}).
Since the IR contribution for the pressure density $p$ vanishes, as shown in (\ref{pMDIRreginted}),
$p$ evolves according to (\ref{pMDUV}), and the equation of state approaches $w\simeq 0$.

To summarize, the equation of state for the energy density of a quantum 
field starts from $w\simeq -1$ in the 
inflation era, turns into $w\simeq 1/3$ in the RD period,
and then into $w\simeq 0$ in the MD period.
It is interesting that the equation of state of the induced energy
density coincides with that
of the background geometry.
This is consistent with the fact that $\rho$ and $p$  behave as $(H_I H)^2$,
and thus have time dependence of $H^2$.
 
\begin{figure}
\begin{center}
\begin{minipage}{.45\linewidth}
\includegraphics[width=1.1 \linewidth]{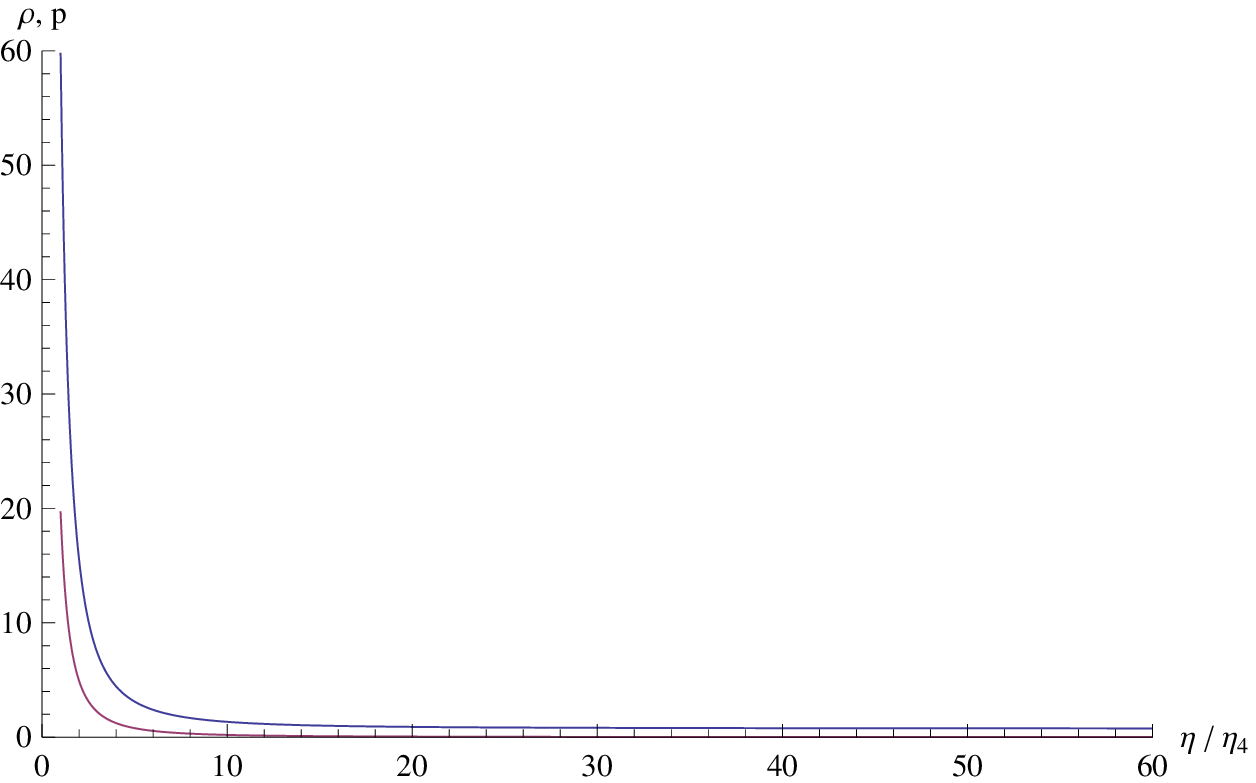}
  \end{minipage}
  \hspace{1.0pc}
\begin{minipage}{.45\linewidth}
\includegraphics[width=1.1 \linewidth]{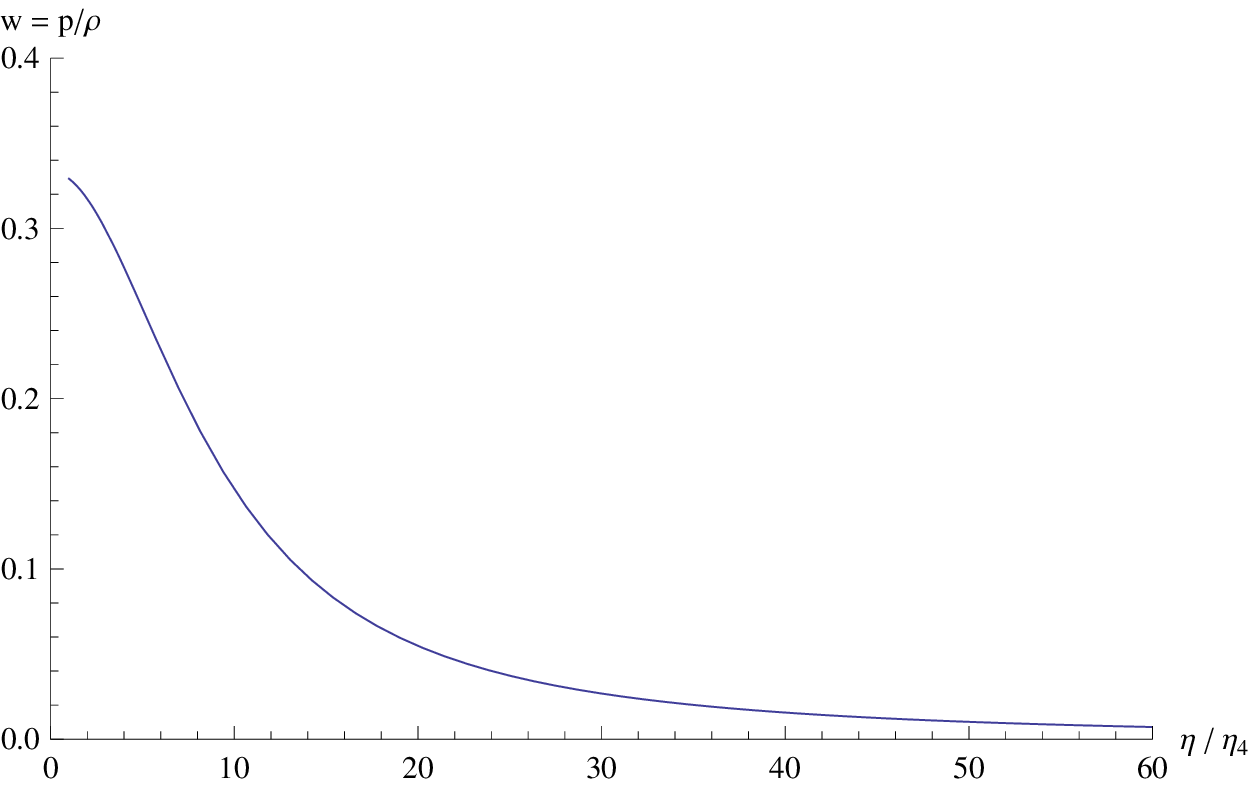}
  \end{minipage}
\end{center}
\caption{Time evolutions  in the MD period.
(Left) The upper and lower curves show 
the time evolutions of $\rho$ and $p$, in the unit of $(H_I H)^2/(8 \pi)^2$.
(Right) The time evolution of the equation of state $w=p/\rho$ is shown.
The parameter is taken to be $\ln{(\eta_4/\eta_2)}=59$.
}
\label{fig:rhopMDtimeevo}
\end{figure}

Finally, the present energy and pressure densities at $\eta=\eta_0$ become\footnote{
These expressions agree with the ones obtained in \cite{Glavan:2013mra}.}
\beqa
\rho_0 &\simeq& \rho_{\rm MD}^{\rm IR}\simeq \frac{3}{32 \pi^2}(H_I H_0)^2 \ , 
\label{rhopreval} \\
p_0 &\simeq& p_{\rm MD}^{\rm UV} \simeq 
\frac{1}{24\pi^2}\left(\frac{a_{\rm eq}}{a_0}\right)\ln{\left(\frac{\eta_4}{\eta_2}\right)}~(H_I H_0)^2 
< 7.0\times 10^{-5} (H_I H_0)^2 \ ,~~ \label{ppreval}
\eeqa
where $a_0/a_{\rm eq} \sim 3.4 \times 10^{3}$ has been used.
If the Hubble parameter at the inflation period were of the order of 
the Planck scale, $H_I \sim M_P$,
the present energy density (\ref{rhopreval}) would be about the observed value $(M_P H_0)^2$.
However, since the CMB fluctuation data
give the constraint $H_I <3.6\times 10^{-5} M_p$,
(\ref{rhopreval}) takes a much smaller value,
$\rho_0 < \frac{3}{32\pi^2}(3.6 \times 10^{-5})^2 (M_P H_0)^4 \sim (6.3 \times 10^{-3}{\rm meV})^4$,
than the desired one for the dark energy density.
Moreover,  unless we include possible effects of interactions,
(\ref{ppreval}) cannot explain the negative pressure of the dark energy.
In the next section, we try to solve the first issue, namely, a possibility to enhance the 
present energy density to the desired value for the dark energy.
The second issue of  emergence of negative pressure due to interactions is left
for future investigations.

\section{Models for Planckian era before inflation}
\label{sec:double}   
\setcounter{equation}{0}

Let us now pursue a possibility that the energy density of
a quantum field is enhanced 
due to the fluctuations created before inflation.

If the inflation period has a finite duration, {\it i.e.}, a finite e-folding,
the long-wavelength modes, 
which are out of causal horizon throughout the inflation period,
are not necessarily specified by the Bunch-Davies vacuum.
If the universe was created quantum gravitationally, it is natural that 
quantum fields had fluctuations of the order of the Planck scale,
and that the long-wavelength modes survive the inflation period.
For instance, in the context of eternal inflation, our universe is generally surrounded by the
region with the larger Hubble parameter, where larger fluctuations are generated.

As a simple example,
we consider a model of a cosmic history with two inflation periods,
{\it i.e.}, the ordinary inflation with the Hubble parameter $H_I$ and another one
with $H_P \sim M_P$ before the ordinary inflation.\footnote{
CMB spectra in particular models that
lead to two stages of inflation have been studied in 
\cite{Jain:2008dw,Jain:2009pm}.}
Because of the large Hubble parameter,
the IR modes of the wave functions are enhanced to an order of $H_P \sim M_P$, and 
give the present energy density with an almost desired value, $(M_P H_0)^2$.
In order to connect the scale factor $a$ and its derivative $a'$ 
continuously between the two inflationary periods, we need to insert 
another period between them. 
A simple model is to insert the RD stage. The analysis of the energy and pressure
densities of this model is performed in Appendix~\ref{sec:chasmmodel}.
In the following, we study another model in which the curvature-dominated period
is sandwiched between the two inflation periods.

\subsection{Double inflation model}
\label{sec:plateaumodel}

We consider a model with  two inflation periods, connected by
an intermediate curvature-dominated (CD) stage.
We call the first inflation a pre-inflation.
The scale factor is given by
\beq
a(\eta)=\left\{ \begin{array}{lll}
a_{\rm PI}(\eta)=-\frac{1}{H_P \eta} &(-\infty<\eta<\tilde{\eta}_1<0) &(\mbox{Pre-Inflation})  \\
a_{\rm CD}(\eta)= e^{\gamma \eta} &(0<\tilde{\eta}_2<\eta<\tilde{\eta}_3) &(\mbox{CD})  \\
a_{\rm Inf}(\eta)=-\frac{1}{H_I \eta} &(\tilde{\eta}_4<\eta<\eta_1<0) &(\mbox{Inflation})
\end{array} \right. \ .
\label{a_doubleinf_pla}
\eeq
The above periods are followed by the big bang universe
(\ref{a_caldera}) with the usual RD and MD periods.
Here, $H_P$ is the Hubble parameter in the pre-inflation period.
In the intermediate stage, one has $a_{\rm CD}(t)=\gamma t$.
In order to satisfy the Friedman equations,
 (\ref{friedmann}) and (\ref{conservationlaw}), we need the energy density
$\rho=3(\gamma M_P)^2 a^{-2}$ and the pressure 
 $p=-\rho/3$, which has the same equation of state as the curvature.
This model is studied  as an example to connect the two inflationary periods continuously,
and we do not discuss here how they are realized.

The matching conditions for the scale factor $a(\eta)$,
{\it i.e.}, continuity of $a$ and $a'$,
between the pre-inflation and CD periods are given by
\beqa
a_{\rm PI}(\tilde{\eta}_1)=a_{\rm CD}(\tilde{\eta}_2)
&:& -\frac{1}{H_P \tilde{\eta}_1}=e^{\gamma \tilde{\eta}_2} \ , \label{macoa1a2_t}\\
a'_{\rm PI}(\tilde{\eta}_1)=a'_{\rm CD}(\tilde{\eta}_2)
&:& \frac{1}{H_P \tilde{\eta}_1^2}= \gamma e^{\gamma \tilde{\eta}_2} \ . \label{macoa1pa2p_t}
\eeqa
Similarly, the conditions between the CD and inflation periods are
\beqa
a_{\rm CD}(\tilde{\eta}_3)=a_{\rm Inf}(\tilde{\eta}_4)
&:& e^{\gamma \tilde{\eta}_3} =  -\frac{1}{H_I \tilde{\eta}_4} \ , \label{macoa3a4_t}\\
a'_{\rm CD}(\tilde{\eta}_3)=a'_{\rm Inf}(\tilde{\eta}_4)
&:& \gamma e^{\gamma \tilde{\eta}_3} =  \frac{1}{H_I \tilde{\eta}_4^2} \ . \label{macoa3pa4p_t}
\eeqa
The conditions are solved as
\beqa
&&\gamma =-\frac{1}{\tilde{\eta}_1}= -\frac{1}{\tilde{\eta}_4} \ , \label{et1et4} \\
&&e^{\gamma(\tilde{\eta}_3-\tilde{\eta}_2)} = \frac{H_P}{H_I} \ . \label{et2et3}
\eeqa

The model has two additional parameters, $H_P$ and $\gamma$.
As mentioned above, $H_P$ is assumed to be close to $M_P$.
$\gamma=-\tilde\eta_4^{-1}$ might be constrained by the condition
that the inflation period has a sufficiently long duration
to solve the horizon and  flatness problems.
The e-folding number of the inflation period is given by
\beq
N_e=\ln \left( \frac{a_{\rm Inf}(\eta_1)}{a_{\rm Inf}(\tilde\eta_4)} \right) =
\ln \left( \frac{\tilde\eta_4}{\eta_1} \right) \ .
\label{efoldin}
\eeq
If we require that the fluctuations within the current horizon
radius must be causally connected when the inflation begins, we have
the condition $\tilde\eta_4/\eta_1 \geq \eta_0/\eta_2.$ 
From (\ref{e2/e0obs}), it gives
$|\tilde\eta_4| \geq 2.1 \times 10^{27} |\eta_1|$, if the Hubble parameter is
$H_I=3.6 \times 10^{-5} M_P.$ 
In fact, $|\tilde\eta_4|$ can be made smaller and 
the beginning of the inflation can be made later if we solve 
the horizon and flatness problems together 
with the pre-inflation. But then the large fluctuations 
generated in the pre-inflation era enter the cosmic horizon earlier and
may become observable in the CMB spectrum. 
We will discuss this issue again 
at the end of this section.

The potential term of  the wave equation (\ref{waveeq}) becomes
\beq
\frac{1}{6}Ra^2 = \frac{a''}{a}
=\left\{\begin{array}{lll}
2/\eta^2 &   (-\infty <\eta < -|\tilde{\eta}_1|) & \mbox{(Pre-Inflation)} \\
\gamma^2 & (\tilde{\eta}_2 <\eta< \tilde{\eta}_3) & \mbox{(CD)} \\
\end{array}
\right. \ ,
\label{potPICD}
\eeq
followed by the potential (\ref{potIRM}) in the inflation, RD, and MD periods.  
It is depicted in Figure~\ref{fig:doubleinf_pot_CD}.
In the CD period, 
the potential has a plateau with the height $\gamma^2$.
The solutions of the wave equations are given by
\beqa
\chi_{\rm PI} &=& \chi_{\rm BD}
\ , \label{chiPI}\\
\chi_{\rm CD} &=& \tilde{A}~\chi_{\rm pl}+\tilde{B}~\tilde{\chi}_{\rm pl} \ ,
\label{chiCD} \\
\chi_{\rm Inf} &=& \tilde C~\chi_{\rm BD} + \tilde D~\chi_{\rm BD}^* 
\ , \label{chiInf2}
\eeqa
in the pre-inflation, CD, and inflation periods, respectively,
and (\ref{chiRD}) and  (\ref{chiMD}) in the RD and MD periods.
Here, $\chi_{\rm pl}$ and $\tilde\chi_{\rm pl}$ are 
\beq
\chi_{\rm pl} = \left \{ \begin{array}{ll}
\frac{1}{\sqrt{2k}} e^{-\kappa\eta} & (k < \gamma)  \\
\frac{1}{\sqrt{2k}} e^{-i \omega\eta} &(k > \gamma)
\end{array} \right. ~,~~~
\tilde{\chi}_{\rm pl} = \left \{ \begin{array}{ll}
\frac{1}{\sqrt{2k}} e^{\kappa\eta} & (k < \gamma)  \\
\frac{1}{\sqrt{2k}} e^{i \omega\eta} &(k > \gamma)
\end{array} \right. \ ,
\eeq
with
\beqa
\kappa&=& \sqrt{\gamma^2-k^2}  \ ,\\
\omega&=& \sqrt{k^2-\gamma^2} \ .
\eeqa
 
\begin{figure}
\begin{center}
\includegraphics[height=5.5cm]{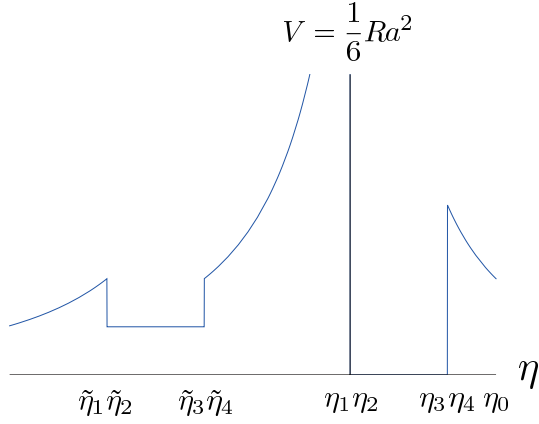}
\end{center}
\caption{Potential of the wave equation in the double inflation model
with an intermediate CD stage. 
It has a plateau with the height $\gamma^2$ in the CD period.
The potential heights at the end of the pre-inflation
and at the beginning of the inflation are the same and given by
$2/\tilde{\eta}_1^2=2/\tilde{\eta}_4^2=2\gamma^2$.
$\eta_0$ is the present time.
}
\label{fig:doubleinf_pot_CD}
\end{figure}

The coefficients $\tilde{A}$, $\tilde{B}$, $\tilde{C}$, and $\tilde{D}$ are determined by
the continuity of $\chi$ and $\chi'$ as
\beq
\begin{pmatrix} \tilde A \cr \tilde B \end{pmatrix}
=
\frac{1}{2}\begin{pmatrix} 
\left[(1+i\frac{k}{\kappa})(1-\frac{i}{k\tilde{\eta}_1})
-i\frac{k}{\kappa}\frac{1}{k^2 \tilde{\eta}_1^2}\right]e^{\kappa\tilde{\eta}_2} \cr
\left[(1-i\frac{k}{\kappa})(1-\frac{i}{k\tilde{\eta}_1})
+i\frac{k}{\kappa}\frac{1}{k^2 \tilde{\eta}_1^2}\right]  e^{-\kappa\tilde{\eta}_2} \end{pmatrix}
 e^{-ik\tilde{\eta}_1} \ ,
 \label{AtBt}
\eeq
\beqa 
&&\begin{pmatrix} \tilde C \cr \tilde D \end{pmatrix}
=\frac{\kappa}{2 i k} 
\left( \begin{array}{c}
\left[(1+i\frac{k}{\kappa})(1+\frac{i}{k\tilde{\eta}_4})
-i\frac{k}{\kappa}\frac{1}{k^2 \tilde{\eta}_4^2}\right] e^{-\kappa\tilde{\eta}_3+ik\tilde{\eta}_4} \\
-\left[(1-i\frac{k}{\kappa})(1-\frac{i}{k\tilde{\eta}_4})
+i\frac{k}{\kappa}\frac{1}{k^2 \tilde{\eta}_4^2}\right] e^{-\kappa\tilde{\eta}_3-ik\tilde{\eta}_4}
\end{array} \right. \n
&&~~~~~~~~~~~~~~~~~~~~~~\left. \begin{array}{c} 
-\left[(1-i\frac{k}{\kappa})(1+\frac{i}{k\tilde{\eta}_4})
+i\frac{k}{\kappa}\frac{1}{k^2 \tilde{\eta}_4^2}\right] e^{\kappa\tilde{\eta}_3+ik\tilde{\eta}_4} \\
\left[(1+i\frac{k}{\kappa})(1-\frac{i}{k\tilde{\eta}_4})
-i\frac{k}{\kappa}\frac{1}{k^2 \tilde{\eta}_4^2}\right] e^{\kappa\tilde{\eta}_3-ik\tilde{\eta}_4}
\end{array} \right)
\begin{pmatrix} \tilde A \cr \tilde B \end{pmatrix} \ ,  \n
\label{AtBtCtDt} 
\eeqa
for $k < \gamma$.
The results for $k > \gamma$ are obtained by replacing $\kappa$ by $i \omega$.
The constants $A$ and  $B$ in (\ref{chiRD})  are determined as
\beq
\begin{pmatrix} A \cr B \end{pmatrix}
=\begin{pmatrix}\left(1-\frac{i}{k\eta_1}-\frac{1}{2k^2 \eta_1^2}\right)e^{-ik(\eta_1-\eta_2)}
&\frac{1}{2k^2 \eta_1^2}e^{ik(\eta_1+\eta_2)}\cr
\frac{1}{2k^2 \eta_1^2}e^{-ik(\eta_1+\eta_2)}
&\left(1+\frac{i}{k\eta_1}-\frac{1}{2k^2 \eta_1^2}\right)e^{ik(\eta_1-\eta_2)} \end{pmatrix}
\begin{pmatrix} \tilde C \cr \tilde D \end{pmatrix} \ ,
\label{ABCpDp} 
\eeq
instead of (\ref{AB}).
The coefficients $C$ and $D$ in (\ref{chiMD}) are determined as (\ref{ABCD}).

We first examine the IR behaviors of the wave function,
and energy and pressure densities.
As shown in Appendix~\ref{sec:wfIRplateau},
the wave function in the IR region 
behaves as
\beq
u^{\rm IR}=\frac{i}{\sqrt{2}} H_P k^{-3/2} + {\cal O}(k^{1/2}) 
\label{u_IR_doubleinf}
\eeq
in all the periods.
As expected,
(\ref{u_IR_doubleinf}) is amplified 
by the Hubble parameter $H_P$ in the pre-inflation period.
Note that (\ref{u_IR_doubleinf}) is valid in the IR region $k< \gamma$ and $k <\eta^{-1}$. 
Substituting the IR wave function  (\ref{u_IR_doubleinf}) into the expressions
(\ref{rhogenu}) and (\ref{pgenu}),
the energy and pressure densities become
\beqa
\rho^{\rm IR} &=& \frac{H_P^2}{8\pi^2 a^2} \int_0 dk \left[k+{\cal O}(k^3)\right]  \ , 
\label{rhoIRdi}\\
p^{\rm IR} &=& -\frac{1}{3}\frac{H_P^2}{8\pi^2 a^2} \int_0 dk \left[k+{\cal O}(k^3)\right]  \ .
\label{pIRdi}
\eeqa
Compared to (\ref{rhoIR}) and  (\ref{pIR}),
they are enhanced by the factor $(H_P/H_I)^2$.

We next examine the UV behavior of the wave function.
For $k > \gamma$, $k$ is above the plateau of the potential in the CD period.
Hence, wave functions with $k> \gamma$ are not affected much by this potential,
and reduce to the previous one without the pre-inflation and the CD periods.
Indeed, eqs.~(\ref{AtBt}) and (\ref{AtBtCtDt}), with $\kappa$ replaced by $i\omega$, show that
the coefficients approach $\tilde C=1$ and $\tilde D=0$ for large $k$.
If we smoothly connect the potential in different periods, $\tilde D$ approaches $0$ much faster.
Consequently, the IR amplification of the wave function in the pre-inflation period
 terminates at $k \sim \gamma$.

Figure~\ref{fig:rhopindRD_DIp} shows the integrands $\rho(k)$ and $p(k)$.
The parameters are taken to be the same as those in the third row of Figure  \ref{fig:rhopintegrandRD}.
The upper figures show that $\rho(k)$ and $p(k)$ acquire a new peak
in the IR region $0<k<\gamma$, which is generated by the pre-inflation.
It is also confirmed in the lower figures.
The upper figures also 
indicate that the integrands reduce to the previous ones for larger $k$.

It is important to note that the new peak is located in the region $k \in [0,\gamma]$,
and independent of $\eta.$ The behavior is different from the peaks generated in the 
ordinary inflation, which are dependent on $\eta$.
The difference comes from the difference of scales 
$|\tilde\eta_1|  \gg |\eta_1|$.
Since the potential peak of the ordinary inflation $2/|\eta_1|^{2}$ is high,
the modes up to large $k = |\eta_1|^{-1}$ are amplified. 
As we discussed before, 
the integrands $\rho(k)$ and $p(k)$
decrease and oscillate with the period $\Delta k=\pi/\eta$,
and this $\eta$-dependent behavior shows up in the  region $k \in [0,|\eta_1|^{-1}]$
since $\pi/\eta < |\eta_1|^{-1}$.
On the contrary, only the modes with $k < \gamma = |\tilde\eta_1|^{-1}$ are 
affected by the pre-inflation.
Then the $\eta$-dependent behavior does not show up 
in the region $k \in [0,|\tilde\eta_1|^{-1}]$
if $\pi/\eta > |\tilde\eta_1|^{-1}$.
In other words, the modes enhanced in the inflationary period are
entering the horizon after the big bang, but the modes enhanced in the pre-inflation period
are still out of the horizon. It causes the difference of the behaviors 
of the integrands $\rho(k)$ and $p(k)$,
and consequently the big difference of the time evolution of the 
energy densities as explained
in Section~\ref{sec8.2}.
 
\begin{figure}
\begin{center}
\begin{minipage}{.45\linewidth}
\includegraphics[width=1.1 \linewidth]{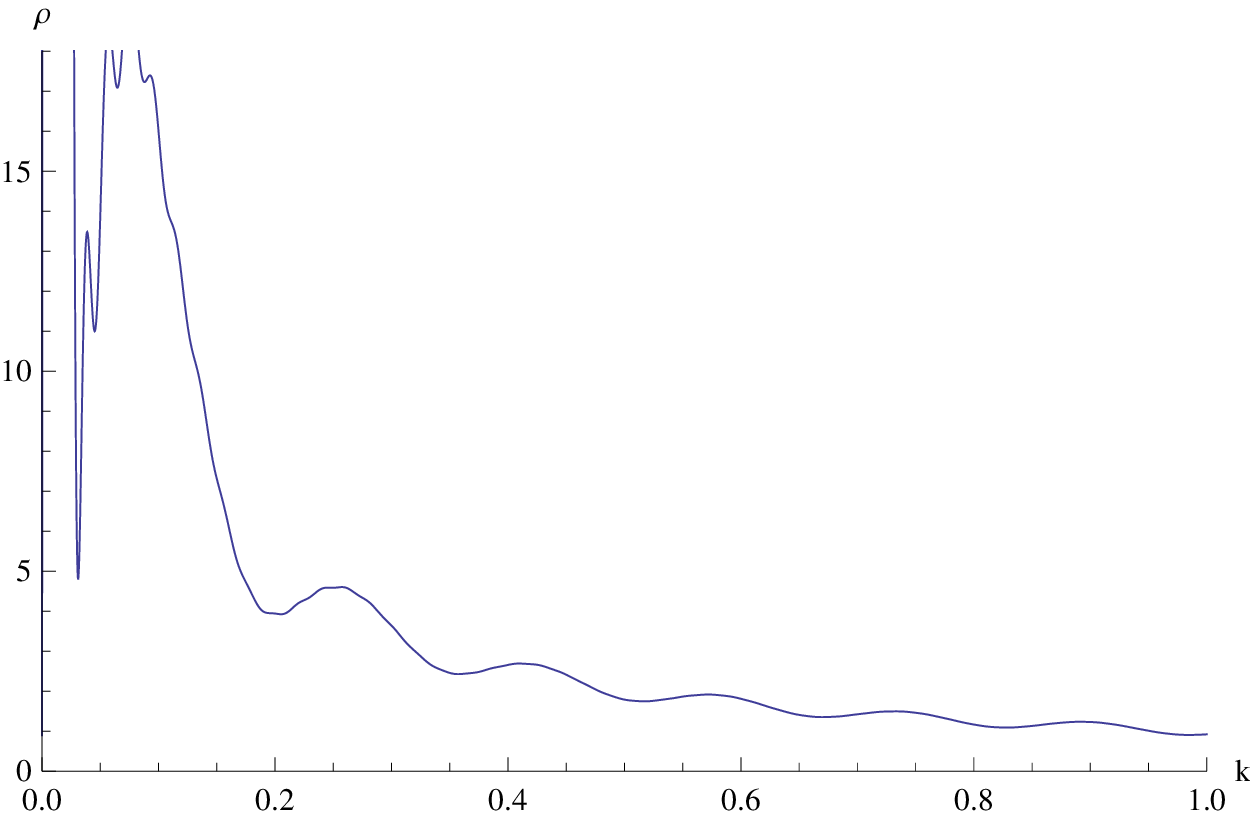}
  \end{minipage}
  \hspace{1.0pc}
\begin{minipage}{.45\linewidth}
\includegraphics[width=1.1 \linewidth]{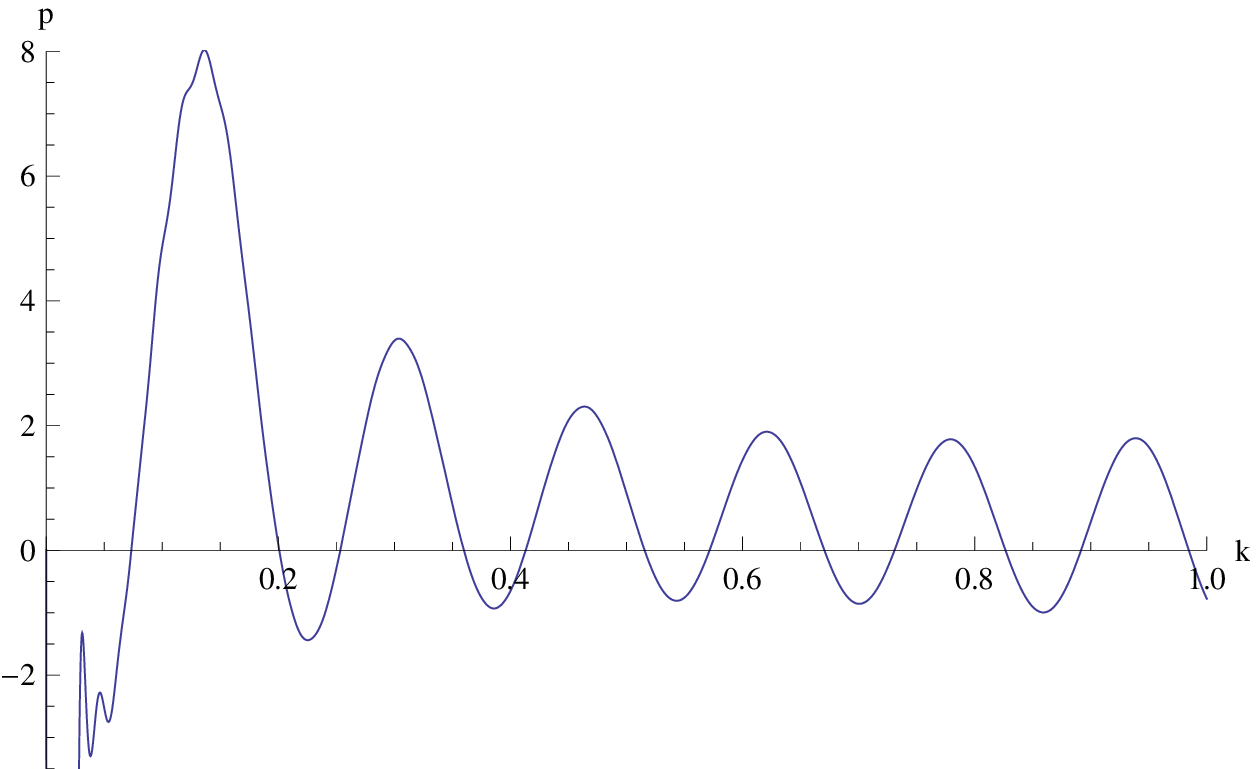}
  \end{minipage}
  \begin{minipage}{.45\linewidth}
\includegraphics[width=1.1 \linewidth]{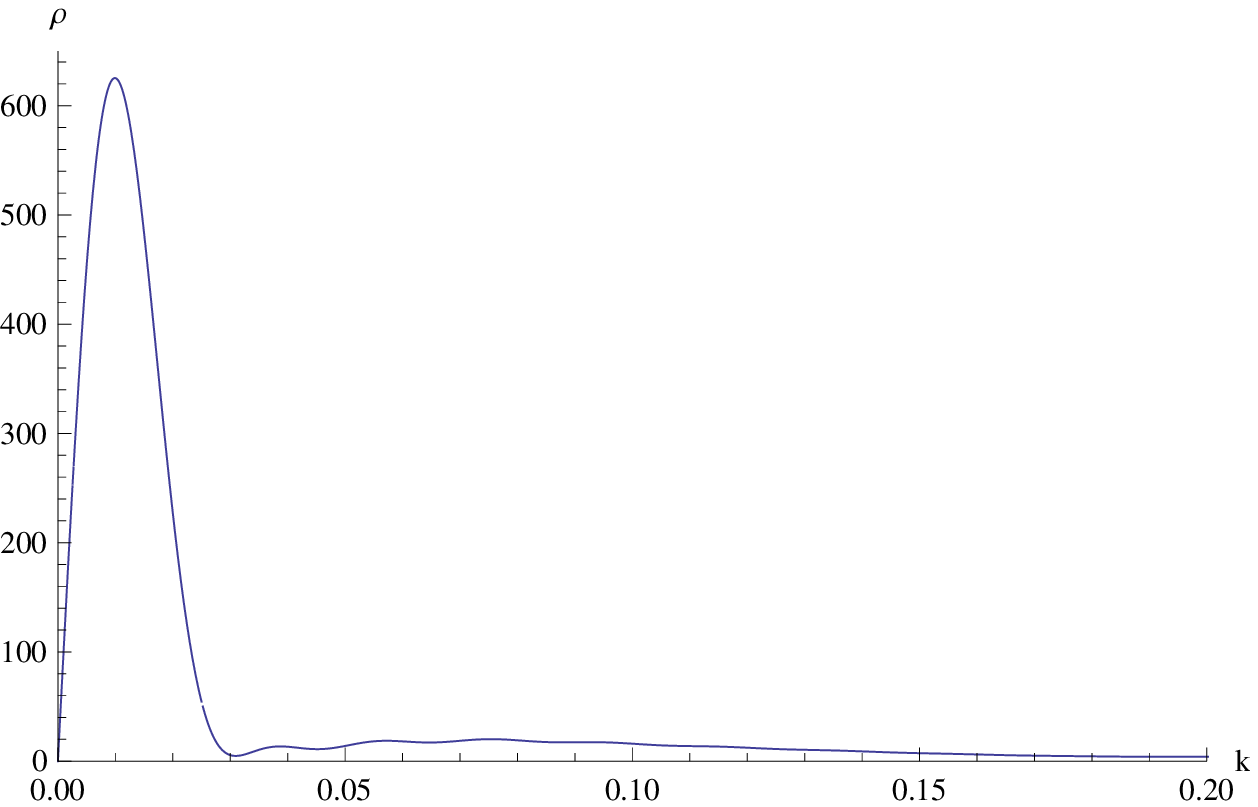}
  \end{minipage}
  \hspace{1.0pc}
\begin{minipage}{.45\linewidth}
\includegraphics[width=1.1 \linewidth]{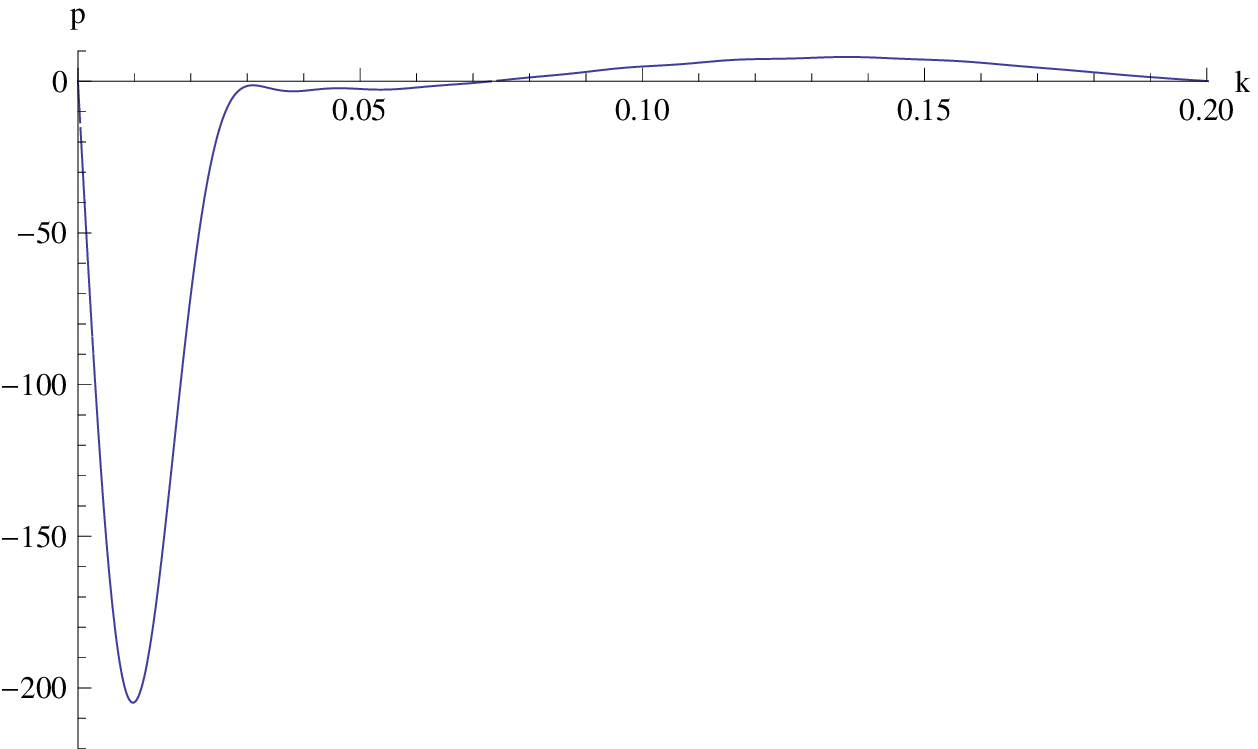}
  \end{minipage}
\end{center}
\caption{The integrands of $\rho(k)$ and $p(k)$ in the RD period,
caused by the double inflation model.
As in the third row of Figure~\ref{fig:rhopintegrandRD}, we take
 $\eta_2=-\eta_1=1$ and $\eta=20$. 
The other parameters are set to be
$\gamma^{-1}=-\tilde\eta_1=-\tilde\eta_4=40$ and  $H_P/H_I =16$.
Accordingly, $\tilde\eta_3-\tilde\eta_2=\gamma^{-1} \ln (H_P/H_I) =40 \ln (16) \sim 111$.
Note that these values of $\gamma$ and $H_P$ are not realistic, but just chosen here
to draw the figures.
In the upper figures, the scale of the axes is taken to be
the same as in the third row of Figure  \ref{fig:rhopintegrandRD}.
The lower figures magnify the IR region of the upper figures.
We can see modifications of $\rho$ and $p$ in the IR region by the pre-inflation.
}
\label{fig:rhopindRD_DIp}
\end{figure}

We now perform the integration of $\rho(k)$ and $p(k)$ over $k.$
When $\eta<|\tilde\eta_1|$, 
the integrations are dominated by the large IR peak
generated by the pre-inflation period and given by
\beqa
\rho^{\rm IRpeak}
&\sim& \frac{H_P^2}{8\pi^2 a^2} 2 \int_0^{\gamma/2} dk~k 
=\frac{H_P^2}{32\pi^2 a^2} \gamma^2 \ ,
\label{rhoDIplIRpeak} \\
p^{\rm IRpeak}
&=& -\frac{1}{3} \rho^{\rm IRpeak} \ ,
\label{pDIplIRpeak}
\eeqa
where we approximated the position of the peak  at $\gamma/2$.
The factor $2$ comes from an approximation that
the peak has a form of an isosceles triangle.  
They give the equation of state $w=p/\rho=-1/3$. 
This is consistent with the fact that the time dependence of (\ref{rhoDIplIRpeak})
and (\ref{pDIplIRpeak})  is $a^{-2}$.
Neither the integrand nor the upper bound $\gamma/2$ depends on $\eta$.
It is due to the condition of $\eta< |\tilde\eta_1|$.
Namely, the amplified waves in the pre-inflation period are still out of the horizon 
and are not yet affected by the oscillatory behaviors of quantum waves.

When time passes and $\eta$ becomes larger than $\gamma^{-1}=|\tilde\eta_1|$,
the oscillating and decreasing behavior of the integrand with the period $\Delta k =\pi/\eta$
enters the interval $0 < k < \gamma$ of the IR peak,
and $\eta$ dependence  other than $a^{-2}$ arises.
The results approach those in Section~\ref{sec:timeevoMD},
with $H_I$ replaced by $H_P$,
if the background geometry of the MD period continues.

The energy density (\ref{rhoDIplIRpeak}) depends on the two parameters of the model,
$H_P$ and  $\gamma=|\tilde\eta_4|^{-1}$. 
We consider natural values for these parameters. 
Since we consider the Planckian era, $H_P$ is close to $M_P$.
The other parameter, $\gamma=|\tilde\eta_4|^{-1}$, is related to
the e-folding number (\ref{efoldin}). 
The necessary e-folding number 
to solve the horizon and flatness problems
can be gained together with the pre-inflation era.
However, if we chose $\gamma>  2\pi \eta_0^{-1}$, 
the enormously amplified wave functions by the pre-inflation,
which lie in the region $k \in [0, \gamma]$, 
would enter the current horizon,
which corresponds to the comoving wave number $k=2\pi\eta_0^{-1}$.
Such modes within the horizon could be detected.
For instance, they may generate additional CMB fluctuations and 
receive strong constraints from the observed data.
We then choose $\gamma \leq  2\pi \eta_0^{-1}$

To conclude this section,
if we set the parameters of the model as $H_P = M_P$ and $\gamma = 2\pi\eta_0^{-1}$,
(\ref{rhoDIplIRpeak}) gives $\rho_0=(M_P H_0)^2/32$ at present,
where $H_0 = 2(a_0\eta_0)^{-1}$ is used.
It becomes close to the desired order of magnitude of the dark energy 
$\sim (M_P H_0)^2$  at present.

\subsection{Time evolution of energy and pressure densities}
\label{sec8.2}

We then study how the energy and pressure densities,
caused by the double inflation model with 
$H_P = M_P$ and $\gamma = 2\pi\eta_0^{-1}$, evolve with time.

The energy density generated by the pre-inflation is estimated in (\ref{rhoDIplIRpeak}).
Substituting $H_P = M_P$ and $\gamma = 2\pi\eta_0^{-1}$, it becomes
\beq
\rho_{\rm pre-inf}\simeq \frac{1}{8}M_P^2\frac{1}{a^2\eta_0^2}
= \left\{\begin{array}{ll}
\frac{1}{32}(M_P H)^2 \left(\frac{a_{\rm eq}}{a_0}\right) \left(\frac{a_{\rm BB}}{a_{\rm eq}}\right)^2
\left(\frac{a_{\rm BB}}{a}\right)^2  & \mbox{(Inflation)} \\
\frac{1}{32}(M_P H)^2 \left(\frac{a_{\rm eq}}{a_0}\right) \left(\frac{a}{a_{\rm eq}}\right)^2 & \mbox{(RD)} \\
\frac{1}{32}(M_P H)^2 \left(\frac{a}{a_0} \right) & \mbox{(MD)}
\end{array}\right. \ .
\label{rhoPreInfevo}
\eeq  
In the last equality, we used (\ref{eta0H0a0}) and 
 factored out $H^2$, which is proportional to 
$a^{-3}$, $a^{-4}$, and $a^0$ at the MD, RD, and inflation periods, respectively.
Here, BB stands for the big bang, the time when the RD period begins.
eq and 0 stand for 
the matter-radiation equality and the present.
The equation of state is given by $w=-1/3.$

On the other hand, 
the energy density produced by the standard inflation is given
by (\ref{rhoRDUV}) in the RD period, 
and the sum of (\ref{rhoMDIRreginted}) and (\ref{rhoMDUV}) in the MD period.
It is written as
\beq
\rho_{\rm inf}\simeq \left\{\begin{array}{ll}
\frac{1}{8\pi^2}(H_I H)^2 \ln{(\frac{a}{a_{\rm BB}})}& \mbox{(RD)} \\
\frac{1}{8\pi^2}(H_I H)^2 \left[\frac{3}{4} + \ln{(\frac{a_{\rm eq}}{a_{\rm BB}})~ 
\left(\frac{a_{\rm eq}}{a}\right)}\right]& \mbox{(MD)} \\
\end{array}\right. \ .
\label{rhoInfevo}
\eeq
The equation of state is $w=1/3$ in the RD period and approaches $w=0$ in the MD period.

Figure~\ref{fig:timeevorhos} shows the time evolution of 
the ratios of the energy densities,
(\ref{rhoPreInfevo}) and (\ref{rhoInfevo}),  to the
critical energy density
\beq
\rho_{\rm cr}=3 (M_P H)^2 \ 
\label{rhocritical}
\eeq
in the RD and MD periods.
The ratios are depicted as a function of  $a/a_{\rm BB}$ in the logarithmic scale.

\begin{figure}
\begin{center}
\includegraphics[height=5.5cm]{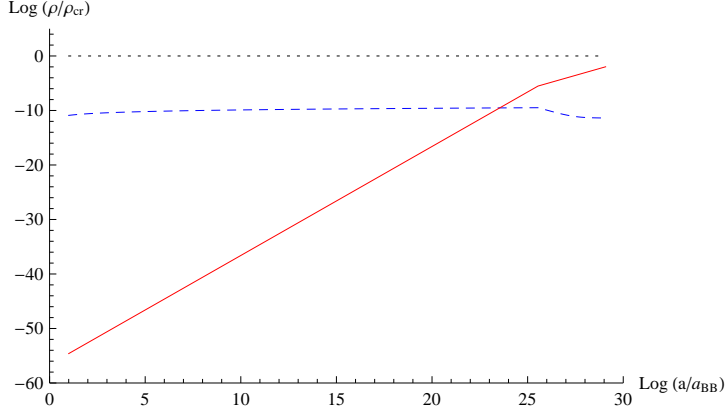}
\end{center}
\caption{
Time evolution of the energy density of a quantum field
in the RD and MD periods. The energy density 
divided by the critical energy is depicted against 
the scale factor $a/a_{\rm BB}$
in the logarithmic scale.
The solid and dashed lines represent the ratios of the energy
density generated in the pre-inflation and inflation, respectively,
to the critical energy density.
The dotted line corresponds to the critical density.
The parameters are taken to be $H_I/M_P=3.6\times 10^{-5}$,
$a_{\rm eq}/a_{\rm BB}=3.5 \times10^{25}$, and $a_0/a_{\rm BB}=1.2 \times10^{29}$.
}
\label{fig:timeevorhos}
\end{figure}

The ratio $\rho_{\rm inf}/\rho_{\rm cr}$ is almost constant since both 
$\rho_{\rm inf}$ and $\rho_{\rm cr}$ have a time dependence of $H^2$.
$\rho_{\rm inf}$ is always smaller 
than the critical value $\rho_{\rm cr}$ by the factor of $(H_I/M_P)^2$.
On the contrary, $\rho_{\rm pre-inf}$ has the time dependence 
shown in (\ref{rhoPreInfevo}).
The ratio  $\rho_{\rm pre-inf}/\rho_{\rm cr}$ decreases in the inflationary period and
takes a small value when the RD period starts.
But it grows after that.       
$\rho_{\rm pre-inf}$ becomes larger than $\rho_{\rm inf}$
at $a_* = 3.5 \times 10^{23} a_{\rm BB} =2.9\times10^{-6} a_0$,
and eventually approaches $\rho_{\rm cr}$.
Hence, the energy density of a quantum field is 
dominated by the contribution $\rho_{\rm inf}$ before $a=a_*$
and by $\rho_{\rm pre-inf}$ after that.
The equation of state for the sum of these two contributions
changes from 
$w_{\rm vac}=1/3$ to $w_{\rm vac}=-1/3$ around $a=a_*.$ 

In the  future, as we mentioned below (\ref{rhoDIplIRpeak}),
the wave functions amplified in the pre-inflation enter the horizon and
the energy density
evolves as $(M_P H)^2/32$.
If the background geometry of the MD period continues, the equation of state eventually
becomes $w_{\rm vac}=0$. 
But if the induced energy dominates the critical density, we need to take back reactions into
account, and the equation of state will also be changed accordingly.

If we take $(H_P, \gamma)=(M_P, 2\pi \eta_0^{-1})$, 
the energy density of a quantum field
becomes $(M_P H_0)^2/32$ at present.
It is still 96 times smaller than the critical value (\ref{rhocritical}).
In order to obtain a larger value,
we may make either $H_P$ larger or $\gamma=|\tilde\eta_4|^{-1}$ larger.
For simplicity,  we fix $H_I=3.6 \times 10^{-5} M_P$, and, accordingly, the conformal time
at the end of the standard inflation $\eta_1$ is fixed. 
If we write $|\tilde\eta_4|^{-1}=2\pi \eta_0^{-1} x$, 
$x$ is related to the e-folding number during the ordinary inflation as
$x=\frac{1}{2\pi}\frac{\eta_0}{\eta_2}\frac{\eta_1}{\tilde\eta_4}=e^{61.1-N_e}$. 
$x=\frac{1}{2\pi}$ corresponds to the lower value of the e-folding in 
the ordinary scenario of the inflation.
Then the energy density (\ref{rhoDIplIRpeak}) at present becomes
\beq
\rho_0 = \frac{e^{2(61.1-N_e)}}{32} (H_P H_0)^2 \ .
\eeq
In order to make it comparable to the critical density, we need $x H_P/M_P \sim 10$.
The condition can be satisfied by setting, {\it e.g.}, $(H_P, N_e)=(M_P, 58.8)$, $(5 M_P, 60.4)$,
or $(10M_P, 61.1)$.
These numbers should not be taken at face value since they are sensitive to 
details of the cosmic history such as the intermediate
stage between the two inflations, or interactions neglected in the analysis of this paper.

\section{Conclusions and discussion}
\label{sec:discussion}   
\setcounter{equation}{0}

In this paper, we calculated the time evolution of the energy-momentum
tensor of a minimally-coupled massless scalar field throughout
the history of the universe. 
We considered two types of cosmic histories.
The first one is the standard cosmology model, starting from the 
inflation and followed by the RD and MD periods. To perform 
quantization, the Bunch-Davies initial condition is imposed on
the field in the inflation period. Due to the fact that inflation
produces fluctuations of the order $H_I$, the energy
density of a quantum field becomes of the order
$\rho_{\rm inf}\sim (H_I H)^2$ where $H$ and $H_I$ are the 
Hubble parameters at each moment and in the inflation period.
The evolution of  $\rho_{\rm inf}$ is given by eq.~(\ref{rhoInfevo})
and its ratio to the critical density is depicted as the dashed line in 
Figure~\ref{fig:timeevorhos}. The ratio is almost constant but the
magnitude 
is much smaller than the critical density.
The equation of state evolves from a negative value to $w=1/3$ in the RD period,
and from $1/3$ to $w=0$ in the MD period.

The second type of model we considered 
is a double inflation model, in which
we assume another inflation, which we call a pre-inflation, exists
before the ordinary inflation starts.
The present energy density of the fluctuations 
generated in the pre-inflation is enhanced to an order of
$\rho_{\rm pre-inf} \sim (H_P H_0)^2$ and is comparable to the critical density if $H_P \sim M_P$. 
 Here, $H_P$ is the Hubble parameter of the pre-inflation and $M_P$ is the Planck scale.
The evolution of $\rho_{\rm pre-inf}$, given by eq.~(\ref{rhoPreInfevo}), is  different
from $\rho_{\rm inf}$ because the amplified wave functions in the pre-inflation
have larger wavelengths and have not entered the horizon yet. 
The ratio to the critical density is depicted by the solid line in Figure~\ref{fig:timeevorhos}.
The equation of state is given by 
$w=p/\rho = -1/3$. It is negative, but cannot drive the acceleration of our universe
within the free field approximation.

Our result suggests that fluctuations generated before 
the ordinary inflation could have an important effect on the present 
universe. We have considered the double inflation models as a simple
explicit example in this paper. There are theoretically well-motivated models 
in the context of eternal inflation. In a theory with
a metastable vacuum, universes are created by  
bubble nucleation~\cite{CDL}; such universes are 
surrounded by de Sitter space 
(the ancestor vacuum) with the Hubble parameter being supposedly much higher 
than the one for the ordinary inflation $H_I$. The vacuum state in 
such a universe can be defined by the Euclidean prescription,
and is different from the Bunch-Davies vacuum for ordinary 
inflation~\cite{FSSY}. It is an important problem to find the 
renormalized energy-momentum tensors in such a universe, 
extending the analysis of this paper.

In this work, we have not considered back reactions from the
quantum fluctuations to the geometry. In recent 
papers~\cite{Habara1, Habara2}, in which 
the effect of a large number of fields (Kaluza-Klein and string
states) on the CMB fluctuations was studied, the authors made
an interesting proposal that the vacuum energy from these
fluctuations itself drives the acceleration in inflation.
Back reactions to the geometry
in this context are being studied by these authors.
In our work, the energy density of the fluctuations is 
much smaller than the critical value
 during most of the periods in the cosmic history.
However, it becomes relevant at the very
early universe when the Hubble parameter is close to the Planck scale,
and also at times later than the present where the vacuum
fluctuation found in this paper starts to be dominant. 
It would be a very interesting theoretical problem to 
study the late time behavior to understand the fate of the 
universe. 
In this region, the interplay among the
scale factor, the IR behavior of the  wave function, and 
the Bunch-Davies initial condition should
determine the dynamics of the universe self-consistently.

We solved the evolution equation in the free field
approximation. The system is very simple, yet the evolution of 
the energy density of a quantum field 
summarized in Figure~\ref{fig:timeevorhos} is  nontrivial. 
Then what will happen if we include  interactions among them? 
Although there have been 
debates on the physical effect of quantum loops in the de Sitter background 
(see, {\it e.g.}, \cite{Tanaka} and references therein), it would be 
reasonable
to assume that massive (including light but not exactly massless) 
fields will reach the interacting Hartle-Hawking vacuum~\cite{Marolf}. 
If this is the case, the effect of loops modifies the numerical coefficients
in front of $\rho$, but not their qualitative behaviors.

As a final remark, we mention a possible scenario of
how interactions change the equation of state.
As is well known in the Bogoliubov theory of superfluidity,
quantum Bose gases fall into a macroscopic state in the presence
of interactions. If the interaction is attractive, the system is unstable
with negative pressure. Let us assume here that condensation occurs
in our system.  
Indeed, the scalar field has gravitational interaction.
 Since the gravitational interaction is very weak 
and the energy density is also very low, 
it will take a long time for the condensation to occur. 
Once the condensation occurs,
the interaction energy dominates the kinetic energy and
 the equation of state will be changed from the free case considered in 
 the present paper to the  interaction dominated form.
Since the gravitational interaction is attractive, 
the negative pressure is expected to arise. 
It is interesting to understand  the time scale of the condensation dynamics and 
 evolutions of the energy-momentum tensor modified by gravitational
interactions.
We will investigate these issues in future publications. 

\section*{Acknowledgements}
The authors would like to thank Yoshinobu Habara, 
Hikaru Kawai, Hiroyuki Kitamoto, Yoshihisa Kitazawa, 
and Masao Ninomiya for valuable comments and discussions, 
and the participants in the KEK theory workshop
held on February 18-21, 2014, for valuable discussions.
This work is supported
in part by Grant-in-Aid for Scientific Research
(No. 23244057, 23540329, 24540279 and 24540293) from 
the Japan Society for the Promotion of Science.
 This work is also partially supported by
``The Center for the Promotion of Integrated Sciences (CPIS)'' of Sokendai.

\appendix

\section{Double inflation model with an intermediate RD stage}
\label{sec:chasmmodel}
\setcounter{equation}{0}

In this appendix, we consider a double inflation model, as in Section~\ref{sec:plateaumodel},
but with the RD period as an intermediate stage.
We call the fist inflation a pre-inflation period,
and the subsequent intermediate RD stage a pre-RD period.
The scale factor $a(\eta)$ is given by
\beq
a(\eta)=\left\{ \begin{array}{lll}
a_{\rm PI}(\eta)=-\frac{1}{H_P \eta} &(-\infty<\eta<\eta'_1<0) &(\mbox{Pre-Inflation})  \\
a_{\rm PR}(\eta)=\al' \eta &(0<\eta'_2<\eta<\eta'_3) &(\mbox{Pre-RD})  \\
a_{\rm Inf}(\eta)=-\frac{1}{H_I \eta} &(\eta'_4<\eta<\eta_1<0) &(\mbox{Inflation})
\end{array} \right. \ ,
\label{a_preinf}
\eeq
instead of (\ref{a_doubleinf_pla}).
The matching conditions,
{\it i.e.}, the continuity of $a$ and $a'$, between the pre-inflation and pre-RD periods are given by
\beqa
a_{\rm PI}(\eta'_1)=a_{\rm PR}(\eta'_2)&:& -\frac{1}{H_P \eta'_1}=\al' \eta'_2 \ , \label{macoa1a2_p}\\
a'_{\rm PI}(\eta'_1)=a'_{\rm PR}(\eta'_2)&:& \frac{1}{H_P \eta'_1{}^2}=\al' \ , \label{macoa1pa2p_p}
\eeqa
and those between the pre-RD and inflation periods are
\beqa
a_{\rm PR}(\eta'_3)=a_{\rm Inf}(\eta'_4)&:& \al' \eta'_3 =  -\frac{1}{H_I \eta'_4} \ , \label{macoa3a4_p}\\
a'_{\rm PR}(\eta'_3)=a'_{\rm Inf}(\eta'_4)&:& \al' =  \frac{1}{H_I \eta'_4{}^2} \ . \label{macoa3pa4p_p}
\eeqa
They are solved as
\beqa
-\eta'_1&=& \eta'_2 \ , \label{ep1ep2}\\
\eta'_3&=&-\eta'_4 \ , \label{ep3ep4}\\
\frac{\eta'_1}{\eta'_4} &=& \sqrt{\frac{H_I}{H_P}} \ . \label{ep1ep4}
\eeqa
Note that (\ref{ep1ep4}) gives the constraint $\eta_4'/\eta'_1 > 1.7 \times 10^2 \gg 1$.

The potential for the wave equation (\ref{waveeq}) becomes
\beq
\frac{1}{6}Ra^2 = \frac{a''}{a}
=\left\{\begin{array}{ll}
2/\eta^2 & \mbox{(Pre-Inflation)} \\
0 & \mbox{(Pre-RD)} \\
\end{array}
\right. \ ,
\label{potPIPR}
\eeq
followed by (\ref{potIRM}). It is depicted in Figure~\ref{fig:doubleinf_pot_RD}.
Since the potential vanishes in the RD period,
the pre-RD period gives a chasm in the potential.
The solution for the wave equation is given as
\beq
\chi_{\rm PR} = A'~\chi_{\rm PW}+B'~\chi_{\rm PW}^*
\label{chiPR}
\eeq
in the pre-RD period.
The wave functions in the pre-inflation and inflation periods are
given by (\ref{chiPI}) and 
(\ref{chiInf2}), with $\tilde C$ and $\tilde D$ replaced by $C'$ and $D'$,
and those in the RD and MD periods
are given by (\ref{chiRD}) and  (\ref{chiMD}).

\begin{figure}
\begin{center}
\includegraphics[height=5.5cm]{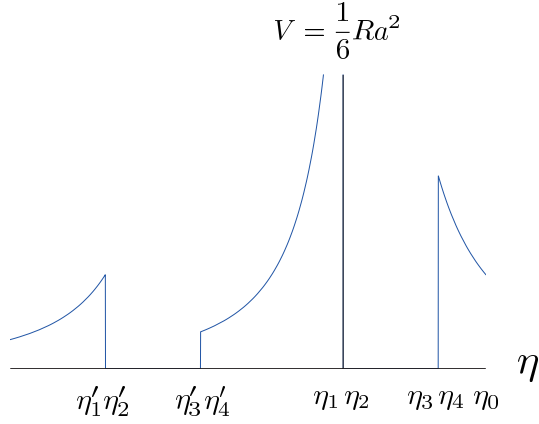}
\end{center}
\caption{
Potential for the wave equation in the double inflation model
with the intermediate RD stage.
$\eta'_1$ to $\eta'_4$ and $\eta_1$ to $\eta_4$ are
the edges of the pre-inflation, pre-RD, inflation, RD, and MD periods.
$\eta_0$ is the present time.
}
\label{fig:doubleinf_pot_RD}
\end{figure}

The constants $A'$, $B'$, $C'$, and $D'$ are determined by
the matching conditions for the wave functions, {\it i.e.},
the continuity of $\chi$ and $\chi'$,
as
\beq
\begin{pmatrix} A' \cr B' \end{pmatrix}
=\begin{pmatrix} \left(1-\frac{i}{k\eta'_1}-\frac{1}{2k^2 \eta'_1{}^2}\right)e^{ik\eta'_2} \cr
 \frac{1}{2k^2 \eta'_1{}^2} e^{-ik\eta'_2} \end{pmatrix}
 e^{-ik\eta'_1} \ ,
 \label{ApBp}
 \eeq
 \beq
\begin{pmatrix} C' \cr D' \end{pmatrix}
=\begin{pmatrix}\left(1+\frac{i}{k\eta'_4}-\frac{1}{2k^2 \eta'_4{}^2}\right)e^{ik(\eta'_4-\eta'_3)}
&-\frac{1}{2k^2 \eta'_4{}^2}e^{ik(\eta'_4+\eta'_3)}\cr
-\frac{1}{2k^2 \eta'_4{}^2}e^{-ik(\eta'_4+\eta'_3)}
&\left(1-\frac{i}{k\eta'_4}-\frac{1}{2k^2 \eta'_4{}^2}\right)e^{-ik(\eta'_4-\eta'_3)} \end{pmatrix}
\begin{pmatrix} A' \cr B' \end{pmatrix} \ ,
\label{ApBpCpDp} 
\eeq
which have exactly the same form as  (\ref{AB})  and (\ref{ABCD})
with $A$, $B$, $C$, $D$, and $\eta_1, \dots, \eta_4$ replaced by those with a prime.
The coefficients $A$, $B$, $C$, and $D$ here are determined by 
(\ref{ABCpDp}), with $\tilde C$ and $\tilde D$ replaced by $C'$ and $D'$, and (\ref{ABCD}).

The IR behaviors of the wave function can be estimated as before.
As will be shown in Appendix \ref{sec:wfIRchasm},
the IR behavior (\ref{u_IR_doubleinf}) is held in all the periods. 
The enhancement of the integrand for the energy and pressure densities, 
(\ref{rhoIRdi}) and (\ref{pIRdi}), is obtained accordingly.

The UV behavior, on the other hand, has a slightly complicated structure.
Let us recall that  (\ref{ep1ep4}) gives $|\eta'_1|^{-1}  \gg |\eta'_4|^{-1}$.
It follows that the potential height at the end of the pre-inflation and that at the beginning of the inflation 
in Figure~\ref{fig:doubleinf_pot_RD} are largely different.
For $k <  |\eta'_4|^{-1}$, the above estimations of the IR behaviors are valid,
and the large enhancement (\ref{rhoIRdi}) and (\ref{pIRdi}) is obtained.
For $k > |\eta'_1|^{-1}$, since $k$ is above the potential throughout the pre-inflation and pre-RD periods,
the results are not affected by those periods,
and are reduced to the original ones without the pre-inflation.
However, in the present model, 
there exist modes with $ |\eta'_4|^{-1} < k<  |\eta'_1|^{-1}$,
which are modified from the original ones, but are not enhanced as largely as the IR modes.

We now examine the modes with $ |\eta'_4|^{-1} < k <  |\eta'_1|^{-1}$.
For $\eta_3^{\prime -1} = |\eta'_4|^{-1} < k <   |\eta'_1|^{-1}= \eta_2^{\prime -1}$,
by using the leading terms in (\ref{ApBp}), (\ref{ApBpCpDp}), and  (\ref{ABCpDp}),
one obtains
\beqa
\begin{pmatrix} A \cr B \end{pmatrix} &\sim&
\frac{-1}{2 k^2\eta_1^2}\begin{pmatrix} 1 & -1 \cr-1 & 1 \end{pmatrix}
\begin{pmatrix} e^{ik(\eta'_4-\eta'_3)} & \cr & e^{-ik(\eta'_4-\eta'_3)} \end{pmatrix}
\frac{-1}{2 k^2\eta_1^{\prime 2}} \begin{pmatrix} 1 \cr -1 \end{pmatrix} \ \ \\
&=& \frac{1}{2k^4\eta_1^2\eta_1^{\prime 2}}
\begin{pmatrix} \cos{k(\eta'_4-\eta'_3)} \cr -\cos{k(\eta'_4-\eta'_3)} \end{pmatrix} \ .
\eeqa
In the transitions from the pre-inflation to pre-RD period, and from the inflation to the RD period,
the complete scattering takes place due to the high potential barrier,
while in the transition from the pre-RD to inflation period,
the scattering amplitude receives the phases
that differ much from unity.
Then the wave function (\ref{chiRD}) becomes
\beqa
\chi_{\rm RD} &\sim&
\frac{1}{2k^4 \eta_1^2\eta_1^{\prime 2}} \cos{(k(\eta'_4-\eta'_3))} \frac{-2i}{\sqrt{2k}} \sin{(k\eta)} \\
&=& -(k\eta'_4)^{-2}\cos{(k(\eta'_4-\eta'_3))} \cdot
\frac{1}{\sqrt{2k}}\frac{i \eta_4^{' 2}}{k^2 \eta_1^2\eta_1^{' 2}} \sin{(k\eta)} \ ,
\label{chiRD_DIch_tail}
\eeqa
where the last factor represents the wave function 
for $k < |\eta'_4|^{-1}$,
which involves the enhancement factor $(\eta'_4/\eta'_1)^2$ caused by the pre-inflation,
as shown in (\ref{chiIRRDchasm}).
Eq.~(\ref{chiRD_DIch_tail}) involves an extra factor, $-(k\eta'_4)^{-2}\cos (k(\eta'_4-\eta'_3))$.
Hence, the integrand for the energy and pressure densities
in $|\eta'_4|^{-1} < k < |\eta'_1|^{-1}$
receive the extra factor $(k\eta'_4)^{-4}\cos^{2} (k(\eta'_4-\eta'_3))$,
compared to the one in the IR region $k <|\eta'_4|^{-1} $.
This extra factor decreases as $k^{-4}$,
and oscillates with the period $\pi/|\eta'_4-\eta'_3|=\pi/|2\eta'_4|$.

Figure~\ref{fig:rhopindRD_DIc} shows the integrands $\rho(k)$ and $p(k)$.
The parameters are taken to be the same as those in the third row of Figure~\ref{fig:rhopintegrandRD}.
From the lower figures, one can see that the large IR peak 
lies in $0<k<\frac{\pi}{2}|\eta'_4|^{-1}$,
but the small modification of the integrand continues until $k \sim 5|\eta'_1|^{-1}$.
The upper figures show that, above $k \sim 5|\eta'_1|^{-1}$, the integrands reduce to the ones
in the third row of Figure~\ref{fig:rhopintegrandRD}.
If the  background geometry is smoothly connected,
the UV tail of $\rho(k)$ and $p(k)$ will  decrease more rapidly above $k=|\eta'_1|^{-1}$.
However, the energy and pressure densities in the region $k \in [|\eta'_4|^{-1}, |\eta'_1|^{-1}]$
behave quite non-trivially.
Due to the inequality $|\eta'_1|^{-1}  \gg |\eta'_4|^{-1}$ from (\ref{ep1ep4}), 
this region is very large and the total energy and pressure receive substantial modifications.

\begin{figure}
\begin{center}
\begin{minipage}{.45\linewidth}
\includegraphics[width=1.1 \linewidth]{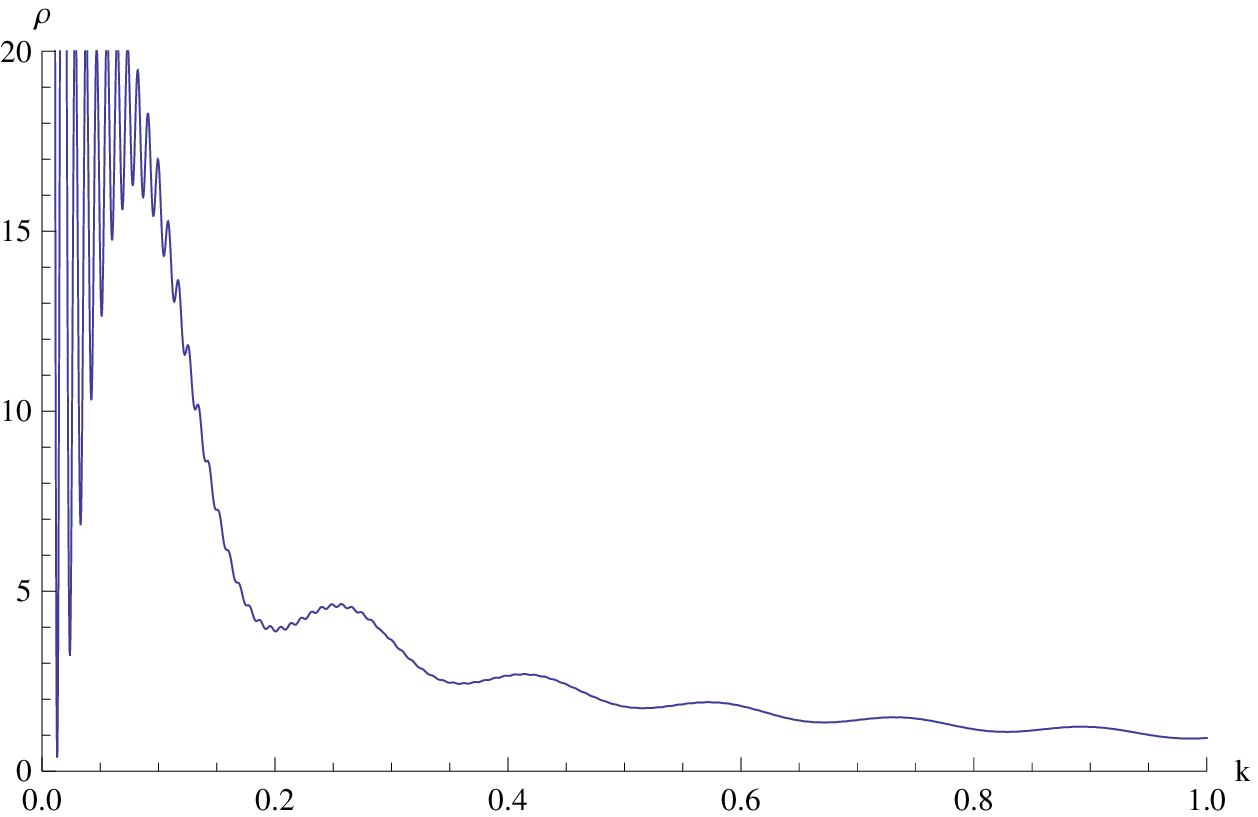}
  \end{minipage}
  \hspace{1.0pc}
\begin{minipage}{.45\linewidth}
\includegraphics[width=1.1 \linewidth]{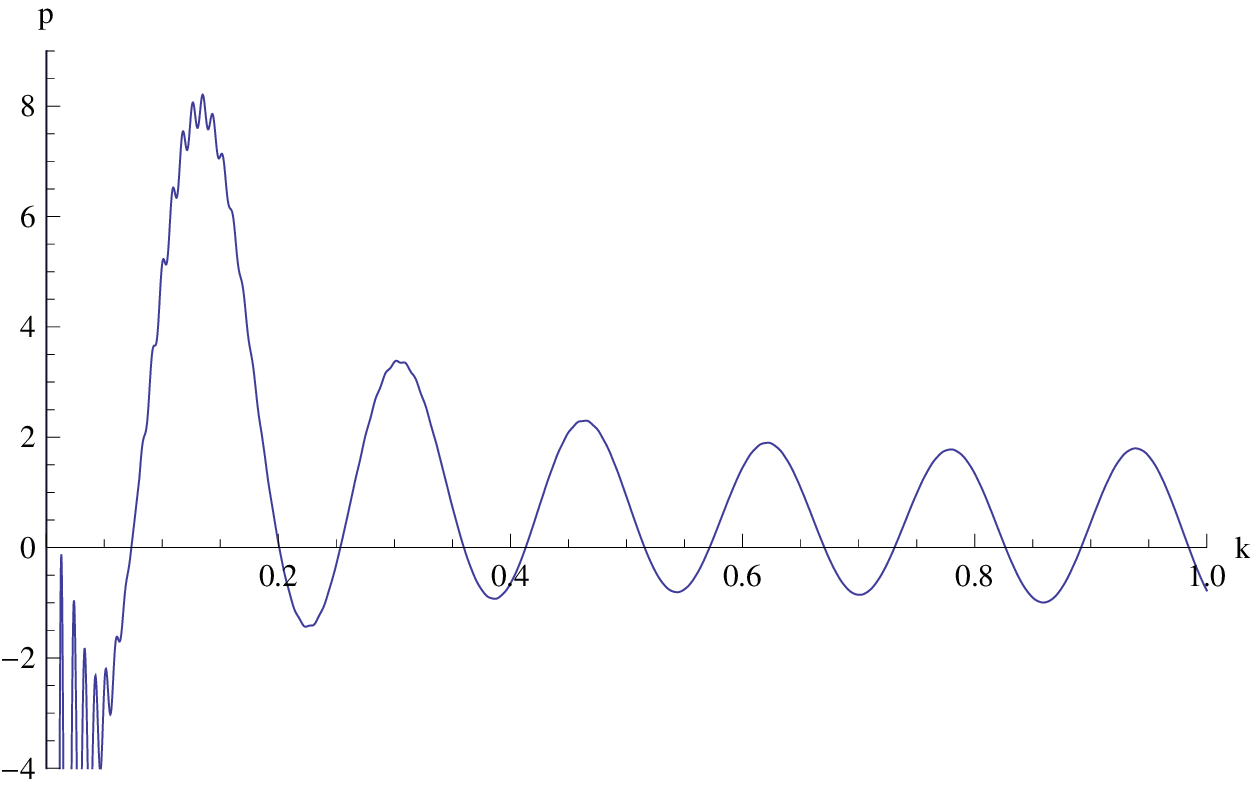}
  \end{minipage}
\begin{minipage}{.45\linewidth}
\includegraphics[width=1.1 \linewidth]{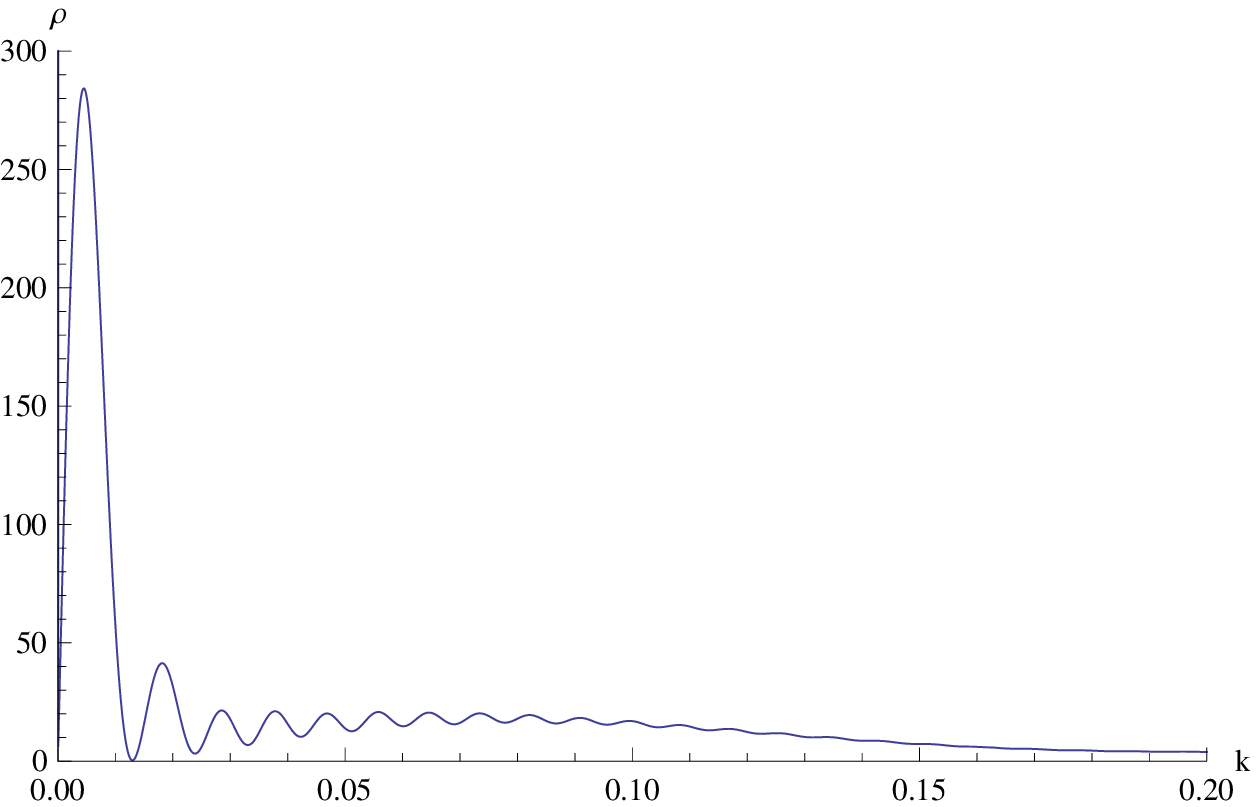}
  \end{minipage}
  \hspace{1.0pc}
\begin{minipage}{.45\linewidth}
\includegraphics[width=1.1 \linewidth]{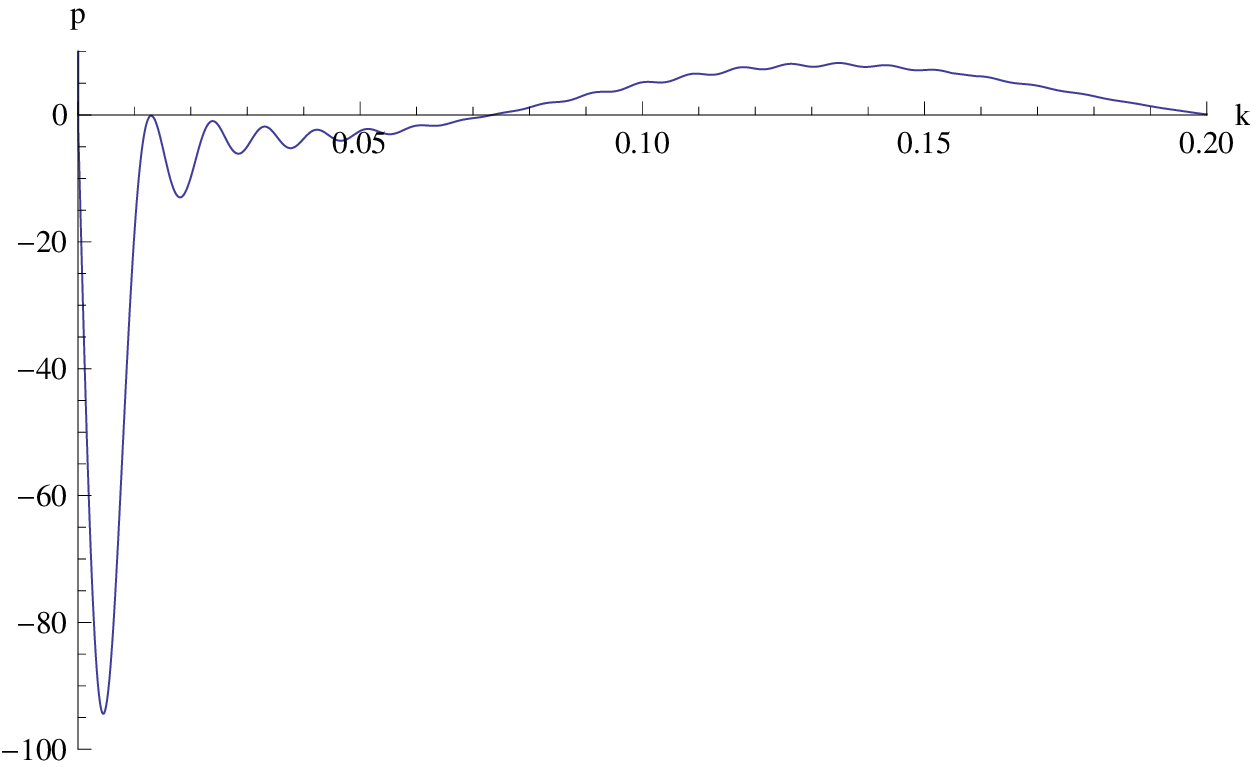}
  \end{minipage}
\end{center}
\caption{Integrands $\rho(k)$ and $p(k)$ in the RD period,
caused by the double inflation model with the intermediate RD period.
As in the third row of Figure~\ref{fig:rhopintegrandRD},
 $\eta_2=-\eta_1=1$ and $\eta=20$ are taken.
The other parameters are set to be
$\eta'_2=-\eta'_1=40$ and $\eta'_3=-\eta'_4=160$.
In the upper figures, the scales of the axes are taken to be same as those in 
the third row of Figure~\ref{fig:rhopintegrandRD}, while
in the lower figures the IR region is magnified.
}
\label{fig:rhopindRD_DIc}
\end{figure}

We finally discuss natural values of the parameters in the model.
If we set $|\eta'_4| \sim \eta_0$,
where $\eta_0$ is the present time,
the large enhancement of (\ref{rhoIRdi}) and (\ref{pIRdi})
continues until $k \sim |\eta'_4|^{-1} \sim \eta_0^{-1}$.
Then the present energy density becomes the desired value
of the order $\rho_0 \sim (M_P H_0)^2$, if we set $H_P \sim M_P$.
However, the modification of the wave function continues until 
$k \sim |\eta'_1|^{-1} > 1.7\times 10^{2} |\eta'_4|^{-1} \sim 1.7\times 10^{2} \eta_0^{-1} $.
Since they are deep within the current horizon, they could be detected by observations.
For instance,
they may generate the CMB fluctuations,
which contradict the observations.

If we set $|\eta'_1|^{-1}< \eta_0^{-1}$ instead,
the modes within the current horizon are not modified by the pre-inflation at all.
However,
the contribution of the large IR peak to the energy density 
terminates at a too small value of $k$,
giving
\beq
\rho^{\rm IRpeak}_0
\sim \frac{H_P^2}{8\pi^2 a^2_0} 2 \int_0^{\pi |4\eta'_4|^{-1}} dk~k 
=\frac{H_P^2}{8\pi^2 a^2_0} \left(\frac{\pi}{4|\eta'_4|}\right)^{2}
<\frac{1}{512} H_P H_I H_0^2 \ ,
\label{rho0chasm}
\eeq
where we have approximated that the IR peak is located at $\pi |4\eta'_4|^{-1}$
and has form of an isosceles triangle with the slope given by (\ref{rhoIRdi}). 
In the last inequality,
$|\eta'_4|^{-1} = \sqrt{\frac{H_I}{H_P}} |\eta'_1|^{-1}$, $|\eta'_1|^{-1}<\eta_0^{-1}$,
and $H_0=2(a_0 \eta_0)^{-1}$ have been used.
While it is enhanced from  (\ref{rhopreval}),
it is still smaller than the desired value $(M_P H_0)^2$.

The above problem has been caused by the modes with $|\eta'_4|^{-1} <k< |\eta'_1|^{-1}$.
The present model has the constraint $|\eta'_1|^{-1} \gg |\eta'_4|^{-1} $
given by  (\ref{ep1ep4}).
In other models with  $|\eta'_1|^{-1} \sim |\eta'_4|^{-1}$,
such as the one studied in Section~\ref{sec:plateaumodel},
the problem is resolved.

\section{IR behaviors of the wave functions}
\label{sec:IRbhvwf}
\setcounter{equation}{0}

In this appendix, we examine  IR behaviors of the wave functions
and confirm that the IR form (\ref{u_IR_scaleinv}) holds in all the periods in the cosmic history.

\subsection{Single inflation model}
\label{sec:wfIRMD}

We first study the case in the MD period in the single inflation model,
studied in Section~\ref{sec:wavefunc}.
Eq. (\ref{ABCD}) is rewritten as
\beqa
\begin{pmatrix} C \cr D \end{pmatrix}
&=&\frac{-1}{2k^2 \eta_4^2}  \left( \begin{array}{c}
\left[1-e^{2ik\eta_4} \sum_{n=3}^{\infty}\frac{1}{n!}(-2ik\eta_4)^n \right] e^{-ik(\eta_4+\eta_3)} \\
e^{-ik(\eta_4+\eta_3)}
\end{array}\right. \n
&&\hspace{80pt} \left. \begin{array}{c}
e^{ik(\eta_4+\eta_3)} \\
\left[1-e^{-2ik\eta_4} \sum_{n=3}^{\infty}\frac{1}{n!}(2ik\eta_4)^n \right] e^{ik(\eta_4+\eta_3)}
\end{array}\right)
\begin{pmatrix} A \cr B \end{pmatrix}  \ , \n
\label{ABCDrew} 
\eeqa
where the terms with $k^1$ and $k^2$ cancel in the square bracket.
Plugging in (\ref{ABrew}), one obtains
\beqa
C&=& \frac{1}{4k^4\eta_1^2\eta_4^2} 
\Bigg(\left[1-e^{-2ik\eta_1}\sum_{n=3}^\infty \frac{1}{n!}(2ik\eta_1)^n \right]
\left[1-e^{2ik\eta_4} \sum_{n=3}^{\infty}\frac{1}{n!}(-2ik\eta_4)^n \right] \n
&&\times e^{ik(\eta_1+\eta_2-\eta_3-\eta_4)}  
-e^{-ik(\eta_1+\eta_2-\eta_3-\eta_4)} \Bigg) \ , 
\eeqa
\beqa
D&=& \frac{1}{4k^4\eta_1^2\eta_4^2} 
\Bigg(\left[1-e^{-2ik\eta_1}\sum_{n=3}^\infty \frac{1}{n!}(2ik\eta_1)^n \right] e^{ik(\eta_1+\eta_2-\eta_3-\eta_4)} \n
&&-\left[1-e^{-2ik\eta_4} \sum_{n=3}^{\infty}\frac{1}{n!}(2ik\eta_4)^n \right] e^{-ik(\eta_1+\eta_2-\eta_3-\eta_4)} \Bigg) \ . 
\eeqa

Expanding in terms of $k$, one finds
\beqa
C&=& \frac{\eta_1+\eta_2-\eta_3-\eta_4}{2\eta_1^2 \eta_4^2} i k^{-3} 
+{\cal O}(k^{-1}) \ , \label{Cexp}\\
D&=& \frac{\eta_1+\eta_2-\eta_3-\eta_4}{2\eta_1^2 \eta_4^2} i k^{-3} 
+{\cal O}(k^{-1}) \ , \label{Dexp}
\eeqa
where the leading terms with $k^{-4}$ and also the terms with $k^{-2}$ cancel.
The coefficients of $C$ and $D$ coincide at low powers of $k$,
and begin to differ at the term with $k^{0}$ as
\beq
C-D = \frac{(2\eta_1+2\eta_2-2\eta_3+\eta_4)\eta_4}{3\eta_1^2}
+{\cal O}(k^{2}) \ , \label{C-Dexp}
\eeq
where the terms with $k^1$ also cancel.
By using the matching relations  (\ref{eta1eta2}) and (\ref{eta3eta4}), 
the first term in (\ref{C-Dexp}) vanishes, and the second term gives
\beq
C-D = \left(-\frac{4\eta_2 \eta_4}{9}+\frac{\eta_4^4}{180\eta_2^2} \right)k^2 
+{\cal O}(k^{3}) \ . \label{C-Dexp2}
\eeq

The wave function (\ref{chiMD}) is written as
\beq
\chi_{\rm MD}=\frac{C+D}{2}\left(\chi_{\rm BD}+\chi_{\rm BD}^*\right)
+\frac{C-D}{2}\left(\chi_{\rm BD}-\chi_{\rm BD}^*\right) \ ,
\label{chiMDre}
\eeq 
where
\beqa
\chi_{\rm BD}+\chi_{\rm BD}^* &=& \frac{2}{\sqrt{2k}}\left(\cos{(k\eta)}-\frac{\sin{(k\eta)}}{k\eta}\right) 
\label{chiIpchiIst}\\
&=&\frac{2}{\sqrt{2k}}\left(-\frac{1}{3}(k\eta)^2 +{\cal O}((k\eta)^{4})\right) \ , 
\label{chiMpIR}
\eeqa
\beqa
\chi_{\rm BD}-\chi_{\rm BD}^* 
&=& \frac{-2i}{\sqrt{2k}}\left(\sin{(k\eta)}+\frac{\cos{(k\eta)}}{k\eta}\right) \label{chiBD-st1} \\
&=&\frac{-2i}{\sqrt{2k}}\left((k\eta)^{-1} +{\cal O}(k\eta)\right) \ .
\label{chiBD-st2}
\eeqa

Then the first and second terms in (\ref{chiMDre}) behave as
$(C+D)(\chi_{\rm BD}+\chi_{\rm BD}^*) \sim k^{-3}\cdot k^{3/2} = k^{-3/2}$ and 
$(C-D)(\chi_{\rm BD}-\chi_{\rm BD}^*) \sim k^{2}\cdot k^{-3/2} = k^{1/2}$ in the IR limit.
From (\ref{Cexp}), (\ref{Dexp}), and (\ref{chiMpIR}),
the next-to-leading-order term in $(C+D)(\chi_{\rm BD}+\chi_{\rm BD}^*)$ has $k^{1/2}$.
Hence, (\ref{chiMDre}) behaves as
\beq
\chi_{\rm MD}^{\rm IR} = 
\frac{\eta_1+\eta_2-\eta_3-\eta_4}{2\eta_1^2 \eta_4^2} i k^{-3}
 \cdot
\frac{2}{\sqrt{2k}}\left(-\frac{1}{3}\right)(k\eta)^2
+{\cal O}(k^{1/2})
\label{chiMDIR}
\eeq
in the IR region.
The leading term in (\ref{chiMDIR}) gives
\beq
u^{\rm IR}_{\rm MD}=\frac{\eta_1+\eta_2-\eta_3-\eta_4}{2\eta_1^2 \eta_4^2} i k^{-3}\cdot
\frac{2}{\sqrt{2k}}\left(-\frac{1}{3}\right)(k\eta)^2\cdot
\frac{2\eta_4 H_I \eta_1^2}{\eta^2}
=\frac{i}{\sqrt{2}}H_I k^{-3/2} \ ,
\eeq
where (\ref{uchi4}) with (\ref{a_caldera}), (\ref{macoa1pa2p}), and 
(\ref{macoa3pa4p}) has been used in the first equality.
In the second equality, the matching relations
(\ref{eta1eta2}) and (\ref{eta3eta4}) have been used.
The same result with (\ref{u_IR_scaleinv}) is obtained again
in the MD region.
Note that the leading IR behavior $k^{-3/2}$ is much better than
the naive estimation $k^{-11/2}$ mentioned above (\ref{ABrew}).

\subsection{Double inflation model with the intermediate CD stage}
\label{sec:wfIRplateau}

We next study the case in the double inflation model with the intermediate  CD stage,
studied in Section~\ref{sec:plateaumodel}.

In the pre-inflation period, the wave function (\ref{chiPI})
behaves as (\ref{u_IR_doubleinf}) in the IR region,
as can be seen by the same argument in Section~\ref{sec:wavefunc}.
In the CD period, by expanding (\ref{AtBt}) in terms of $k$, one obtains
\beqa
\tilde A &=& \left( \frac{i}{4}\frac{k}{\gamma} 
+ {\cal O}(k^2) \right) e^{\gamma \tilde\eta_2} \ , \\
\tilde B &=& \left( i\frac{\gamma}{k} +\frac{i}{4}(1+2\gamma\tilde\eta_2)\frac{k}{\gamma} 
+ {\cal O}(k^2) \right) e^{-\gamma \tilde\eta_2} \ .
\eeqa
Then the wave function (\ref{chiCD}) becomes
\beq
\chi^{\rm IR}_{\rm CD} 
= i\frac{\gamma}{k} \frac{1}{\sqrt{2k}} e^{\gamma(\eta-\tilde\eta_2)} + {\cal O}(k^{1/2}) \ .
\eeq
The leading term gives
\beq
u^{\rm IR}_{\rm CD} =  
i\frac{\gamma}{k} \frac{1}{\sqrt{2k}} e^{\gamma(\eta-\tilde\eta_2)} \frac{1}{e^{\gamma \eta}}
=\frac{i}{\sqrt{2}} H_P k^{-3/2} \ ,
\eeq
where  (\ref{uchi4}) with (\ref{a_doubleinf_pla}) has been used in the first equality.
In the second equality, (\ref{macoa1a2_t}) and (\ref{et1et4}) have been used.
Eq.~(\ref{u_IR_doubleinf}) is obtained again.

In the subsequent inflation period, 
by plugging (\ref{AtBt}) into (\ref{AtBtCtDt}),
and expanding in terms of $k$, one obtains
\beqa
\tilde C +\tilde D&=&
\frac{i}{2}\frac{\gamma}{k}\left(e^{\gamma(\tilde\eta_3-\tilde\eta_2)} -e^{-\gamma(\tilde\eta_3-\tilde\eta_2)}\right)
+{\cal O}(k^0) \ , \\
\tilde C - \tilde D&=& e^{\gamma(\tilde\eta_3-\tilde\eta_2)} + {\cal O}(k^2) \ . 
\eeqa
Then, when the wave function (\ref{chiInf2})
is rewritten as in (\ref{chiMDre}), with $C$ and $D$ replaced by $\tilde C$ and $\tilde D$,
the second term  dominates over the first term, giving
\beq
\chi^{\rm IR}_{\rm Inf} 
=\frac{1}{2}  e^{\gamma(\tilde\eta_3-\tilde\eta_2)} \frac{-2i}{\sqrt{2k}}(k\eta)^{-1}
+{\cal O}(k^{1/2}) \ .
\eeq
The leading term gives
\beq
u^{\rm IR}_{\rm Inf} =  
\frac{1}{2}  e^{\gamma(\tilde\eta_3-\tilde\eta_2)} \frac{-2i}{\sqrt{2k}}(k\eta)^{-1} \frac{1}{-1/(H_I \eta)}
=\frac{i}{\sqrt{2}} H_P k^{-3/2} \ ,
\eeq
where  (\ref{uchi4}) with (\ref{a_doubleinf_pla}) has been used in the first equality,
and (\ref{et2et3}) in the second equality.
Then eq.~(\ref{u_IR_doubleinf}) is obtained again.

In the subsequent RD period,
by plugging (\ref{AtBt}) and (\ref{AtBtCtDt}) into (\ref{ABCpDp}),
and expanding in terms of $k$, one obtains
\beqa
A+B&=&\frac{i}{6} \eta_2 k\left[(4-3\gamma\eta_2)e^{\gamma(\tilde\eta_3-\tilde\eta_2)} 
+3\gamma\eta_2e^{-\gamma(\tilde\eta_3-\tilde\eta_2)}\right]
+{\cal O}(k^2) \ , \\
A-B&=& -(\eta_2 k)^{-2} e^{\gamma(\tilde\eta_3-\tilde\eta_2)}
+{\cal O}(k^0) \ .
\eeqa
Then, in the wave function (\ref{chiRD}), rewritten as 
\beqa
\chi_{\rm RD} &=& \frac{A+B}{2}(\chi_{\rm PW} + \chi^*_{\rm PW})
+\frac{A-B}{2}(\chi_{\rm PW} - \chi^*_{\rm PW})  \label{chiRDrew1} \\
&=&\frac{A+B}{2} \frac{2}{\sqrt{2k}} \cos{(k\eta)}
+\frac{A-B}{2} \frac{-2i}{\sqrt{2k}} \sin{(k\eta)} \ , \label{chiRDrew2}
\eeqa
the second term dominates over the first term, giving
\beq
\chi_{\rm RD}^{\rm IR} 
= -\frac{1}{2} (\eta_2 k)^{-2} e^{\gamma(\tilde\eta_3-\tilde\eta_2)} 
\frac{-2i}{\sqrt{2k}} k\eta +{\cal O}(k^{1/2}) \ .
\eeq
The leading term gives
\beq
u^{\rm IR}_{\rm RD} =  
-\frac{1}{2} (\eta_2 k)^{-2} e^{\gamma(\tilde\eta_3-\tilde\eta_2)} 
\frac{-2i}{\sqrt{2k}} k\eta 
 \frac{1}{\al \eta}
=\frac{i}{\sqrt{2}} H_P k^{-3/2} \ ,
\eeq
where  (\ref{uchi4}) with (\ref{a_caldera}) has been used in the first equality.
In the second equality,
(\ref{macoa1pa2p}), (\ref{eta1eta2}),
and (\ref{et2et3}) have been used.
Eq.~(\ref{u_IR_doubleinf}) is obtained again.

In the subsequent MD period, 
by plugging (\ref{AtBt}), (\ref{AtBtCtDt}), and (\ref{ABCpDp}) into (\ref{ABCD}),
and expanding in terms of $k$, one obtains
\beqa
C+D &=& -\frac{3i}{2\eta_2^2 \eta_4}e^{\gamma(\tilde\eta_3-\tilde\eta_2)}~k^{-3}
+{\cal O}(k^{-1}) \ , \\
C-D &=& \frac{\eta_4}{180 \eta_2^2 \gamma^3} 
\Big[\left(60(\gamma\eta_2)^4-80(\gamma\eta_2)^3+(\gamma\eta_4)^3\right)e^{\gamma(\tilde\eta_3-\tilde\eta_2)} \n
&&-60(\gamma\eta_2)^4e^{-\gamma(\tilde\eta_3-\tilde\eta_2)}\Big] k^2 
+{\cal O}(k^{3}) \ .
\eeqa
Then, when the wave function (\ref{chiMD}) is written as (\ref{chiMDre}),
the first term dominates over the second term, giving
\beq
\chi^{\rm IR}_{\rm MD} = \frac{1}{2} \frac{-3i}{2\eta_2^2 \eta_4}e^{\gamma(\tilde\eta_3-\tilde\eta_2)}~k^{-3}
\frac{2}{\sqrt{2k}}\left(-\frac{1}{3}\right)(k\eta)^2
+{\cal O}(k^{1/2}) \ .
\eeq
The leading term gives
\beq
u^{\rm IR}_{\rm MD} = 
\frac{1}{2} \frac{-3i}{2\eta_2^2 \eta_4}e^{\gamma(\tilde\eta_3-\tilde\eta_2)}~k^{-3}
\frac{2}{\sqrt{2k}}\left(-\frac{1}{3}\right)(k\eta)^2
 \frac{1}{\beta\eta^2}
=\frac{i}{\sqrt{2}} H_P k^{-3/2} \ ,
\eeq
where  (\ref{uchi4}) with (\ref{a_caldera}) has been used in the first equality.
In the second equality, we have used (\ref{et2et3}) and
$2\be\eta_2^2\eta_4=H_I^{-1}$, which is obtained 
by (\ref{macoa1pa2p}), (\ref{macoa3pa4p}), and (\ref{eta1eta2}).
Then eq.~(\ref{u_IR_doubleinf}) is obtained again.

\subsection{Double inflation model with the intermediate RD stage}
\label{sec:wfIRchasm}

We finally  study the case in the double inflation model with the intermediate RD stage,
studied in Appendix \ref{sec:chasmmodel}.

In the pre-inflation and pre-RD periods, 
the wave functions (\ref{chiPI}) and ({\ref{chiPR})
behave as (\ref{u_IR_doubleinf}) in the IR region,
as can be seen by the same calculations in Section~\ref{sec:wavefunc}.

In the inflation period, 
since (\ref{ApBpCpDp}) has the same form as (\ref{ABCD}) with primes added,
the coefficients $C'$ and $D'$ behave again  as
(\ref{Cexp}), (\ref{Dexp}), and (\ref{C-Dexp}) in the IR regions.
However, because of the matching relations (\ref{ep1ep2}) and (\ref{ep3ep4}),
the leading term in (\ref{Cexp}) and (\ref{Dexp}) vanishes,
but that in (\ref{C-Dexp}) does not vanish in this case.
Then the second term dominates over the first term in (\ref{chiMDre}).
The leading term gives 
\beq
u^{\rm IR}_{\rm Inf} =
\frac{1}{2}
\frac{(2\eta_1+2\eta_2-2\eta_3+\eta_4)\eta_4}{3\eta_1^2}
\frac{-2i}{\sqrt{2k}} (k\eta)^{-1}  (-H_I \eta)
=\frac{i}{\sqrt{2}}H_P k^{-3/2} \ ,
\eeq
where (\ref{C-Dexp}), (\ref{chiBD-st2}), and (\ref{uchi4}) with (\ref{a_preinf})
have been used in the first equality.
In the second equality,
the matching relations (\ref{ep1ep2}), (\ref{ep3ep4}), and (\ref{ep1ep4}) have been used.
Eq.~(\ref{u_IR_doubleinf}) is obtained again.
 
In the subsequent RD period, 
by plugging (\ref{ApBp}) and (\ref{ApBpCpDp}) into (\ref{ABCpDp}),
and expanding in terms of $k$, one obtains
\beq
A-B=\frac{2\eta_1^3(\eta'_1+\eta'_2-\eta'_3-\eta'_4)-\eta_4^{'3}(2\eta'_1+2\eta'_2-2\eta'_3+\eta'_4)}
{3 \eta_1^2 \eta_1^{\prime 2} \eta_4^{\prime 2} } k^{-2} +{\cal O}(k^0) \ ,
\label{A-B_chasm}
\eeq
\beqa
A+B&=&\frac{-1}{3\eta_1^2\eta_1^{'2} \eta_4^{'2}}
\Big[\eta_1^3(\eta_1-2\eta_2)(\eta'_1+\eta'_2-\eta'_3-\eta'_4) \n
&&+(\eta_1+\eta_2)(2\eta'_1+2\eta'_2-2\eta'_3+\eta'_4)\eta_4^{'3}\Big] i k^{-1}
+{\cal O} (k^1) \ .
\label{A+B_chasm}
\eeqa
The leading terms with $k^{-6}$ to $k^{-3}$ have canceled,
which can be checked, for instance, by rewriting  (\ref{ApBp}), (\ref{ApBpCpDp}), and (\ref{ABCpDp})
as in  (\ref{ABrew}) and (\ref{ABCDrew}).
Substituting the matching relations (\ref{eta1eta2}), (\ref{ep1ep2}), and (\ref{ep3ep4}),
Eqs.~(\ref{A-B_chasm}) and (\ref{A+B_chasm}) become
\beqa
A-B &=& -\frac{\eta_4^{'2}}{\eta_1^2\eta_1^{'2}} k^{-2} + {\cal O}(k^0) \ , \\
A+B &=&  {\cal O}(k^1) \ .
\eeqa
Then, when the wave function (\ref{chiRD}) is rewritten as (\ref{chiRDrew2}), 
the second term dominates over the first term and gives
\beq
\chi^{\rm IR}_{\rm RD} = \frac{1}{2} \frac{-\eta_4^{'2}}{\eta_1^2\eta_1^{'2}} k^{-2}
 \frac{-2i}{\sqrt{2k}} k\eta + {\cal O}(k^{1/2})  \ .
 \label{chiIRRDchasm}
\eeq 
The leading term gives
\beq
u^{\rm IR}_{\rm RD} =  \frac{1}{2} \frac{-\eta_4^{'2}}{\eta_1^2\eta_1^{'2}} k^{-2}
 \frac{-2i}{\sqrt{2k}} k\eta \frac{1}{\alpha \eta}
=\frac{i}{\sqrt{2}} H_P k^{-3/2} \ ,
\label{uIR_chasm}
\eeq
where  (\ref{uchi4}) with (\ref{a_caldera}) has been used in the first equality.
In the second equality, (\ref{macoa1pa2p}) and (\ref{ep1ep4}) 
have been used.
Eq.~(\ref{u_IR_doubleinf}) is obtained again.

In the subsequent MD period, 
by plugging (\ref{ApBp}), (\ref{ApBpCpDp}), and (\ref{ABCpDp}) into (\ref{ABCD}),
and expanding in terms of $k$, one obtains
\beqa
C+D&=& \frac{1}{3\eta_1^2\eta_4^2\eta_1^{'2}\eta_4^{'2}}\Big[
\eta_1^3(\eta_1-2\eta_2+2\eta_3+2\eta_4)(\eta'_1+\eta'_2-\eta'_3-\eta'_4) \n
&&\hspace{50pt}+(\eta_1+\eta_2-\eta_3-\eta_4)(2\eta'_1+2\eta'_2-2\eta'_3+\eta'_4)\eta_4^{'3}
\Big]i k^{-3} \n 
&&+ {\cal O}(k^{-1})  \ , 
\label{C+D_chasm}
\eeqa
\beqa
C-D &=& \frac{\eta_4}{9\eta_1^2\eta_1^{'2} \eta_4^{'2}}\Big[
2\eta_1^3(\eta_1-2\eta_2+2\eta_3-\eta_4)(\eta'_1+\eta'_2-\eta'_3-\eta'_4) \n
&&\hspace{50pt}+(2\eta_1+2\eta_2-2\eta_3+\eta_4)(2\eta'_1+2\eta'_2-2\eta'_3+\eta'_4)\eta_4^{'3}
\Big] k^{0} \n
&&+ {\cal O}(k^{2})  \ .
\label{C-D_chasm}
\eeqa
Substituting the matching relations (\ref{eta1eta2}), (\ref{eta3eta4}), (\ref{ep1ep2}), and (\ref{ep3ep4}),
Eqs.~(\ref{C+D_chasm}) and (\ref{C-D_chasm}) become
\beqa
C+D&=& -\frac{3\eta_4^{'2}}{2\eta_1^2\eta_4\eta_1^{'2}} i k^{-3} + {\cal O}(k^{-1})  \ , \\
C-D&=&{\cal O}(k^{2})  \ .
\eeqa
Then in the wave function (\ref{chiMDre}), the first term dominates over the second term, giving
\beq
\chi^{\rm IR}_{\rm MD} = \frac{1}{2}  \frac{-3\eta_4^{'2}}{2\eta_1^2\eta_4\eta_1^{'2}} i k^{-3} 
\frac{2}{\sqrt{2k}}\left(-\frac{1}{3}\right)(k\eta)^2
+{\cal O}(k^{1/2}) \ .
\eeq
The leading term gives
\beq
u^{\rm IR}_{\rm MD} = \frac{1}{2}  \frac{-3\eta_4^{'2}}{2\eta_1^2\eta_4\eta_1^{'2}} i k^{-3} 
\frac{2}{\sqrt{2k}}\left(-\frac{1}{3}\right)(k\eta)^2 \frac{1}{\beta\eta^2}
=\frac{i}{\sqrt{2}} H_P k^{-3/2} \ ,
\label{uIRMDcha}
\eeq
where  (\ref{uchi4}) with (\ref{a_caldera}) has been used in the first equality.
In the second equality, we have used (\ref{ep1ep4}) and
$2\be\eta_1^2\eta_4=H_I^{-1}$,
which is obtained by (\ref{macoa1pa2p}) and (\ref{macoa3pa4p}) .
Then eq.~(\ref{u_IR_doubleinf}) is obtained again.


\end{document}